\documentclass[10pt,a4paper,onecolumn,oneside]{article}

\usepackage{utils}
\usepackage{multirow}
\usepackage[normalem]{ulem}
\usepackage{lineno}
\usepackage{comment}

\usepackage[UKenglish]{babel}

\usepackage{natbib}

\setcitestyle{round,comma,authoryear,aysep={},yysep={,}}

\usepackage[nottoc]{tocbibind}

\usepackage{authblk}

\usepackage{tikz}
\usetikzlibrary{arrows,decorations,shapes,automata,positioning,decorations.pathmorphing,calc,patterns,backgrounds}

\pgfdeclaredecoration{penciline}{initial}{
  \state{initial}[width=+\pgfdecoratedinputsegmentremainingdistance,
  auto corner on length=1mm,]{
    \pgfpathcurveto%
    {
        \pgfqpoint{\pgfdecoratedinputsegmentremainingdistance}
                  {\pgfdecorationsegmentamplitude}
    }
    {
    \pgfmathrand
    \pgfpointadd{\pgfqpoint{\pgfdecoratedinputsegmentremainingdistance}{0pt}}
                {\pgfqpoint{-\pgfdecorationsegmentaspect
                 \pgfdecoratedinputsegmentremainingdistance}%
                           {\pgfmathresult\pgfdecorationsegmentamplitude}
                }
    }
    {
    \pgfpointadd{\pgfpointdecoratedinputsegmentlast}{\pgfpoint{1pt}{1pt}}
    }
  }
  \state{final}{}
}

\tikzset{
  every picture/.style = {
    thick,
    ->,
    >=stealth',
  }
  ,
  within/.style = {
    fill = white,
    inner sep = 2pt
  }
  ,
  topicclass/.style = {
    -,
    draw=\colorfilltopicclass!75!black,
    fill=\colorfilltopicclass,
    fill opacity=0.2
  }
  ,
  cross line/.style={
    preaction = {
      draw=white,
      -,
      line width=4pt}
  }
  ,
  frame rectangle/.style = {
    framed,
    draw=black,
    rounded corners,
  }
  ,
  mundo/.style = {
    rectangle,
    rounded corners = 5,
    draw = black!100,
    fill = lightgray!40,
    inner sep = 4pt,
    minimum size = 12pt,
    font = \footnotesize,
  }
  ,
  etiqMundo/.style = {
    font = \scriptsize,
    outer sep = 0pt,
    inner sep = 1pt
  }
  ,
  flecha/.style = {
    black,
    solid
  }
  ,
  etiqFlecha/.style = {
    font = \scriptsize
  }
}

\newcommand{\colorfilltopicclass}{yellow}

\newcommand{\ladocubo}{0.2em}

\usepackage{hyperref}

\newcommand{\titulo}{Communication between agents in dynamic epistemic logic}
\newcommand{\autor}{Fernando R. Vel{\'a}zquez-Quesada}

\definecolor{DarkBlue}{RGB}{0,0,192}

\newcommand{\colorlink}{blue}
\newcommand{\colorurl}{Emerald}
\newcommand{\colorcite}{RedOrange}

\hypersetup{
  pdftitle          = {\titulo},
  pdfauthor         = {\textcopyright \autor},
  pdfstartview      = {FitH},
  pdfpagelayout     = {OneColumn},
  bookmarksnumbered = {true},
  naturalnames      = {true},
  colorlinks        = {true},           
  linkcolor         = \colorlink,       
  urlcolor          = \colorurl,        
  citecolor         = \colorcite        
}

\usepackage[hyperpageref]{backref}

\renewcommand*{\backref}[1]{}

\renewcommand*{\backrefalt}[4]{
  \ifcase #1
    No cited.
  \or
    Cited on page \textbf{#2}.
  \else
    Cited on pages \textbf{#2}.
  \fi
}

\addto\extrasUKenglish{%

}

\newcommand{\autorefp}[1]{(\autoref{#1})}

\newcommand{\itemfautoref}[2]{\autoref{#1}\hyperref[#2]{.\itmformat \ref{#2}}\xspace}
\newcommand{\itemfautorefp}[2]{(\itemfautoref{#1}{#2})}

\usepackage{enumitem}

\newcommand{\itmformat}{\bfseries \itshape}

\setlist[enumerate,1]{label={\itmformat (\roman*)}}
\setlist[enumerate,2]{label*={\itmformat \arabic*}}

\newlist{todolist}{enumerate}{2}
\setlist[todolist,1]{label=\arabic*.-}

\newlist{inlineenum}{enumerate*}{1}
\setlist[inlineenum,1]{label={\itmformat (\roman*)}}

\newlist{compactenumerate}{enumerate}{1}
\setlist[compactenumerate,1]{label={\itmformat (\roman*)}, leftmargin=1.5em, rightmargin=0em, topsep=0.25em, itemsep=0.25em}

\newlist{enumeratetr}{enumerate}{1}
\setlist[enumeratetr,1]{label={\itmformat ($\tr$\arabic*)}, leftmargin=2.5em, rightmargin=0em, topsep=0.25em, itemsep=0.25em}

\newlist{inlineenumtr}{enumerate*}{1}
\setlist[inlineenumtr,1]{label={\itmformat ($\tr$\arabic*)}}

\setlist[itemize,1]{label=\textbullet}
\setlist[itemize,2]{label=--}

\newlist{compactitemize}{itemize}{3}
\setlist[compactitemize,1]{label=\textbullet, leftmargin=1.5em, rightmargin=0em, topsep=0.25em, itemsep=0.25em}
\setlist[compactitemize,2]{label=--, leftmargin=1.25em, rightmargin=0em, topsep=0.25em, itemsep=0.25em}
\setlist[compactitemize,3]{label=$\cdot$, leftmargin=0.25em, rightmargin=0em, topsep=0.25em, itemsep=0.25em}

\newlist{itemizecom}{enumerate}{1}
\setlist[itemizecom,1]{label=--, leftmargin=1em, rightmargin=0em, topsep=0.25em, itemsep=0.25em, before*=\small}

\renewcommand{\arraystretch}{1.3}

\newcommand{\comp}[1]{\ov{#1}}

\renewcommand{\ag}{\ensuremath{\formatagent{A}}\xspace}
\renewcommand{\aga}{\formatagent{a}}
\renewcommand{\agb}{\formatagent{b}}
\renewcommand{\agc}{\formatagent{c}}
\newcommand{\sag}{\ensuremath{\formatagent{G}}}
\newcommand{\sen}{\ensuremath{\formatagent{S}}}
\newcommand{\rec}{\ensuremath{\formatagent{R}}}

\newcommand{\opK}{\operatorname{K}}

\newcommand{\opgDist}{\operatorname{D}}

\newcommand{\marm}{\ensuremath{\mb{M}_{\ag}}\xspace}

\newcommand{\dom}[1]{\mathfrak{D}(#1)}
\newcommand{\R}{R}
\newcommand{\sbi}[1]{{#1}_{\agi}}
\newcommand{\sbj}[1]{{#1}_{\agj}}
\newcommand{\sbk}[1]{{#1}_{\agk}}
\newcommand{\sba}[1]{{#1}_{\aga}}
\newcommand{\sbb}[1]{{#1}_{\agb}}
\newcommand{\sbc}[1]{{#1}_{\agc}}
\newcommand{\sbDs}[2]{{#1}^{\opgDist}_{#2}}
\newcommand{\sbDg}[1]{{#1}^{\opgDist}_{\sag}}

\newcommand{\modK}[1]{\mathop{\opK_{#1}}}

\newcommand{\mmKi}[1]{\modK{\agi}#1}
\newcommand{\mmKj}[1]{\modK{\agj}#1}
\newcommand{\mmKa}[1]{\modK{\aga}#1}
\newcommand{\mmKb}[1]{\modK{\agb}#1}
\newcommand{\mmKc}[1]{\modK{\agc}#1}

\newcommand{\modDs}[1]{\mathop{\opgDist_{#1}}}
\newcommand{\mmDs}[2]{\modDs{#1}#2}
\newcommand{\modDg}{\modDs{\sag}}
\newcommand{\mmDg}[1]{\modDg#1}

\newcommand{\modDsr}[2]{\mathop{\opgDist^{#2}_{#1}}}
\newcommand{\mmDsr}[3]{\modDsr{#1}{#2}#3}

\newcommand{\fullig}[2]{{\not\sim^{#1}_{#2}}}
\newcommand{\knonfu}[2]{{\sim^{#1}_{#2}}}

\newcommand{\LAd}{\ensuremath{\LA_{\opgDist}}\xspace}

\newcommand{\LAde}[1]{\ensuremath{\LA_{\opgDist, \,#1}}\xspace}
\newcommand{\LAdEEE}{\LAde{\modEEE}}
\newcommand{\LAdSEE}{\LAde{\modSEE{\sen}}}
\newcommand{\LAdSSE}{\LAde{\modSSE{\sen}{\chi}}}

\renewcommand{\ax}[3]{\textnormal{\tsf{#1}}$^{#2}_{#3}$\xspace}

\renewcommand{\LO}{\ensuremath{\mathsf{L}}\xspace}

\newcommand{\dsrMP}{\ax{MP}{}{}}

\newcommand{\LOd}{\ensuremath{\LO_{\opgDist}}\xspace}
\newcommand{\LOde}[1]{\ensuremath{\LO_{\opgDist, \,#1}}\xspace}

\newcommand{\LOdEEE}{\LOde{\modEEE}}
\newcommand{\dsaEEE}[1]{\ax{A}{#1}{\symbEEE}}
\newcommand{\dsreEEE}{\ax{RE}{}{\symbEEE}}
\newcommand{\vdashLOdEEE}{\vdash}

\newcommand{\LOdSEE}{\LOde{\modSEE{\sen}}}
\newcommand{\dsaSEE}[1]{\ax{A}{#1}{\symbSEE{\sen}}}
\newcommand{\dsreSEE}{\ax{RE}{}{\symbSEE{\sen}}}
\newcommand{\vdashLOdSEE}{\vdash}

\newcommand{\LOdSSE}{\LOde{\modSSE{\sen}{\chi}}}
\newcommand{\dsaSSE}[1]{\ax{A}{#1}{\symbSSE{\sen}{\chi}}}
\newcommand{\dsreSSE}{\ax{RE}{}{\symbSSE{\sen}{\chi}}}
\newcommand{\vdashLOdSSE}{\vdash}

\newcommand{\EEE}{\ensuremath{\qa\qa\qa}\xspace}
\newcommand{\symbEEE}{!} 
\newcommand{\mopEEE}[1]{#1^{\symbEEE}} 
\newcommand{\modEEE}{\mmBox{\symbEEE}{}} 
\newcommand{\mmEEE}[1]{\modEEE#1} 

\newcommand{\SEE}{\ensuremath{\qe\qa\qa}\xspace}
\newcommand{\symbSEE}[1]{{#1}!}
\newcommand{\mopSEE}[2]{#1^{\symbSEE{#2}}}
\newcommand{\modSEE}[1]{\mmBox{\symbSEE{#1}}{}}
\newcommand{\mmSEE}[2]{\modSEE{#1}#2}

\newcommand{\SSE}{\ensuremath{\qe\qe\qa}\xspace}
\newcommand{\symbSSE}[2]{{#1}_{#2}!}
\newcommand{\mopSSE}[3]{#1^{\symbSSE{#2}{#3}}}
\newcommand{\modSSE}[2]{\mmBox{\symbSSE{#1}{#2}}{}}
\newcommand{\mmSSE}[3]{\modSSE{#1}{#2}#3}

\newcommand{\SES}{\ensuremath{\qe\qa\qe}\xspace}

\newcommand{\SSS}{\ensuremath{\qe\qe\qe}\xspace}

\newcommand{\ssub}{\on{ssub}}

\newcommand{\tr}{\operatorname{\tau}}
\newcommand{\com}{\operatorname{c}}
\newcommand{\nscom}{\operatorname{nsc}}
\newcommand{\ndcom}{\operatorname{ndc}}

\newcommand{\colorextras}{teal}

\begin{document}

\title{\titulo\notaVersionPlus{To appear in Revista Mexicana de L{\'o}gica 1(1).}}
\author{\autor}
\affil{Department of Information Science and Media Studies, Universitetet i Bergen\\\texttt{Fernando.VelazquezQuesada@uib.no}}

\date{}
\maketitle

\begin{abstract}
  This manuscript studies actions of \emph{communication} between epistemic logic agents. It starts by looking into actions through which \emph{all}/\emph{some} agents share all their information, defining the model operation that transforms the model, discussing its properties, introducing a modality for describing it and providing an axiom system for the latter. The main part of the manuscript focusses on an action through which some agents share \emph{part} of their information: they share all that they know about a topic defined by a given formula. Once again, the manuscript defines the model operation that transforms the model, discusses its properties, introduces a modality for describing it and provides an axiom system for the latter.

\end{abstract}

\section{Introduction}\label{sec:intro}

Epistemic logic (\ti{EL}; \citealp{Hintikka1962}) is a logical system for reasoning about the knowledge a set of agents might have. On the syntactic side, its language extends propositional logic with a modality $\modK{\agi}$ for every agent $\agi$, with formulas of the form $\mmKi{\varphi}$ read as ``agent $\agi$ knows that $\varphi$ is the case''. On the semantic side, it typically relies on relational `Kripke' models, assigning to each agent an \emph{indistinguishability relation} among epistemic possibilities.\footnote{There are other alternatives; see \autoref{ftn:othermodels}.} The crucial idea is that knowledge is defined in terms of \emph{uncertainty}: agent $\agi$ knows that $\varphi$ is the case when $\varphi$ holds in all situations she considers possible.\footnote{This is the ``information as \emph{range}'' discussed in \citet{vanBenthemMartinez2008}.} Despite its simplicity (or maybe because of it), \ti{EL} has become a widespread tool, contributing to the formal study of complex multi-agent epistemic notions in philosophy \citep{PhilStu:EpisLog}, computer science \citep{FaginHalpernMosesVardi1995,MeyervanDerHoek1995elaics} and economics \citep{deBruin2010,Perea2012}.

\medskip

One of the reasons for the success of \ti{EL} and its variations is that it allows a natural representation of \emph{actions} that affect the agents' information (e.g., knowledge and beliefs). The two paradigmatic examples are \emph{public announcements} \citep{Plaza1989,GerbrandyGroeneveld1997}, representing the effect of agents receiving truthful information, and \emph{belief revision} \citep{vanDitmarsch2005,vanBenthem2007dlbr,BaltagSmets2008tlg}, representing actions of agents receiving information that is reliable and yet potentially fallible. These two frameworks are part of what is known as \emph{dynamic epistemic logic} (\ti{DEL}; \citealp{vanDitmarschEtAl2007,vanBenthem2011ldii}), a field whose main feature is that actions are semantically represented not as relations (as done, e.g., in \emph{propositional dynamic logic}, \citealp{HarelKozenTiuryn2000}), but rather as operations that transform the underlying semantic model.

\smallskip

The mentioned \ti{DEL} frameworks have been used for representing communication \emph{between agents} (e.g., \citealp{AgotnesEtAl2010,vanDitmarschLying2013,BaltagSmets2013,GalimullinAlechina2017}). Yet, they were originally designed to represent the effect of \emph{external} communication, with the information's source being some entity that is not part of the system. This can be observed by noticing that, in these settings, the incoming information $\chi$ does not need to be known/believed by any of the involved agents.

\medskip

This manuscript studies epistemic actions in which the information that is being shared is information some of the agents already have. In this sense, the actions studied here are true actions of \emph{inter-agent} communication. For this, the crucial notion is that of \emph{distributed knowledge} \citep{Hilpinen1977,HalpernMoses1984,HalpernMoses1985,HalpernMoses1990}, representing what a group of agents \emph{would} know by putting all their information together. Distributed knowledge thus `pre-encodes' the information a group of agents would have if they were to share their individual pieces. Then, the actions studied here can be seen as (variations of) actions that fulfil this promise, doing so by defining the model that is obtained \emph{after} communication takes place.

\smallskip

In defining these communication actions, it is important to emphasise that, under relational `Kripke' models, epistemic logic defines knowledge in terms of uncertainty. This is because these models only represent the epistemic uncertainty of the agent, without `explaining' why some uncertainty (i.e., epistemic possibility) has been discarded and why some other remains. This has two important consequences.
\begin{compactitemize}
  \item First, as discussed in \citet{vanderHoekEtAl1999}, distributed knowledge does not satisfy the ``principle of full communication'': there are situations in which a group knows distributively a formula $\varphi$, and yet $\varphi$ does not follow from the individual knowledge of the groups' members. Thus, \emph{under relational models}, distributed knowledge is better understood as what a group of agents would know (in the ``information as range'' sense) if they indicated to one another \emph{which epistemic possibilities they have already discarded}.

  \item Second, recall that an agent's uncertainty is represented by her indistinguishability relation. Thus, although changes in uncertainty can be represented by changing what each epistemic possibility describes (technically, by changing the model's atomic valuation), they are more naturally represented by changes in the relation itself.\footnote{Note that changing the model's domain (removing worlds, as when representing public announcements, or adding them, as when representing non-public forms of communication) is an indirect way of changing indistinguishability relations.}
\end{compactitemize}

\medskip

This text is organised as follows. \autoref{sec:basics} recalls the basics of \ti{EL}, including the semantic model representing the agents' uncertainty, the formal language used for describing them and an axiom system characterising validities. Then, while \autoref{sec:XEE} discusses communication actions through which \emph{all}/\emph{some} agents share all their information with everybody (comparing them with proposals in the literature), \autoref{sec:SSE} discusses a novel action through which some agents share \emph{part} of their information with everybody. \autoref{sec:XXS} is a brief discussion of the issues arising when only \emph{some agents receive} the shared information. Finally, \autoref{sec:end} summarises the work, discussing also further research lines. While the proofs of propositions are found within the text, the proofs of theorems can be found in the appendix.

\section{Basic system}\label{sec:basics}

Throughout this text, let \ag be a finite non-empty set of agents, and let \pa be a non-empty enumerable set of atomic propositions.

\begin{definicion}[Multi-agent relational model]\label{def:model}
  ~\!\!A \emph{multi-agent relational model} (or, simply, a model) is a tuple $M = \tupla{W, \R, V}$ where $W$ (also denoted as $\dom{M}$) is a non-empty set of objects called \emph{possible worlds}, $\R = \set{\sbi{\R} \subseteq W \times W \mid \agi \in \ag}$ contains a binary \emph{indistinguishability} relation on $W$ for each agent in $\ag$, and $V:\pa \to \power{W}$ is the atomic valuation indicating the set of possible worlds in which each atom holds. The class of (multi-agent relational) models is denoted by \marm. A pair $(M, w)$ with $M$ in \marm and $w \in \dom{M}$ is called a \emph{pointed \marm model} (or, simply, a pointed model), with $w$ being the evaluation point.

  \smallskip

  Let $M = \tupla{W, \R, V}$ be a model. For $\sag \subseteq \ag$, define $\sbDg{\R} := \bigcap_{\agk \in \sag} \sbk{\R}$, with edges in $\sbDg{\R}$ called $\sag$-edges. For $S \subseteq W \times W$ and $w \in W$, define $S(w) := \set{u \in W \mid Swu}$.
\end{definicion}

Note: in a model, the indistinguishability relations are arbitrary binary relations. In particular, they need to be neither reflexive nor symmetric nor Euclidean nor transitive, and hence knowledge here is neither truthful nor positively/negatively introspective. The notion of knowledge used here corresponds simply to ``what is true in all the agent's epistemic possibilities''.

\smallskip

Pointed models are described by the following language.

\begin{definicion}[Language \LAd]
  Formulas $\varphi, \psi$ of the language \LAd are given by
  \[ \varphi, \psi ::= p \mid \lnot \varphi \mid \varphi \land \psi \mid \mmDg{\varphi} \]
  for $p \in \pa$ and $\emptyset \subset \sag \subseteq \ag$. Boolean constants ($\top, \bot$) and other Boolean operators ($\lor, \limp, \ldimp$) are defined as usual. Additionally, define $\mmKi{\varphi} := \mmDs{\set{\agi}}{\varphi}$.
\end{definicion}

Note how \LAd contains a modality $\modDg$ for each non-empty set of agents $\sag \subseteq \ag$, thanks to which one can build formulas of the form $\mmDg{\varphi}$, read as ``the agents in $\sag$ have distributed knowledge of $\varphi$''. Thus, $\mmKi{\varphi}$ is read as ``agent $\agi$ has distributed knowledge of $\varphi$'' or, in other words, ``agent $\agi$ knows $\varphi$''.

\smallskip

Formulas of \LAd are semantically interpreted in pointed models.

\begin{definicion}[Interpreting \LAd on pointed models]
  Let $(M, w)$ be a pointed model with $M = \tupla{W, \R, V}$. The satisfiability relation $\Vdash$ between $(M, w)$ and a formula in \LAd is defined inductively. Boolean cases are as usual; for the rest,
  \begin{ltabular}{@{\qquad}l@{\qssidefq}l}
    $(M, w) \Vdash p$              & $w \in V(p)$, \\
    $(M, w) \Vdash \mmDg{\varphi}$ & for all $u \in W$, if $\sbDg{\R}wu$ then $(M, u) \Vdash \varphi$. \\
  \end{ltabular}
  A formula $\varphi$ is valid on \marm (notation: $\Vdash \varphi$) if and only if $(M, w) \Vdash \varphi$ for every $w \in \dom{M}$ of every $M$ in \marm. By defining the truth-set of a formula as $\truthset{M}{\varphi} := \set{w \in W \mid (M, w) \Vdash \varphi}$ (so $\truthset{M}{\varphi}$ is the set of $\varphi$-worlds in $M$, that is, the worlds in $M$ where $\varphi$ holds), one can state equivalently that $\varphi$ is valid on \marm if and only if $\truthset{M}{\varphi} = \dom{M}$ for every $M$ in \marm.
\end{definicion}

The semantic interpretation of $\mmDg{\varphi}$ deserves some comments. Recall: $\sbDg{\R}$ is the \emph{intersection} of the relations of agents in $\sag$. Thus, $\sbDg{\R}wu$ holds if and only if $\sbi{\R} wu$ holds for \emph{every} $\agi$ in $\sag$, that is, if and only if \emph{every} agent in $\sag$ considers $u$ possible when at $w$ or, equivalently, if and only if \emph{no agent} in $\sag$ can discard $u$ when at $w$. Using the notation $\truthset{}{\cdot}$, the semantic interpretation of $\mmDg{\varphi}$ is equivalently stated as
\begin{ltabular}{@{\qquad}l@{\qssidefpq}l}
  $(M, w) \Vdash \mmDg{\varphi}$ & $\sbDg{\R}(w) \subseteq \truthset{M}{\varphi}$.
\end{ltabular}
Note also that the abbreviation $\mmKi{\varphi}$ behaves as expected:
\[
  (M, w) \Vdash \mmKi{\varphi}
  \qssiq
  (M, w) \Vdash \mmDs{\set{\agi}}{\varphi}
  \qssiq
  \sbDs{\R}{\set{\agi}}(w) \subseteq \truthset{M}{\varphi}
  \qssiq
  \sbi{\R}(w) \subseteq \truthset{M}{\varphi},
\]
so agent $\agi$ knows $\varphi$ at $(M, w)$ if and only if every world she cannot distinguish from $w$ is a $\varphi$-world.

\medskip

\begin{ejemplo}\label{eje:marl}
  Here are some examples of this setting.
  \begin{compactenumerate}
    \item\label{eje:marl:sym} Take $\ag = \set{\aga,\agb,\agc}$ and $\pa = \set{p, q, r}$. Consider $M_1 = \tupla{\set{w_0, w_1, w_2, w_3}, \R, V}$, a model whose indistinguishability relations and valuation function are as in the diagram below (each world shows exactly the atoms true at it); take $w_0$ to be the evaluation point (double-circled in the diagram).
    \begin{ctabular}{c@{}c@{}c}
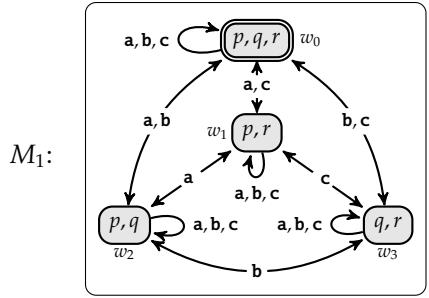
      \begin{tabular}{r}
        $M_1$:
      \end{tabular}
      &
      \begin{tabular}{c}
        \begin{tikzpicture}[node distance = 2em and 3em, frame rectangle, scale = 0.75]
          \node [mundo, label = {[etiqMundo, outer sep = 2pt]right:$w_0$}, double] (w0) {$p, q, r$};
          \node [mundo, label = {[etiqMundo]left:$w_1$}, below = of w0] (w1) {$p,r$};
          \node [mundo, label = {[etiqMundo]below:$w_2$}, below left = of w1] (w2) {$p,q$};
          \node [mundo, label = {[etiqMundo]below:$w_3$}, below right = of w1] (w3) {$q,r$};

          \path (w0) edge [flecha, loop left] node [etiqFlecha] {$\aga,\agb,\agc$} ()
                     edge [flecha, <->] node [etiqFlecha, within] {$\aga,\agc$} (w1)
                     edge [flecha, <->, bend right = 25] node [etiqFlecha, within] {$\aga, \agb$} (w2)
                     edge [flecha, <->, bend left = 25] node [etiqFlecha, within] {$\agb,\agc$} (w3)
                (w1) edge [flecha, loop below] node [etiqFlecha] {$\aga,\agb,\agc$} ()
                     edge [flecha, <->] node [etiqFlecha, within] {$\aga\;$} (w2)
                     edge [flecha, <->] node [etiqFlecha, within] {$\;\agc$} (w3)
                (w2) edge [flecha, loop right] node [etiqFlecha] {$\aga,\agb,\agc$} ()
                     edge [flecha, <->, bend right] node [etiqFlecha, within] {$\agb$} (w3)
                (w3) edge [flecha, loop left] node [etiqFlecha] {$\aga,\agb,\agc$} ();
        \end{tikzpicture}
      \end{tabular}
      &
      \begin{tabular}{l}
        \phantom{$M_1$:}
      \end{tabular}
    \end{ctabular}
    At $(M_1, w_0)$ all atoms are true; yet, no agent knows this. First, agent $\aga$ knows that $p$ holds, but knows the truth value of neither $q$ nor $r$:
    \begin{ltabular}{@{\quad}l}
      $(M_1, w_0) \Vdash \mmKa{p} \land (\lnot \mmKa{q} \land \lnot \mmKa{\lnot q}) \land (\lnot \mmKa{r} \land \lnot \mmKa{\lnot r})$.
    \end{ltabular}
    Then, $\agb$ knows that $q$ holds, but knows the truth value of neither $p$ nor $r$:
    \begin{ltabular}{@{\quad}l}
      $(M_1, w_0) \Vdash (\lnot \mmKb{p} \land \lnot \mmKb{\lnot p}) \land \mmKb{q} \land (\lnot \mmKb{r} \land \lnot \mmKb{\lnot r})$.
    \end{ltabular}
    Finally, $\agc$ knows that $r$ holds, but knows the truth value of neither $p$ nor $q$:
    \begin{ltabular}{@{\quad}l}
      $(M_1, w_0) \Vdash (\lnot \mmKc{p} \land \lnot \mmKc{\lnot p}) \land (\lnot \mmKc{q} \land \lnot \mmKc{\lnot q}) \land \mmKc{r}$.
    \end{ltabular}
    Still, each agent knows that $\aga$ knows $p$'s truth-value, that $\agb$ knows $q$'s truth-value, and that $\agc$ knows $r$'s truth-value:
    \begin{ltabular}{@{\quad}l}
      $\displaystyle (M_1, w_0) \Vdash \bigwedge_{\agi \in \set{\aga,\agb,\agc}} \mmKi{\big( (\mmKa{p} \lor \mmKa{\lnot p}) \land (\mmKb{q} \lor \mmKb{\lnot q}) \land (\mmKc{r} \lor \mmKc{\lnot r}) \big)}$.
    \end{ltabular}
    Finally, agents would benefit from sharing their individual information. In particular, if they all shared, they would know which the real situation is:
    \begin{ltabular}{@{\quad}l}
      $(M_1, w_0) \Vdash \mmDs{\set{\aga,\agb}}{(p \land q)} \land \mmDs{\set{\aga,\agc}}{(p \land r)} \land \mmDs{\set{\agb,\agc}}{(q \land r)} \land \mmDs{\set{\aga,\agb,\agc}}{(p \land q \land r)}$.
    \end{ltabular}

    \item\label{eje:marl:asym} Let $\ag$ and $\pa$ be as in \autoref{eje:marl:sym}; consider the pointed model depicted below.
    \begin{ctabular}{c@{}c@{}c}
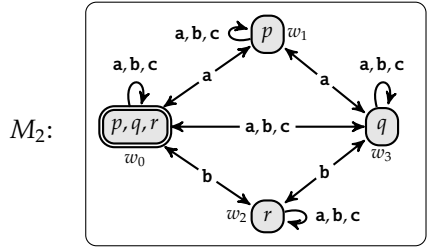
      \begin{tabular}{r}
        $M_2$:
      \end{tabular}
      &
      \begin{tabular}{c}
        \begin{tikzpicture}[node distance = 2em and 3em, frame rectangle, scale = 0.75]
          \node [mundo, label = {[etiqMundo, outer sep = 2pt]below:$w_0$}, double] (w0) {$p, q, r$};
          \node [mundo, label = {[etiqMundo]right:$w_1$}, above right = of w0] (w1) {$p$};
          \node [mundo, label = {[etiqMundo]left:$w_2$}, below right = of w0] (w2) {$r$};
          \node [mundo, label = {[etiqMundo]below:$w_3$}, below right = of w1] (w3) {$q$};

          \path (w0) edge [flecha, loop above] node [etiqFlecha] {$\aga,\agb,\agc$} ()
                     edge [flecha, <->] node [etiqFlecha, within] {$\aga$} (w1)
                     edge [flecha, <->] node [etiqFlecha, within] {$\agb$} (w2)
                     edge [flecha, <->] node [etiqFlecha, within] {$\aga,\agb,\agc$} (w3)
                (w1) edge [flecha, loop left] node [etiqFlecha] {$\aga,\agb,\agc$} ()
                     edge [flecha, <->] node [etiqFlecha, within] {$\aga$} (w3)
                (w2) edge [flecha, loop right] node [etiqFlecha] {$\aga,\agb,\agc$} ()
                     edge [flecha, <->] node [etiqFlecha, within] {$\agb$} (w3)
                (w3) edge [flecha, loop above] node [etiqFlecha] {$\aga,\agb,\agc$} ();
        \end{tikzpicture}
      \end{tabular}
      &
      \begin{tabular}{l}
        \phantom{$M_2$:}
      \end{tabular}
    \end{ctabular}
    Again, all atoms are true in the real situation; yet, no agent knows this. On the one hand, $\aga$ knows $p \lor q$ without knowing the truth-value of $p$ or $q$,
    \begin{ltabular}{@{\quad}l}
      $(M_2, w_0) \Vdash \mmKa{(p \lor q)} \land (\lnot \mmKa{p} \land \lnot \mmKa{\lnot p}) \land (\lnot \mmKa{q} \land \lnot \mmKa{\lnot q})$.
    \end{ltabular}
    On the other hand, $\agb$ knows $q \lor r$ without knowing the truth-value of $q$ or $r$:
    \begin{ltabular}{@{\quad}l}
      $(M_2, w_0) \Vdash \mmKb{(q \lor r)} \land (\lnot \mmKb{q} \land \lnot \mmKb{\lnot q}) \land (\lnot \mmKb{r} \land \lnot \mmKb{\lnot r})$.
    \end{ltabular}
    Agent $\agc$ has slightly more information, as she knows that $q$ is true but still ignores the truth-value of $p$ and $r$:
    \begin{ltabular}{@{\quad}l}
      $(M_2, w_0) \Vdash (\lnot \mmKc{p} \land \lnot \mmKc{\lnot p}) \land \mmKc{q} \land (\lnot \mmKc{r} \land \lnot \mmKc{\lnot r}) $.
    \end{ltabular}
    This time, while communicating would help $\aga$ and $\agb$, it would not help $\agc$. In fact, collectively, the agents do not have enough information to find out which the real situation is:
    \begin{ltabular}{@{\quad}l}
      $(M_2, w_0) \Vdash \mmDs{\set{\aga,\agb}}{q} \land \mmDs{\set{\aga, \agc}}{q} \land \mmDs{\set{\agb,\agc}}{q} \land \lnot \mmDs{\set{\aga,\agb,\agc}}{(p \land r)}$.
    \end{ltabular}

    \item\label{eje:marl:noequiv} Take $\ag = \set{\aga, \agb}$ and $\pa = \set{p,q}$; consider $(M_3, w_0)$ depicted below.
    \begin{ctabular}{c@{}c@{}c}
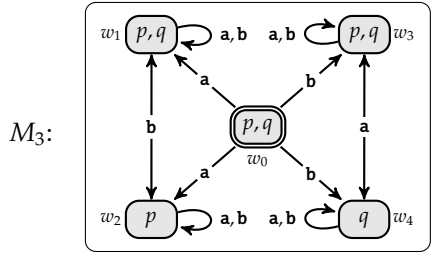
      \begin{tabular}{r}
        $M_3$:
      \end{tabular}
      &
      \begin{tabular}{c}
        \begin{tikzpicture}[node distance = 2em and 2em, frame rectangle, scale = 0.75]
          \node [mundo, label = {[etiqMundo, outer sep = 2pt]below:$w_0$}, double] (w0) {$p, q$};
          \node [mundo, label = {[etiqMundo]left:$w_1$}, above left = of w0] (w1) {$p,q$};
          \node [mundo, label = {[etiqMundo]left:$w_2$}, below left = of w0] (w2) {$\;\,p\,\;$};
          \node [mundo, label = {[etiqMundo]right:$w_3$}, above right = of w0] (w3) {$p,q$};
          \node [mundo, label = {[etiqMundo]right:$w_4$}, below right = of w0] (w4) {$\;\,q\,\;$};

          \path (w0) edge [flecha] node [etiqFlecha, within] {$\aga$} (w1)
                     edge [flecha] node [etiqFlecha, within] {$\aga$} (w2)
                     edge [flecha] node [etiqFlecha, within] {$\agb$} (w3)
                     edge [flecha] node [etiqFlecha, within] {$\agb$} (w4)
                (w1) edge [flecha, loop right] node [etiqFlecha] {$\aga,\agb$} ()
                     edge [flecha, <->] node [etiqFlecha, within] {$\agb$} (w2)
                (w2) edge [flecha, loop right] node [etiqFlecha] {$\aga,\agb$} ()
                (w3) edge [flecha, loop left] node [etiqFlecha] {$\aga,\agb$} ()
                     edge [flecha, <->] node [etiqFlecha, within] {$\aga$} (w4)
                (w4) edge [flecha, loop left] node [etiqFlecha] {$\aga,\agb$} ();
        \end{tikzpicture}
      \end{tabular}
      &
      \begin{tabular}{l}
        \phantom{$M_3$:}
      \end{tabular}
    \end{ctabular}
    In the pointed model, both $p$ and $q$ are true. Both agents have partial information about this: while agent $\aga$ knows $p$ but does not know whether $q$, agent $\agb$ does not know whether $p$ but knows $q$:
    \begin{ltabular}{@{\quad}l}
      $(M_3, w_0) \Vdash \mmKa{p} \land (\lnot \mmKa{q} \land \lnot \mmKa{\lnot q}) \land (\lnot \mmKb{p} \land \lnot \mmKb{\lnot p}) \land \mmKb{q}$.
    \end{ltabular}
    However, both agents have misleading information about what the other knows: $\aga$ thinks $\agb$ knows $p$ without knowing whether $q$, and $\agb$ thinks $\aga$ does not know whether $p$ but knows $q$:
    \begin{ltabular}{@{\quad}l}
      $(M_3, w_0) \Vdash \mmKa{(\mmKb{p} \land (\lnot \mmKb{q} \land \lnot \mmKb{\lnot q}))} \land \mmKb{((\lnot \mmKa{p} \land \lnot \mmKa{\lnot p}) \land \mmKa{q})}$.
    \end{ltabular}
    If they were to share their (partially misleading) information, they would believe inconsistencies:
    \begin{lbtabular}{@{\quad}l}
      $(M_3, w_0) \Vdash \mmDs{\set{\aga, \agb}}{\bot}$.
    \end{lbtabular}
  \end{compactenumerate}
\end{ejemplo}

\bsparagraph{Axiom system} As the reader can imagine, there are other alternatives for defining a logical framework describing the individual and distributed knowledge a set of agents might have. The epistemic logic framework recalled in this section is based on some concrete choices (e.g., relying on a model that represents uncertainty [via relations], and then defining knowledge in terms of it), and these choices define the properties of the notions of knowledge that arise. Still, some properties might not be easy to identify just from the model and the language's semantic interpretation. In such cases, it is useful to look for an axiom system characterising the formulas in \LAd that are valid in pointed (multi-agent relational) models. By doing so, the axiom system provides a list of the essential laws governing the behaviour and interaction of individual and distributed knowledge in this setting.

\medskip

It is well-known (e.g., \citealp{HalpernMoses1985,FaginHalpernMosesVardi1995,BaltagSmets2020}) that the axiom system \LOd, whose axioms and rules are shown on \autoref{tbl:LOd}, characterise the formulas in \LAd that are valid in pointed \marm-models. While \ax{PR}{}{} and \ax{MP}{}{} characterise the behaviour of Boolean operators, axioms \ax{K}{}{\opgDist}, \ax{M}{}{\opgDist} and rule \ax{G}{}{\opgDist} characterise distributed knowledge: $\modDg$ contains all validities (rule \ax{G}{}{\opgDist}), it is closed under modus ponens (axiom \ax{K}{}{\opgDist}) and it is monotone on the set of agents (axiom \ax{M}{}{\opgDist}: if $\varphi$ is distributively known by the agents in $\sag$, then it is also distributively known by any larger group $\sag'$).

\begin{table}[ht]
  \begin{smallctabular}{l@{\;}ll@{\;}l}
    \toprule
    \ax{PR}{}{}:         & $\vdash \varphi$ \;\; for $\varphi$ a propositionally valid scheme &
    \dsrMP:              & If $\vdash \varphi$ and $\vdash \varphi \limp \psi$ then $\vdash \psi$ \\
    \midrule
    \ax{K}{}{\opgDist}: & $\vdash \mmDs{\sag}{(\varphi \limp \psi)} \limp (\mmDs{\sag}{\varphi} \limp \mmDs{\sag}{\psi})$ &
    \ax{G}{}{\opgDist}: & If $\vdash \varphi$ then $\vdash \mmDs{\sag}{\varphi}$ \\
    \ax{M}{}{\opgDist}: & $\vdash \mmDs{\sag}{\varphi} \limp \mmDs{\sag'}{\varphi}$ \;\; for $\sag \subseteq \sag'$ \\
    \bottomrule
  \end{smallctabular}
  \caption{Axiom system \LOd, for \LAd w.r.t. models in \marm.}
  \label{tbl:LOd}
\end{table}

\begin{teorema}\label{teo:LOd}
  The axiom system \LOd \autorefp{tbl:LOd} is sound and strongly complete for formulas in \LAd w.r.t. models in \marm.\footnote{Recall that $\modDs{\emptyset}$ is \emph{not} a modality in \LAd. If it were, then \LOd would need further axioms and rules. In fact, since $\sbDs{\R}{\emptyset}(w) = \dom{M}$ for any $w \in \dom{M}$ and any $M \in \marm$, it follows that $(M,w) \Vdash \mmDs{\emptyset}{\varphi}$ if and only if $(M, u) \Vdash \varphi$ for all $u \in \dom{M}$. In other words, $\mmDs{\emptyset}$ is nothing but the \emph{global universal modality} \citep{GorankoPassy1992}. Because of this, an axiom system for $\mmDs{\emptyset}$ requires not only extending \ax{K}{}{\opgDist}, \ax{G}{}{\opgDist} and \ax{M}{}{\opgDist} to allow $\sag = \emptyset$, but also the use of three additional axioms: $\vdash \mmDs{\emptyset}{\varphi} \limp \varphi$, $\vdash \mmDs{\emptyset}{\mmDs{\emptyset}{\varphi}} \limp \mmDs{\emptyset}{\varphi}$ and $\vdash \mmDs{\emptyset}{\lnot \mmDs{\emptyset}{\varphi}} \limp \lnot \mmDs{\emptyset}{\varphi}$.}
\end{teorema}

\section{Sharing \emph{everything} with \emph{everybody}}\label{sec:XEE}

As discussed before, the concept of distributed knowledge relies on the idea of agents communicating their individual information. Indeed, the fact that a set of agents $\sag$ has distributed knowledge of $\varphi$ ($\mmDg{\varphi}$) encodes the intuition that, if they all would share their information, then afterwards each one of them would know that $\varphi$ is the case. This `encoding' can be made explicit by following the \ti{DEL} approach: define an operation that, when receiving an initial model representing the agents' individual information, returns the model that results from the agents sharing their information. The current section explores two variations of this idea: an operation representing an action through which \emph{all} agents share all her information with everybody \autorefp{sbs:EEE}, and an operation representing an action through which \emph{some} agents share all her information with everybody \autorefp{sbs:SEE}.

\subsection{\emph{Everybody} shares everything with everybody}\label{sbs:EEE}

The simplest form of communication among \ti{EL} agents is one through which \emph{every} agent shares everything with everybody. This will be called an act of \EEE-communication.

\bsparagraph{Operation and modality} Recall that a multi-agent relational model represents not the knowledge of each agent, but rather her uncertainty: the worlds she considers possible from a given one. Thus, in this setting, an action through which every agent shares everything she knows with everybody corresponds, formally, to a model operation through which every agent discards every possibility that has been already discarded by any agent. Such a operation is straightforward: the indistinguishability relation of each agent in the new model is defined as the intersection of the relations of all agents in the original model.

\begin{definicion}[\EEE-communication]\label{def:EEEmo}
  Let $M = \tupla{W, \R, V}$ be in \marm.  The relations in the \marm-model $\mopEEE{M} = \tuplan{W, \mopEEE{\R}, V}$ are given, for each $\agi \in \ag$, as
  \[ \sbi{\mopEEE{\R}} := \sbDs{\R}{\ag}. \]
\end{definicion}

Thus, after the operation an agent $\agi$ cannot distinguish $u$ from $w$ (that is, $\sbi{\mopEEE{\R}} wu$) if and only if, before the operation, \emph{no agent} in \ag could distinguish $u$ from $w$ (that is, $\sbj{\R} wu$ for all $\agj \in \ag$).

\medskip

It should be emphasised that, despite the given intuition, the just defined operation is not one through which the agents share \emph{the information that has allowed them to discard certain possibility}. As discussed in the introduction, relational models only represent the epistemic uncertainty of the agents (the edges). Thus, an agent's communication amounts to sharing the epistemic possibilities she has already discarded (by indicating which edges are \emph{not} in her indistinguishability relation), so others can discard them too.\footnote{\label{ftn:othermodels}Other proposals for representing an agent's knowledge do keep track of the justifications \citep{Artemov2008rsl}, evidence \citep{vanBenthemPacuit2011,BaltagEtAl2016evid,Ozgun2017} or arguments \citep{ShiEtAl2017,ShiEtAl2020} this knowledge is based on. In such settings, the `reasons' each agent has for her knowledge are present, and thus one can define 
model operations through which this information is shared.}

\smallskip

Here are some small yet useful observations.
\begin{itemize}
  \item Each new relation $\sbi{\mopEEE{\R}}$ can be equivalently defined as $\sbi{\R} \cap \sbDs{\R}{\ag\setminus\set{\agi}}$.
  \item Since $\sbi{\mopEEE{\R}}(w) = \sbDs{\R}{\ag}(w)$ is a subset of $\sbi{\R}(w)$ for any world $w$ in any model $M$, it follows that the action of \EEE-communication can only \emph{reduce} the uncertainty of each agent.
  \item The operation preserves universal relational properties: if all relations in $\set{\sbj{\R} \mid \agj \in \ag}$ are, e.g., reflexive/transitive/symmetric/Euclidean, then so is each resulting $\sbi{\mopEEE{\R}}$. Thus, in particular, the operation preserves equivalence relations.
\end{itemize}

{\medskip}

Here is a modality for describing the effects of \EEE-communication.

\begin{definicion}[Modality $\modEEE$; language \LAdEEE]\label{def:EEEmm}
  The language \LAdEEE extends \LAd with a modality $\modEEE$, semantically interpreted in a pointed model $(M, w)$ as
  \begin{lbtabular}{l@{\qssidefq}l}
    $(M, w) \Vdash \mmEEE{\varphi}$ & $(\mopEEE{M}, w) \Vdash \varphi$.
  \end{lbtabular}
\end{definicion}

Using $\truthset{}{\cdot}$, observe how $w \in \truthset{M}{\mmEEE{\varphi}}$ if and only if $w \in \truthset{\mopEEE{M}}{\varphi}$. Thus,
\[ \truthsetn{M}{\mmEEE{\varphi}} = \truthset{\mopEEE{M}}{\varphi}. \]

\medskip

\begin{ejemplo}\label{eje:EEE}
  Here are examples of this operation at work.
  \begin{compactenumerate}
    \item\label{eje:EEE:sym} Recall the model $M_1$ from \itemfautoref{eje:marl}{eje:marl:sym} (diagram below on the left).
    \begin{ctabular}{@{}c@{\quad}c@{\quad}c@{}}
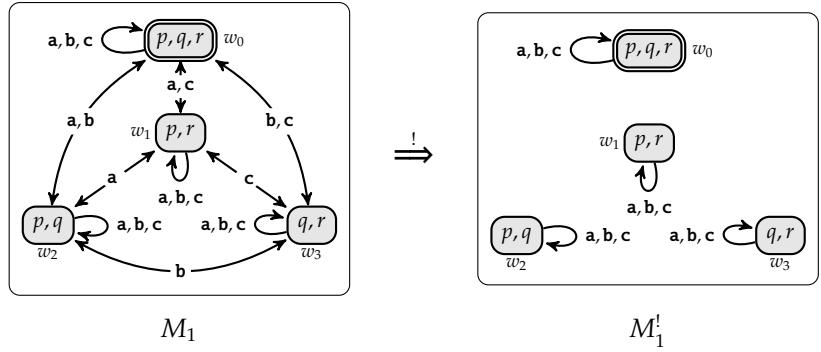
      \begin{tabular}{c}
        \begin{tikzpicture}[node distance = 2em and 3em, frame rectangle]
          \node [mundo, label = {[etiqMundo, outer sep = 2pt]right:$w_0$}, double] (w0) {$p, q, r$};
          \node [mundo, label = {[etiqMundo]left:$w_1$}, below = of w0] (w1) {$p,r$};
          \node [mundo, label = {[etiqMundo]below:$w_2$}, below left = of w1] (w2) {$p,q$};
          \node [mundo, label = {[etiqMundo]below:$w_3$}, below right = of w1] (w3) {$q,r$};

          \path (w0) edge [flecha, loop left] node [etiqFlecha] {$\aga,\agb,\agc$} ()
                     edge [flecha, <->] node [etiqFlecha, within] {$\aga,\agc$} (w1)
                     edge [flecha, <->, bend right = 25] node [etiqFlecha, within] {$\aga, \agb$} (w2)
                     edge [flecha, <->, bend left = 25] node [etiqFlecha, within] {$\agb,\agc$} (w3)
                (w1) edge [flecha, loop below] node [etiqFlecha] {$\aga,\agb,\agc$} ()
                     edge [flecha, <->] node [etiqFlecha, within] {$\aga\;$} (w2)
                     edge [flecha, <->] node [etiqFlecha, within] {$\;\agc$} (w3)
                (w2) edge [flecha, loop right] node [etiqFlecha] {$\aga,\agb,\agc$} ()
                     edge [flecha, <->, bend right] node [etiqFlecha, within] {$\agb$} (w3)
                (w3) edge [flecha, loop left] node [etiqFlecha] {$\aga,\agb,\agc$} ();
        \end{tikzpicture}
      \end{tabular}
      & $\overset{\symbEEE}{\bs{\Longrightarrow}}$ &
      \begin{tabular}{c}
        \begin{tikzpicture}[node distance = 2em and 3em, frame rectangle]
          \node [mundo, label = {[etiqMundo, outer sep = 2pt]right:$w_0$}, double] (w0) {$p, q, r$};
          \node [mundo, label = {[etiqMundo]left:$w_1$}, below = of w0] (w1) {$p,r$};
          \node [mundo, label = {[etiqMundo]below:$w_2$}, below left = of w1] (w2) {$p,q$};
          \node [mundo, label = {[etiqMundo]below:$w_3$}, below right = of w1] (w3) {$q,r$};

          \path (w0) edge [flecha, loop left] node [etiqFlecha] {$\aga,\agb,\agc$} ()
                (w1) edge [flecha, loop below] node [etiqFlecha] {$\aga,\agb,\agc$} ()
                (w2) edge [flecha, loop right] node [etiqFlecha] {$\aga,\agb,\agc$} ()
                (w3) edge [flecha, loop left] node [etiqFlecha] {$\aga,\agb,\agc$} ();
        \end{tikzpicture}
      \end{tabular}
      \\
      $M_1$ &  & $\mopEEE{M_1}$
    \end{ctabular}
    As shown in the diagram above on the right, the \EEE operation removes those edges that were not in the indistinguishability relation of \emph{every} agent (i.e., it preserves only $\ag$-edges). Thus,
    \begin{ltabular}{@{\quad}l@{}}
      $\displaystyle (M_1, w_0) \Vdash \mmDs{\set{\aga,\agb,\agc}}{(p \land q \land r)} \;\land\; \mmEEE{\bigwedge_{\agi \in \set{\aga,\agb,\agc}} \mmKi{(p \land q \land r)}}$.
    \end{ltabular}

    \item\label{eje:EEE:asym} Analogously, recall $M_2$ from \itemfautoref{eje:marl}{eje:marl:asym}, shown below on the left.
    \begin{ctabular}{@{}c@{\quad}c@{\quad}c@{}}
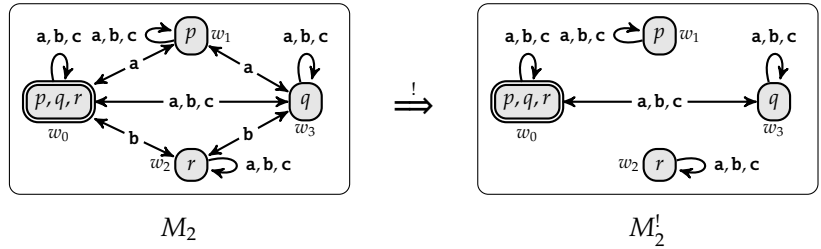
      \begin{tabular}{c}
        \begin{tikzpicture}[node distance = 1em and 3em, frame rectangle]
          \node [mundo, label = {[etiqMundo, outer sep = 2pt]below:$w_0$}, double] (w0) {$p, q, r$};
          \node [mundo, label = {[etiqMundo]right:$w_1$}, above right = of w0] (w1) {$p$};
          \node [mundo, label = {[etiqMundo]left:$w_2$}, below right = of w0] (w2) {$r$};
          \node [mundo, label = {[etiqMundo]below:$w_3$}, below right = of w1] (w3) {$q$};

          \path (w0) edge [flecha, loop above] node [etiqFlecha] {$\aga,\agb,\agc$} ()
                     edge [flecha, <->] node [etiqFlecha, within] {$\aga$} (w1)
                     edge [flecha, <->] node [etiqFlecha, within] {$\agb$} (w2)
                     edge [flecha, <->] node [etiqFlecha, within] {$\aga,\agb,\agc$} (w3)
                (w1) edge [flecha, loop left] node [etiqFlecha] {$\aga,\agb,\agc$} ()
                     edge [flecha, <->] node [etiqFlecha, within] {$\aga$} (w3)
                (w2) edge [flecha, loop right] node [etiqFlecha] {$\aga,\agb,\agc$} ()
                     edge [flecha, <->] node [etiqFlecha, within] {$\agb$} (w3)
                (w3) edge [flecha, loop above] node [etiqFlecha] {$\aga,\agb,\agc$} ();
        \end{tikzpicture}
      \end{tabular}
      & $\overset{\symbEEE}{\bs{\Longrightarrow}}$ &
      \begin{tabular}{c}
        \begin{tikzpicture}[node distance = 1em and 3em, frame rectangle]
          \node [mundo, label = {[etiqMundo, outer sep = 2pt]below:$w_0$}, double] (w0) {$p, q, r$};
          \node [mundo, label = {[etiqMundo]right:$w_1$}, above right = of w0] (w1) {$p$};
          \node [mundo, label = {[etiqMundo]left:$w_2$}, below right = of w0] (w2) {$r$};
          \node [mundo, label = {[etiqMundo]below:$w_3$}, below right = of w1] (w3) {$q$};

          \path (w0) edge [flecha, loop above] node [etiqFlecha] {$\aga,\agb,\agc$} ()
                     edge [flecha, <->] node [etiqFlecha, within] {$\aga,\agb,\agc$} (w3)
                (w1) edge [flecha, loop left] node [etiqFlecha] {$\aga,\agb,\agc$} ()
                (w2) edge [flecha, loop right] node [etiqFlecha] {$\aga,\agb,\agc$} ()
                (w3) edge [flecha, loop above] node [etiqFlecha] {$\aga,\agb,\agc$} ();
        \end{tikzpicture}
      \end{tabular}
      \\
      $M_2$ &  & $\mopEEE{M_2}$
    \end{ctabular}
    Hence, $\displaystyle (M_2, w_0) \Vdash \mmDs{\set{\aga,\agb,\agc}}{q} \;\land\; \mmEEE{\bigwedge_{\agi \in \set{\aga,\agb,\agc}} \mmKi{q}}$.

    \item\label{eje:EEE:noequiv} Finally, recall $M_3$ \itemfautorefp{eje:marl}{eje:marl:noequiv}, shown below on the left.
    \begin{ctabular}{@{}c@{\quad}c@{\quad}c@{}}
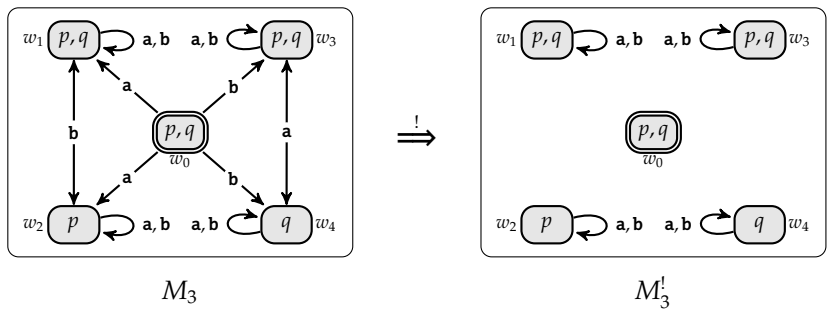
      \begin{tabular}{c}
        \begin{tikzpicture}[node distance = 2em and 2em, frame rectangle]
          \node [mundo, label = {[etiqMundo]below:$w_0$}, double] (w0) {$p, q$};
          \node [mundo, label = {[etiqMundo]left:$w_1$}, above left = of w0] (w1) {$p,q$};
          \node [mundo, label = {[etiqMundo]left:$w_2$}, below left = of w0] (w2) {$\;\,p\,\;$};
          \node [mundo, label = {[etiqMundo]right:$w_3$}, above right = of w0] (w3) {$p,q$};
          \node [mundo, label = {[etiqMundo]right:$w_4$}, below right = of w0] (w4) {$\;\,q\,\;$};

          \path (w0) edge [flecha] node [etiqFlecha, within] {$\aga$} (w1)
                     edge [flecha] node [etiqFlecha, within] {$\aga$} (w2)
                     edge [flecha] node [etiqFlecha, within] {$\agb$} (w3)
                     edge [flecha] node [etiqFlecha, within] {$\agb$} (w4)
                (w1) edge [flecha, loop right] node [etiqFlecha] {$\aga,\agb$} ()
                     edge [flecha, <->] node [etiqFlecha, within] {$\agb$} (w2)
                (w2) edge [flecha, loop right] node [etiqFlecha] {$\aga,\agb$} ()
                (w3) edge [flecha, loop left] node [etiqFlecha] {$\aga,\agb$} ()
                     edge [flecha, <->] node [etiqFlecha, within] {$\aga$} (w4)
                (w4) edge [flecha, loop left] node [etiqFlecha] {$\aga,\agb$} ();
        \end{tikzpicture}
      \end{tabular}
      & $\overset{\symbEEE}{\bs{\Longrightarrow}}$ &
      \begin{tabular}{c}
        \begin{tikzpicture}[node distance = 2em and 2em, frame rectangle]
          \node [mundo, label = {[etiqMundo]below:$w_0$}, double] (w0) {$p, q$};
          \node [mundo, label = {[etiqMundo]left:$w_1$}, above left = of w0] (w1) {$p,q$};
          \node [mundo, label = {[etiqMundo]left:$w_2$}, below left = of w0] (w2) {$\;\,p\,\;$};
          \node [mundo, label = {[etiqMundo]right:$w_3$}, above right = of w0] (w3) {$p,q$};
          \node [mundo, label = {[etiqMundo]right:$w_4$}, below right = of w0] (w4) {$\;\,q\,\;$};

          \path (w1) edge [flecha, loop right] node [etiqFlecha] {$\aga,\agb$} ()
                (w2) edge [flecha, loop right] node [etiqFlecha] {$\aga,\agb$} ()
                (w3) edge [flecha, loop left] node [etiqFlecha] {$\aga,\agb$} ()
                (w4) edge [flecha, loop left] node [etiqFlecha] {$\aga,\agb$} ();
        \end{tikzpicture}
      \end{tabular}
      \\
      $M_3$ &  & $\mopEEE{M_3}$
    \end{ctabular}
    Thus, $\displaystyle (M_3, w_0) \Vdash \mmDs{\set{\aga,\agb}}{\bot} \;\land\; \mmEEE{\bigwedge_{\agi \in \set{\aga, \agb}} \mmKi{\bot}}$.
  \end{compactenumerate}
\end{ejemplo}

\bsparagraph{Properties} Here are some properties describing the effects of the \EEE operation. First, intuitively, the action turns distributive knowledge of the whole group into individual knowledge. The following proposition shows that this is true only up to a certain extent.

\begin{proposicion}\label{pro:EEE:Moore}
  Let $(M, w)$ be a pointed \marm model; take $\agi \in \ag$. Then, $(M, w) \Vdash \mmDs{\ag}{\varphi} \limp \mmEEE{\mmKi{\varphi}}$ holds when $M$ and $\varphi$ are such that $\truthset{M}{\varphi} \subseteq \truthset{\mopEEE{M}}{\varphi}$ (i.e., when applying ``$\symbEEE$'' to $M$ does not reduce $\varphi$'s truth-set). However, $\not\Vdash \mmDs{\ag}{\varphi} \limp \mmEEE{\mmKi{\varphi}}$.
  \begin{proof}
    By semantic interpretation, $(M, w) \Vdash \mmDs{\ag}{\varphi}$ holds for a given $(M,w)$ if and only if $\sbDs{\R}{\ag}(w) \subseteq \truthset{M}{\varphi}$. But, by the definition of $\sbi{\mopEEE{\R}}$ \autorefp{def:EEEmo}, $\sbi{\mopEEE{\R}}(w) \subseteq \sbDs{\R}{\ag}(w)$ (\EEE can only reduce uncertainty), so $\sbi{\mopEEE{\R}}(w) \subseteq \truthset{M}{\varphi}$. Hence, by the assumption, $\sbi{\mopEEE{\R}}(w) \subseteq \truthset{\mopEEE{M}}{\varphi}$ and therefore $(\mopEEE{M}, w) \Vdash \mmKi{\varphi}$, that is, $(M, w) \Vdash \mmEEE{\mmKi{\varphi}}$. However, $\not\Vdash \mmDs{\ag}{\varphi} \limp \mmEEE{\mmKi{\varphi}}$, as shown by taking $\varphi := \lnot \mmKb{p}$ and observing that, on $M_1$ in \itemfautoref{eje:EEE}{eje:EEE:sym}, $(M_1, w_0) \Vdash \mmDs{\set{\aga,\agb,\agc}}{\lnot \mmKb{p}} \;\land\; \lnot \mmEEE{\mmKa{\lnot \mmKb{p}}}$. In general, there is not guarantee that every world making $\varphi$ true in a given $M$ (a world in $\truthset{M}{\varphi}$) is also a world making $\varphi$ true \emph{in $\mopEEE{M}$} (a world in $\truthset{\mopEEE{M}}{\varphi}$).
  \end{proof}
\end{proposicion}

The fact that $\mmDs{\ag}{\varphi} \limp \mmEEE{\mmKi{\varphi}}$ is not valid should not be taken to mean that \EEE is flawed. First, this formula one intuitively expects to be valid is so in those situations one intuitively considers: those in which $\varphi$ describes \emph{ontic} (i.e., propositional) facts.\footnote{The \EEE-act does not affect atomic valuations, so $\truthset{M}{\gamma} = \truthset{\mopEEE{M}}{\gamma}$ for every $M$ and every propositional $\gamma$. Thus, by \autoref{pro:EEE:Moore}, $\Vdash \mmDs{\ag}{\gamma} \limp \mmEEE{\mmKi{\gamma}}$.} The formula is in fact valid for a larger class of formulas in \LAd including, e.g., those describing knowledge about propositional formulas $\set{\mmKj{\gamma} \mid \gamma \stn{is propositional}}$\footnote{By definition, $w \in \truthsetn{M}{\mmKj{\gamma}}$ if and only if $\sbj{\R}(w) \subseteq \truthset{M}{\gamma}$. But $\sbj{\mopEEE{\R}}(w) \subseteq \sbj{\R}$ and $\truthset{M}{\gamma} = \truthset{\mopEEE{M}}{\gamma}$, so $\sbj{\mopEEE{\R}}(w) \subseteq \truthset{\mopEEE{M}}{\gamma}$ and thus $w \in \truthsetn{\mopEEE{M}}{\mmKj{\gamma}}$. Hence, $\truthsetn{M}{\mmKj{\gamma}} \subseteq \truthsetn{\mopEEE{M}}{\mmKj{\gamma}}$ so $\Vdash \mmDs{\ag}{\mmKj{\gamma}} \limp \mmEEE{\mmKi{\mmKj{\gamma}}}$.} and more. Second, the formula is not valid for an arbitrary $\varphi \in \LAd$ because $\LAd$ can express not only an agent's (distributed) knowledge but also her \emph{ignorance}, which might be reduced by the operation. This is used in the counterexample in \autoref{pro:EEE:Moore}, as $\lnot\mmKb{p}$ states that ``agent $\agb$ does not know that $p$ holds''.

\medskip

Here are two further results.

\begin{proposicion}\label{pro:EEE:mo}
  Let $M = \tupla{W, \R, V}$ be in \marm; let $\mopEEE{M} = \tuplan{W, \mopEEE{\R}, V}$ and $\mopEEE{(\mopEEE{M})} = \tuplan{W, \mopEEE{(\mopEEE{\R})}, V}$ be as indicated in \autoref{def:EEEmo}. Then,
  \begin{multicols}{2}
    \begin{enumerate}
      \item\label{pro:EEE:mo:D} $\sbDg{(\mopEEE{\R})} = \sbDs{\R}{\ag}$ \; for every $\sag \subseteq \ag$;
      \item\label{pro:EEE:mo:EEE} $\sbi{\mopEEE{(\mopEEE{\R})}} = \sbi{\mopEEE{\R}}$ \; for every $\agi \in \ag$.
    \end{enumerate}
  \end{multicols}
  \begin{proof}
    \begin{inlineenum} \item For any $\sag \subseteq \ag$, the relation $\sbDg{(\mopEEE{\R})}$ is $\bigcap_{\agk \in \sag} \sbk{\mopEEE{\R}}$ (def. of $\sbDg{(\mopEEE{\R})}$), equal to $\bigcap_{\agk \in \sag} \sbDs{\R}{\ag}$ (def. of $\sbk{\mopEEE{\R}}$), equal to $\sbDs{\R}{\ag}$. \item For any $\agi \in \ag$, the relation $\sbi{\mopEEE{(\mopEEE{\R})}}$ is $\sbDs{(\mopEEE{\R})}{\ag}$ (def. of $\mopEEE{(\mopEEE{\R})}$), equal to $\sbDs{\R}{\ag}$ (\autoref{pro:EEE:mo:D}), equal to $\sbi{\mopEEE{\R}}$ for any $\agi \in \ag$ (def. of $\sbi{\mopEEE{\R}}$).\end{inlineenum}
  \end{proof}
\end{proposicion}

\autoref{pro:EEE:mo} provides two interesting observations. First, the relation for interpreting $\modDg$ in the new model, $\sbDg{(\mopEEE{\R})}$, is the same as the relation for interpreting $\mmDs{\ag}$ in the original model, $\sbDs{\R}{\ag}$. Thus, for talking about distributed knowledge after \EEE communication (what $\sbDg{(\mopEEE{\R})}$ encodes), the modality $\modEEE$ is not needed: it is enough to use the modality $\modDs{\ag}$ (the one for $\sbDs{\R}{\ag}$) \emph{in the appropriate way} (see the translation of \autoref{def:trans:EEE}, used for the completeness argument).

\smallskip

The second observation states that each relation $\sbi{\mopEEE{(\mopEEE{\R})}}$ in the model after two \EEE acts is exactly as $\sbi{\mopEEE{\R}}$, its matching relation in the model after a single \EEE act. Since $\mopEEE{M}$ and $\mopEEE{(\mopEEE{M})}$ have the same domain and atomic valuation, this implies that the \EEE operation is idempotent: $\mopEEE{M} = \mopEEE{(\mopEEE{M})}$ for any $M$ in \marm.

\medskip

Thus,

\begin{proposicion}\label{pro:EEE}
  \begin{inlineenum} \item $\Vdash \mmEEE{\mmDg{\varphi}} \lldimp \mmDs{\ag}{\mmEEE{\varphi}}$; \;\; \item $\Vdash \mmEEE{\mmEEE{\varphi}} \lldimp \mmEEE{\varphi}$.\end{inlineenum}
  \begin{proof}
    Let $M = \tupla{W, \R, V}$ be in \marm. \begin{inlineenum} \item By definition, $w \in \truthset{M}{\mmEEE{\mmDg{\varphi}}}$ holds if and only if $w \in \truthset{\mopEEE{M}}{\mmDg{\varphi}}$, which holds if and only if $\sbDg{(\mopEEE{\R})}(w) \subseteq \truthset{\mopEEE{M}}{\varphi}$. But, from \itemfautoref{pro:EEE:mo}{pro:EEE:mo:D} and semantic interpretation, the latter is equivalent to $\sbDs{\R}{\ag}(w) \subseteq \truthset{M}{\mmEEE{\varphi}}$, which holds if and only if $w \in \truthset{M}{\mmDs{\ag}{\mmEEE{\varphi}}}$. \item By definition, $w \in \truthset{M}{\mmEEE{\mmEEE{\varphi}}}$ if and only if $w \in \truthset{\mopEEE{M}}{\mmEEE{\varphi}}$, i.e., if and only if $w \in \truthset{\mopEEE{(\mopEEE{M})}}{\varphi}$. Then, as a consequence of \itemfautoref{pro:EEE:mo}{pro:EEE:mo:EEE}, the latter holds if and only if $w \in \truthset{\mopEEE{M}}{\varphi}$, that is, if and only if $w \in \truthset{M}{\mmEEE{\varphi}}$.\end{inlineenum}
  \end{proof}
\end{proposicion}

In particular, the first item of \autoref{pro:EEE} provides a characterisation of the knowledge any agent $\agi$ has after an act of \EEE-communication:
\[ \Vdash \mmEEE{\mmKi{\varphi}} \lldimp \mmDs{\ag}{\mmEEE{\varphi}}. \]

\msparagraph{Axiom system} To axiomatise a modality whose semantic interpretation relies on a model operation, a common \ti{DEL} strategy is to provide \emph{recursion axioms}: valid formulas and validity-preserving rules defining a translation that takes a formula with model-changing modalities (a formula in the `dynamic' language) and returns one without them (a formula in the initial `static' language). To prove soundness within this strategy, it is enough to show that the new axioms and rules are valid and preserve validity; this also shows that a formula and its translation are \emph{semantically} equivalent. To prove completeness, notice that the recursion axioms make a formula and its translation \emph{provably} equivalent, and thus one can rely on the completeness of the axiom system for the `static' language. The reader is referred to \citet[Chapter 7]{vanDitmarschEtAl2007} and \citet{WangCao2013} for an extensive explanation of this technique.

\medskip

For the case at hand, the recursion axioms appear on \autoref{tbl:LOdEEE}. Axiom \dsaEEE{p} states that $\modEEE$ does not affect the truth value of atomic propositions, and axioms \dsaEEE{\lnot} and \dsaEEE{\land} state, respectively, that $\modEEE$ commutes with negation and distributes over conjunction. Axiom \dsaEEE{\opgDist} (\autoref{pro:EEE}) indicates that what a group $\sag$ knows distributively after the operation is exactly what the whole group $\ag$ knew distributively about $\modEEE$'s effects. Finally, \dsreEEE states that replacing logical equivalents within the scope of $\modEEE$ preserves logical equivalence. As detailed in the proof of \autoref{teo:LOdEEE} \autorefp{proof:teo:LOdEEE}, these axioms and rule define an ``inside-out' translation (see \citealp{Plaza1989} and \citealp[Section 4.4]{Gerbrandy1999}) that, when dealing with nested $\modEEE$, works first with the deepest occurrence before dealing with the rest.

\begin{table}[ht]
  \begin{smallctabular}{l@{\qquad}l}
    \toprule
    \begin{tabular}[t]{@{}l@{\;\;}l@{}}
      \dsaEEE{p}:     & $\vdashLOdEEE \mmEEE{p} \lldimp p$ \\
      \dsaEEE{\lnot}: & $\vdashLOdEEE \mmEEE{\lnot \varphi} \lldimp \lnot \mmEEE{\varphi}$ \\
      \dsaEEE{\land}: & $\vdashLOdEEE \mmEEE{(\varphi_1 \land \varphi_2)} \lldimp (\mmEEE{\varphi_1} \land \mmEEE{\varphi_2})$ \\
    \end{tabular}
    &
    \begin{tabular}[t]{@{}l@{\;\;}l@{}}
      \dsaEEE{\opgDist}: & $\vdashLOdEEE \mmEEE{\mmDg{\varphi}} \lldimp \mmDs{\ag}{\mmEEE{\varphi}}$ \\
      \dsreEEE:          & If $\vdashLOdEEE \varphi_1 \ldimp \varphi_2$ then $\vdashLOdEEE \mmEEE{\varphi_1} \ldimp \mmEEE{\varphi_2}$
    \end{tabular} \\
    \bottomrule
  \end{smallctabular}
  \caption{Additional axioms and rules for \LOdEEE, which characterises the formulas in \LAdEEE that are valid on models in \marm.}
  \label{tbl:LOdEEE}
\end{table}

\begin{teorema}\label{teo:LOdEEE}
  The axiom system \LOdEEE (\LOd[\autoref{tbl:LOd}]+\autoref{tbl:LOdEEE}) is sound and strongly complete for formulas in \LAdEEE valid over models in \marm.
\end{teorema}

\subsection{\emph{Somebody} shares everything with everybody}\label{sbs:SEE}

A simple generalisation of the act of \EEE communication is one where only \emph{some} agents share everything they know with everybody. This will be called an act of \SEE-communication.

\bsparagraph{Operation and modality} The operation representing \SEE communication is a simple variation of the \EEE-case. First, worlds and valuation are preserved, as before. Then, the indistinguishability relation of each agent in the new model is defined as the intersection of her original relation with those of the agents that share their information (the \emph{senders}).

\begin{definicion}[\SEE-communication]\label{def:SEEmo}
  Let $M = \tupla{W, \R, V}$ be in \marm; take $\sen \subseteq \ag$. The relations in the \marm-model $\mopSEE{M}{\sen} = \tuplan{W, \mopSEE{\R}{\sen}, V}$ are given, for each $\agi \in \ag$, as
  \[ \sbi{\mopSEE{\R}{\sen}} := \sbi{\R} \cap \sbDs{\R}{\sen}. \]
\end{definicion}

Thus, after the operation, an agent $\agi$ cannot distinguish $u$ from $w$ (that is, $\sbi{\mopSEE{\R}{\sen}} wu$) if and only if, before the operation, neither she nor any agent \emph{in \sen} could distinguish $u$ from $w$ (that is, $\sbi{\R}wu$ and $\sbDs{\R}{\sen}wu$). Here are some small and yet useful observations.
\begin{itemize}
  \item Obviously, $\sbi{\mopSEE{\R}{\sen}} = \sbDs{\R}{\sen \cup \set{\agi}}$ and, moreover, $\agi \in \sen$ implies $\sbi{\mopSEE{\R}{\sen}} = \sbDs{\R}{\sen}$.
  \item For any world $w$ in any model $M$, the set $\sbi{\mopSEE{\R}{\sen}}(w) = \sbi{\R}(w) \cap \sbDs{\R}{\sen}(w)$ is a subset of $\sbi{\R}(w)$. Hence, just as \EEE, the action of \SEE-communication can only \emph{reduce} the uncertainty of each agent.
  \item As one might expect, an \SEE-communication with $\ag$ the communicating agents is exactly an \EEE-communication, as $\sbi{\mopSEE{\R}{\ag}} = \sbi{\mopEEE{\R}}$ for every $\agi \in \ag$ in any model $M = \tupla{W, \R, V}$.
  \item If $\sbi{\R}$ and the relations in $\set{\sbj{\R} \mid \agj \in \sen}$ are all reflexive (resp., transitive, symmetric, Euclidean), then so is the resulting $\sbi{\mopEEE{\R}}$. In particular, the \SEE operation preserves equivalence relations.
\end{itemize}

{\medskip}

Here is the associated modality.

\begin{definicion}[Modality $\modSEE{\sen}$; language \LAdSEE]\label{def:SEEmm}
  The language \LAdSEE extends \LAd with a modality $\modSEE{\sen}$ for each set of agents $\sen \subseteq \ag$. Its semantic interpretation in a pointed model $(M, w)$ is given by
  \begin{lbtabular}{l@{\qssidefq}l}
    $(M, w) \Vdash \mmSEE{\sen}{\varphi}$ & $(\mopSEE{M}{\sen}, w) \Vdash \varphi$.
  \end{lbtabular}
\end{definicion}

Using the alternative notation, $w \in \truthset{M}{\mmSEE{\sen}{\varphi}}$ if and only if $w \in \truthset{\mopSEE{M}{\sen}}{\varphi}$. Thus,
\[ \truthsetn{M}{\mmSEE{\sen}{\varphi}} = \truthset{\mopSEE{M}{\sen}}{\varphi}. \]

\medskip

\begin{ejemplo}\label{eje:SEE}
  Here are examples of this operation at work.
  \begin{compactenumerate}
    \item\label{eje:SEE:sym} Recall the model $M_1$ from \itemfautoref{eje:marl}{eje:marl:sym} (diagram below on the left).
    \begin{ctabular}{@{}c@{\quad}c@{\quad}c@{}}
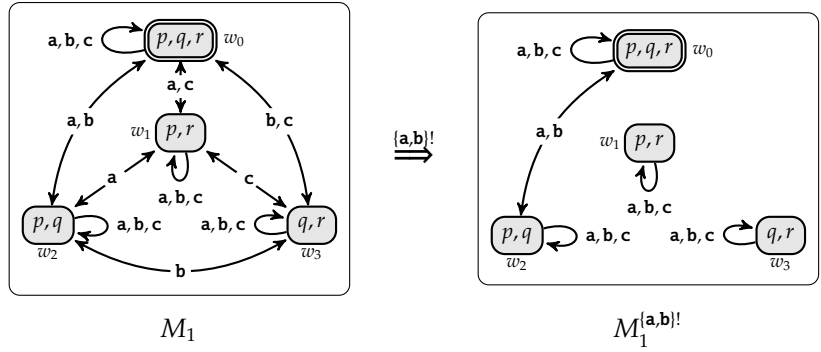
      \begin{tabular}{c}
        \begin{tikzpicture}[node distance = 2em and 3em, frame rectangle]
          \node [mundo, label = {[etiqMundo, outer sep = 2pt]right:$w_0$}, double] (w0) {$p, q, r$};
          \node [mundo, label = {[etiqMundo]left:$w_1$}, below = of w0] (w1) {$p,r$};
          \node [mundo, label = {[etiqMundo]below:$w_2$}, below left = of w1] (w2) {$p,q$};
          \node [mundo, label = {[etiqMundo]below:$w_3$}, below right = of w1] (w3) {$q,r$};

          \path (w0) edge [flecha, loop left] node [etiqFlecha] {$\aga,\agb,\agc$} ()
                     edge [flecha, <->] node [etiqFlecha, within] {$\aga,\agc$} (w1)
                     edge [flecha, <->, bend right = 25] node [etiqFlecha, within] {$\aga, \agb$} (w2)
                     edge [flecha, <->, bend left = 25] node [etiqFlecha, within] {$\agb,\agc$} (w3)
                (w1) edge [flecha, loop below] node [etiqFlecha] {$\aga,\agb,\agc$} ()
                     edge [flecha, <->] node [etiqFlecha, within] {$\aga\;$} (w2)
                     edge [flecha, <->] node [etiqFlecha, within] {$\;\agc$} (w3)
                (w2) edge [flecha, loop right] node [etiqFlecha] {$\aga,\agb,\agc$} ()
                     edge [flecha, <->, bend right] node [etiqFlecha, within] {$\agb$} (w3)
                (w3) edge [flecha, loop left] node [etiqFlecha] {$\aga,\agb,\agc$} ();
        \end{tikzpicture}
      \end{tabular}
      & $\overset{\symbSEE{\set{\aga, \agb}}}{\bs{\Longrightarrow}}$ &
      \begin{tabular}{c}
        \begin{tikzpicture}[node distance = 2em and 3em, frame rectangle]
          \node [mundo, label = {[etiqMundo, outer sep = 2pt]right:$w_0$}, double] (w0) {$p, q, r$};
          \node [mundo, label = {[etiqMundo]left:$w_1$}, below = of w0] (w1) {$p,r$};
          \node [mundo, label = {[etiqMundo]below:$w_2$}, below left = of w1] (w2) {$p,q$};
          \node [mundo, label = {[etiqMundo]below:$w_3$}, below right = of w1] (w3) {$q,r$};

          \path (w0) edge [flecha, loop left] node [etiqFlecha] {$\aga,\agb,\agc$} ()
                     edge [flecha, <->, bend right = 25] node [etiqFlecha, within] {$\aga, \agb$} (w2)
                (w1) edge [flecha, loop below] node [etiqFlecha] {$\aga,\agb,\agc$} ()
                (w2) edge [flecha, loop right] node [etiqFlecha] {$\aga,\agb,\agc$} ()
                (w3) edge [flecha, loop left] node [etiqFlecha] {$\aga,\agb,\agc$} ();
        \end{tikzpicture}
      \end{tabular}
      \\
      $M_1$ &  & $\mopSEE{M_1}{\set{\aga, \agb}}$
    \end{ctabular}
    The action $\symbSEE{\set{\aga,\agb}}$ produces the model depicted by the diagram above on the right. The new uncertainty of each sharing agent is given by the old uncertainty of the sharers (thus, e.g., $\sba{\mopSEE{\R}{\set{\aga,\agb}}} = \sbDs{\R}{\set{\aga,\agb}}$), and the new uncertainty of each non-sharing agent is given by the old uncertainty of the sharers \emph{and herself} (thus, e.g., $\sbc{\mopSEE{\R}{\set{\aga,\agb}}} = \sbDs{\R}{\set{\aga,\agb,\agc}}$). Thus,
    \begin{ltabular}{@{}l@{}}
      $(M_1, w_0) \Vdash
      \displaystyle
      \bigwedge \left\{
                  \begin{array}{l}
                    \mmDs{\set{\aga,\agb}}{(p \land q)} \;\land\; \mmKc{r}, \\
                    \mmSEE{\set{\aga,\agb}}{\Bigg(
                      \bigwedge
                      \left\{
                        \begin{array}{l}
                          \displaystyle \bigwedge_{\agi \in \set{\aga, \agb}} \big( \mmKi{(p \land q)} \land (\lnot \mmKi{r} \land \lnot \mmKi{\lnot r}) \big), \\
                          \mmKc{(p \land q \land r)}
                        \end{array}
                      \right\}
                    \Bigg)}
                  \end{array}
                \right\}$.
    \end{ltabular}
    In words, while $\aga$ and $\agb$ know $p \land q$ distributively, $\agc$ knows $r$. Then, after $\aga$ and $\agb$ share all their information to everyone, they both get to know $p \land q$ but still do not know whether $r$. However, $c$ gets to know what the real situation is. Analogous situations result if the communicating agents are $\set{\aga,\agc}$ or $\set{\agb,\agc}$.

    \item\label{eje:SEE:asym} Now recall $M_2$ from \itemfautoref{eje:marl}{eje:marl:asym} (diagram below on the left).
    \begin{ctabular}{c@{\quad}c@{\quad}c}
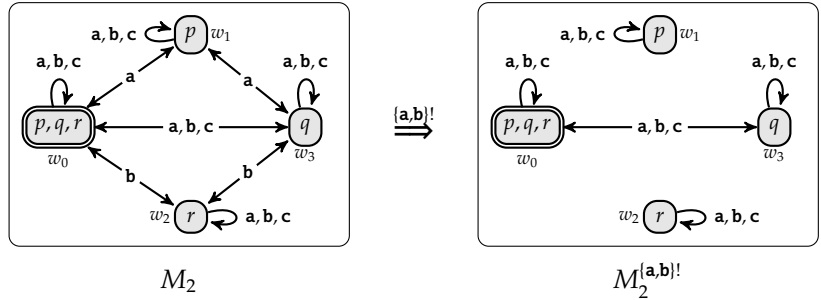
      \begin{tabular}{c}
        \begin{tikzpicture}[node distance = 2em and 3em, frame rectangle]
          \node [mundo, label = {[etiqMundo, outer sep = 2pt]below:$w_0$}, double] (w0) {$p, q, r$};
          \node [mundo, label = {[etiqMundo]right:$w_1$}, above right = of w0] (w1) {$p$};
          \node [mundo, label = {[etiqMundo]left:$w_2$}, below right = of w0] (w2) {$r$};
          \node [mundo, label = {[etiqMundo]below:$w_3$}, below right = of w1] (w3) {$q$};

          \path (w0) edge [flecha, loop above] node [etiqFlecha] {$\aga,\agb,\agc$} ()
                     edge [flecha, <->] node [etiqFlecha, within] {$\aga$} (w1)
                     edge [flecha, <->] node [etiqFlecha, within] {$\agb$} (w2)
                     edge [flecha, <->] node [etiqFlecha, within] {$\aga,\agb,\agc$} (w3)
                (w1) edge [flecha, loop left] node [etiqFlecha] {$\aga,\agb,\agc$} ()
                     edge [flecha, <->] node [etiqFlecha, within] {$\aga$} (w3)
                (w2) edge [flecha, loop right] node [etiqFlecha] {$\aga,\agb,\agc$} ()
                     edge [flecha, <->] node [etiqFlecha, within] {$\agb$} (w3)
                (w3) edge [flecha, loop above] node [etiqFlecha] {$\aga,\agb,\agc$} ();
        \end{tikzpicture}
      \end{tabular}
      & $\overset{\symbSEE{\set{\aga,\agb}}}{\bs{\Longrightarrow}}$ &
      \begin{tabular}{c}
        \begin{tikzpicture}[node distance = 2em and 3em, frame rectangle]
          \node [mundo, label = {[etiqMundo, outer sep = 2pt]below:$w_0$}, double] (w0) {$p, q, r$};
          \node [mundo, label = {[etiqMundo]right:$w_1$}, above right = of w0] (w1) {$p$};
          \node [mundo, label = {[etiqMundo]left:$w_2$}, below right = of w0] (w2) {$r$};
          \node [mundo, label = {[etiqMundo]below:$w_3$}, below right = of w1] (w3) {$q$};

          \path (w0) edge [flecha, loop above] node [etiqFlecha] {$\aga,\agb,\agc$} ()
                     edge [flecha, <->] node [etiqFlecha, within] {$\aga,\agb,\agc$} (w3)
                (w1) edge [flecha, loop left] node [etiqFlecha] {$\aga,\agb,\agc$} ()
                (w2) edge [flecha, loop right] node [etiqFlecha] {$\aga,\agb,\agc$} ()
                (w3) edge [flecha, loop above] node [etiqFlecha] {$\aga,\agb,\agc$} ();
        \end{tikzpicture}
      \end{tabular}
      \\
      $M_2$ &  & $\mopSEE{M_2}{\set{\aga,\agb}}$
    \end{ctabular}
    The model $\mopSEE{M_2}{\set{\aga,\agb}}$, which results from $\aga$ and $\agb$ sharing all their information to everyone, appears above on the right. Note that the same model results if $\agc$ is the only communicating agent (i.e., $\mopSEE{M_2}{\set{\aga,\agb}} = \mopSEE{M_2}{\set{\agc}}$), and also if the communication is \EEE (i.e., $\mopSEE{M_2}{\set{\aga,\agb}} = \mopEEE{M_2}$; see \itemfautoref{eje:EEE}{eje:EEE:asym}). Thus,
    \[
      \displaystyle (M_2, w_0)
      \Vdash
      \big(\mmSEE{\set{\aga,\agb}}{\varphi} \ldimp \mmSEE{\set{\agc}}{\varphi}\big)
      \land
      \big(\mmSEE{\set{\aga,\agb}}{\varphi} \ldimp \mmEEE{\varphi}\big).
    \]
    The reason for this is that what agents in $\set{\aga,\agb}$ know distributively is exactly what $\agc$ knows individually (i.e., $\sbDs{\R}{\set{\aga,\agb}} = \sbc{\R}$).
  \end{compactenumerate}
\end{ejemplo}

\bsparagraph{Properties} What is the effect of a \SEE operation? Intuitively, the action turns distributive knowledge of a group into individual knowledge of the group's members. Just as in the \EEE case, this is true only up to a certain extent.

\begin{proposicion}\label{pro:SEE:Moore}
  Let $(M, w)$ be a pointed \marm model; take $\sen \subseteq \ag$ and $\agi \in \ag$. Then, $(M, w) \Vdash \mmDs{\sen}{\varphi} \limp \mmSEE{\sen}{\mmKi{\varphi}}$ holds when $M$ and $\varphi$ are such that $\truthset{M}{\varphi} \subseteq \truthset{\mopSEE{M}{\sen}}{\varphi}$. However, $\not\Vdash \mmDs{\sen}{\varphi} \limp \mmSEE{\sen}{\mmKi{\varphi}}$.
  \begin{proof}
    As that of \autoref{pro:EEE:Moore}, here using $\sbi{\mopSEE{\R}{\sen}}(w) \subseteq \sbi{\R}(w)$ ($\symbSEE{\sen}$ can only reduce uncertainty). For showing $\not\Vdash \mmDs{\sen}{\varphi} \limp \mmSEE{\sen}{\mmKi{\varphi}}$ for $\agi \in \sen$, take $\varphi := \lnot \mmKb{p}$ and note that, on $M_1$ in \itemfautoref{eje:SEE}{eje:SEE:sym}, $(M_1, w_0) \Vdash \mmDs{\set{\aga,\agb}}{\lnot \mmKb{p}} \;\land\; \lnot \mmSEE{\set{\aga,\agb}}{\mmKa{\lnot \mmKb{p}}}$. Again, there is not guarantee that every world making $\varphi$ true in a given $M$ (a world in $\truthset{M}{\varphi}$) is also a world making $\varphi$ true \emph{in $\mopSEE{M}{\sen}$} (a world in $\truthset{\mopSEE{M}{\sen}}{\varphi}$).
  \end{proof}
\end{proposicion}

As in the \EEE case, $\not\Vdash \mmDs{\sen}{\varphi} \limp \mmSEE{\sen}{\mmKi{\varphi}}$ for $\agi \in \sen$ should not be taken as a drawback for the \SEE operation: the formula is valid for the situations one intuitively considers (i.e., when $\varphi$ is propositional) and also for further fragments of \LAd (see the discussion after \autoref{pro:EEE:Moore}).

\medskip

Here are two further results.

\begin{proposicion}\label{pro:SEE:mo}
  Let $M = \tupla{W, \R, V}$ be in \marm; take $\sen, \sen_1, \sen_2 \subseteq \ag$. Let $\mopSEE{M}{\sen} = \tuplan{W, \mopSEE{\R}{\sen}, V}$ and $\mopSEE{(\mopSEE{M}{\sen})}{\sen'} = \tuplan{W, \mopSEE{(\mopSEE{\R}{\sen})}{\sen'}, V}$ be as indicated in \autoref{def:SEEmo}. Then,
  \begin{multicols}{2}
    \begin{enumerate}
      \item\label{pro:SEE:mo:D} $\sbDg{(\mopSEE{\R}{\sen})} = \sbDs{\R}{\sen \cup \sag}$ \, for $\sag \subseteq \ag$;
      \item\label{pro:SEE:mo:SEE} $\sbi{\mopSEE{(\mopSEE{\R}{\sen_1})}{\sen_2}} = \sbi{\mopSEE{\R}{{(\sen_1 \cup \sen_2)}}}$ \, for $\agi \in \ag$.
    \end{enumerate}
  \end{multicols}
  \begin{proof}
    \begin{inlineenum} \item For any $\sag \subseteq \ag$, the relation $\sbDg{(\mopSEE{\R}{\sen})}$ is $\bigcap_{\agk \in \sag} \sbk{\mopSEE{\R}{\sen}}$ (def. of $\sbDg{(\mopSEE{\R}{\sen})}$), equal to $\bigcap_{\agk \in \sag} \sbDs{\R}{\sen \cup \set{\agk}}$ (def. of $\sbk{\mopSEE{\R}{\sen}}$), equal to $\sbDs{\R}{\sen \cup \sag}$. \item For any $\agi \in \ag$, the relation $\sbi{\mopSEE{(\mopSEE{\R}{\sen_1})}{\sen_2}}$ is $\sbDs{(\mopSEE{\R}{\sen_1})}{\sen_2 \cup \set{\agi}}$ (def. of $\sbi{\mopSEE{(\mopSEE{\R}{\sen_1})}{\sen_2}}$), equal to $\bigcap_{\agk \in \sen_2 \cup \set{\agi}} \sbk{\mopSEE{\R}{\sen_1}}$ (def. of $\sbDs{(\mopSEE{\R}{\sen_1})}{\sen_2 \cup \set{\agi}}$), equal to $\bigcap_{\agk \in \sen_2 \cup \set{\agi}} \sbDs{\R}{\sen_1 \cup \set{\agk}}$ (def. of $\sbk{\mopSEE{\R}{\sen_1}}$), equal to $\sbDs{\R}{\sen_1 \cup \sen_2 \cup \set{\agi}}$, equal to $\sbi{\mopSEE{\R}{{(\sen_1 \cup \sen_2)}}}$ (def. of $\sbi{\mopSEE{\R}{(\sen_1 \cup \sen_2)}}$).\end{inlineenum}
  \end{proof}
\end{proposicion}

Analogous to \autoref{pro:EEE:mo} for the \EEE-case, \autoref{pro:SEE:mo} states two facts. First, it states that the relation for interpreting $\modDg$ in the new model, $\sbDg{(\mopSEE{\R}{\sen})}$, is the same as the relation for interpreting $\modDs{\sen \cup \sag}$ in the original model, $\sbDs{\R}{\sen \cup \sag}$. As in the \EEE case, this provides a validity \itemfautorefp{pro:SEE}{pro:SEE:D} that is crucial for an axiomatisation by translation.

The second fact is relative to how two consecutive \SEE acts can be `comprised' into a single one. More precisely, it indicates how a `\SEE-conversation' in which group $\sen_1$ shares first and group $\sen_2$ shares second can be replaced by a single \SEE-act in which the sharing agents are those in $\sen_1 \cup \sen_2$. Given that $\mopSEE{(\mopSEE{M}{\sen_1})}{\sen_2}$ and $\mopSEE{M}{(\sen_1 \cup \sen_2)}$ have the same domain and atomic valuation, this implies that $\mopSEE{(\mopSEE{M}{\sen_1})}{\sen_2} = \mopSEE{M}{(\sen_1 \cup \sen_2)}$ holds for any $M$ in \marm.

\medskip

Thus,

\begin{proposicion}\label{pro:SEE}
  \begin{inlineenum} \item\label{pro:SEE:D} $\Vdash \mmSEE{\sen}{\mmDg{\varphi}} \ldimp \mmDs{\sen \cup \sag}{\mmSEE{\sen}{\varphi}}$;  \item\label{pro:SEE:SEE} $\Vdash \mmSEE{\sen_1}{\mmSEE{\sen_2}{\varphi}} \ldimp \mmSEE{(\sen_1 \cup \sen_2)}{\varphi}$.\end{inlineenum}
  \begin{proof}
    Let $M = \tupla{W, \R, V}$ be in \marm. \begin{inlineenum} \item By definition, $w \in \truthset{M}{\mmSEE{\sen}{\mmDg{\varphi}}}$ holds if and only if $w \in \truthset{\mopSEE{M}{\sen}}{\mmDg{\varphi}}$, which holds if and only if $\sbDg{(\mopSEE{\R}{\sen})}(w) \subseteq \truthset{\mopSEE{M}{\sen}}{\varphi}$. But, from \itemfautoref{pro:SEE:mo}{pro:SEE:mo:D} and semantic interpretation, the latter is equivalent to $\sbDs{\R}{(\sen \cup \sag)}(w) \subseteq \truthset{M}{\mmSEE{\sen}{\varphi}}$, which holds if and only if $w \in \truthsetn{M}{\mmDs{(\sen \cup \sag)}{\mmSEE{\sen}{\varphi}}}$. \item By definition, $w \in \truthset{M}{\mmSEE{\sen_1}{\mmSEE{\sen_2}{\varphi}}}$ if and only if $w \in \truthset{\mopSEE{M}{\sen_1}}{\mmSEE{\sen_2}{\varphi}}$, i.e., if and only if $w \in \truthset{\mopSEE{(\mopSEE{M}{\sen_1})}{\sen_2}}{\varphi}$. Then, as a consequence of \itemfautoref{pro:SEE:mo}{pro:SEE:mo:SEE}, the latter holds if and only if $w \in \truthset{\mopSEE{M}{(\sen_1 \cup \sen_2)}}{\varphi}$, that is, if and only if $w \in \truthset{M}{\mmSEE{(\sen_1 \cup \sen_2)}{\varphi}}$.\end{inlineenum}
  \end{proof}
\end{proposicion}

Thanks to these validities and previous observations, one can find further validities describing properties of an act of \SEE communication. First, as it was mentioned, the \EEE operation of the previous subsection is the particular case of the \SEE operation in which all agents share. Thus,
\[ \Vdash \mmEEE{\varphi} \ldimp \mmSEE{\ag}{\varphi}.  \]
From \itemfautoref{pro:SEE}{pro:SEE:D}, it follows that the knowledge any agent $\agi$ has after an act of \SEE-communication is what was distributively known among her and the sharing agents about the effects of the action:
\[ \Vdash \mmSEE{\sen}{\mmKi{\varphi}} \lldimp \mmDs{\set{\agi} \cup \sen}{\mmSEE{\sen}{\varphi}}. \]
From \itemfautoref{pro:SEE}{pro:SEE:SEE} it follows that an act of \SEE-communication by the same group is idempotent:
\[ \Vdash \mmSEE{\sen}{\mmSEE{\sen}{\varphi}} \lldimp \mmSEE{\sen}{\varphi}. \]
Moreover, it also follows that, when two groups communicate, the order in which they do so is irrelevant:
\[ \Vdash \mmSEE{\sen_1}{\mmSEE{\sen_2}{\varphi}} \lldimp \mmSEE{\sen_2}{\mmSEE{\sen_1}{\varphi}}. \]
Thus, repeated \SEE-sharing by the same group does not provide anything new, regardless of whether it is immediate or after some other groups have shared:
\[ \Vdash \mmSEE{\sen_1}{\cdots\mmSEE{\sen_m}{\mmSEE{\sen_1}{\varphi}}} \lldimp \mmSEE{\sen_1}{\cdots\mmSEE{\sen_m}{\varphi}}. \]

\bsparagraph{Axiom system} The axiom system for the extended language $\LAdSEE$ relies again on the \ti{DEL} reduction axioms technique, with the axioms and rule for the case at hand being those on \autoref{tbl:LOdSEE}. Axioms \dsaSEE{p}, \dsaSEE{\lnot}, \dsaSEE{\land} and rule \dsreSEE are as in the \EEE case, indicating respectively that $\modSEE{\sen}$ does not affect atomic propositions, commutes with negation, distributes over conjunction and `preserves' logical equivalence. Axiom \dsaSEE{\opgDist} is the one that distinguishes \SEE from \EEE, indicating that what a group $\sag$ knows distributively after $\modSEE{\sen}$ is exactly what the group $\sen \cup \sag$ knew distributively about $\modSEE{\sen}$'s effects. Together, these axioms and rule define an ``inside-out' translation from \LAdSEE to \LAd such that a formula and its translation are both semantically and provably equivalent (see the proof of \autoref{teo:LOdSEE} on \autoref{proof:teo:LOdSEE}).

\begin{table}[ht]
  \begin{smallctabular}{l@{\qquad}l}
    \toprule
    \begin{tabular}[t]{@{}l@{\;\;}l@{}}
      \dsaSEE{p}:     & $\vdashLOdSEE \mmSEE{\sen}{p} \lldimp p$ \\
      \dsaSEE{\lnot}: & $\vdashLOdSEE \mmSEE{\sen}{\lnot \varphi} \lldimp \lnot \mmSEE{\sen}{\varphi}$ \\
      \dsaSEE{\land}: & $\vdashLOdSEE \mmSEE{\sen}{(\varphi \land \psi)} \lldimp (\mmSEE{\sen}{\varphi} \land \mmSEE{\sen}{\psi})$ \\
    \end{tabular}
    &
    \begin{tabular}[t]{@{}l@{\;\;}l@{}}
      \dsaSEE{\opgDist}: & $\vdashLOdSEE \mmSEE{\sen}{\mmDg{\varphi}} \lldimp \mmDs{\sen \cup \sag}{\mmSEE{\sen}{\varphi}}$ \\
      \dsreSEE:          & If $\vdashLOdSEE \varphi_1 \ldimp \varphi_2$ then $\vdashLOdSEE \mmSEE{\sen}{\varphi_1} \ldimp \mmSEE{\sen}{\varphi_2}$
    \end{tabular} \\
    \bottomrule
  \end{smallctabular}
  \caption{Additional axioms and rules for \LOdSEE, which characterises the formulas in \LAdSEE that are valid on models in \marm.}
  \label{tbl:LOdSEE}
\end{table}

\begin{teorema}\label{teo:LOdSEE}
  The axiom system \LOdSEE (\LOd[\autoref{tbl:LOd}]+\autoref{tbl:LOdSEE}) is sound and strongly complete for formulas in \LAdSEE valid over models in \marm.
\end{teorema}

\subsection{Other operations for communication in the literature}\label{sbs:XEE:others}

The operations proposed in this section are not the first ones representing actions of agents sharing their individual information. The action of ``tell us all you know'' of \citet{Baltag2010slides} is one through which \emph{a single agent} $\aga \in \ag$ shares \emph{all her information} with \emph{every agent}. Thus, the new indistinguishability relation of each agent $\agi \in \ag$ is defined as $\sbi{\R} \cap \sba{\R}$. This action can be seen as the particular instance of the act of \SEE-communication \autorefp{sbs:SEE} in which only one agent communicates (thus corresponding to the modality $\modSEE{\set{\agi}}$). Then there is the action for `resolving the distributed knowledge of a group' studied in \citet{AgotnesWang2017}. Through it, agents \emph{in a group $\sag$} share \emph{all their information} only \emph{within $\sag$ itself}. Thus, while the new indistinguishability relation of every agent not in $\sag$ remains exactly as before, that of each agent in $\sag$ is defined as $\sbDs{\R}{\sag}$.\footnote{With this definition, it is interesting to notice that, while agents not in $\sag$ do not receive the information that is being shared, they still get to know that agents in $\sag$ shared their information within themselves. Thus, using the terminology of \citet{BaltagSmets2020}, `resolving distributed knowledge' is a \emph{semi-public} form of communication.} This action can be also seen as an instance of the act of \SEE-communication \autorefp{sbs:SEE} (only agents in $\sag$ share), but note that only agents in $\sag$ receive the information. In the terminology of this manuscript, this action is better described as an act of `\SES-communication' (\autoref{sec:XXS} will discuss briefly \SSS-communication). Finally, there is the more general action of `semi-public reading events' of \citet{BaltagSmets2020}, where each agent $\agi$ gets a set $\alpha(\agi) \subseteq \ag$ satisfying $\agi \in\alpha(\agi)$. Intuitively, $\alpha(\agi)$ contains those agents whose information $\agi$ will receive when communication occurs. Thus, for each agent $\agi$, the operation defines her new indistinguishability relation as $\sbDs{\R}{\alpha(\agi)}$. Note again that this form of communication is semi-public: even though each agent $\agi$ only `hears' what agents in $\alpha(\agi)$ `say', the definition implies that $\agi$ still learns that every agent $\agj$ receives the information provided by $\agj$'s sources (agents in $\alpha(\agj)$). This can be seen as a generalisation of a `\SEE'-communication: while some agents (those in $\alpha(\agi)$ for some agent $\agi$) share all their information, every agent receives information from potentially different agents (each agent $\agi$ only `listen to' those agents in $\alpha(\agi)$).

\section{Sharing \emph{something} with \emph{everybody}}\label{sec:SSE}

\autoref{sec:XEE} discussed model operations for actions through which all/some agents share all the information with everybody; thus, the actions represent acts of full communication (from a subset of agents). As discussed, the operations are small variations of proposals already present in the literature.

\smallskip

This section, which constitutes the core of this contribution, discusses a variation of the \SEE case: one in which the sharing agents communicate \emph{only part of their information}. This action makes the process of communication between agents a more realistic one. Indeed, there are natural restrictions on the `amount' of information an agent can communicate at once, and operations with such restrictions can represent more realistic `conversations'. In defining this operation, probably the most important question is the following: what defines what each agent will communicate? There are indeed several possibilities (see the discussion in \autoref{sec:end}), but a natural one is to assume that the `conversation' is relative to a given \emph{subject}/\emph{topic}. This manuscripts uses this idea, assuming that this subject/topic is defined by a formula $\chi$. Following the previous notation, this action through which \emph{some} agents share \emph{some} of their information (that relative to the given subject $\chi$) with everybody will be called an act of \SSE-communication.

\bsparagraph{Operation and modality} For the definitions, the following will be useful.

\begin{definicion}[Relations $\fullig{}{\chi}$ and $\knonfu{}{\chi}$]\label{def:fullig-knonfu}
  Let $M$ be in \marm; let $\chi$ be a formula that can be evaluated at worlds in $M$. The relation $\fullig{M}{\chi} \subseteq \dom{M}\times \dom{M}$ is given by
  \[
    \fullig{M}{\chi}
    :=
    \left(\truthset{M}{\chi} \times \truthset{M}{\lnot \chi}\right) \cup \left(\truthset{M}{\lnot \chi} \times \truthset{M}{\chi}\right).
  \]
  Its complement, the (note: equivalence) relation $\ov{\fullig{M}{\chi}}$ given by $(\truthset{M}{\chi} \times \truthset{M}{\chi}) \cup (\truthset{M}{\lnot \chi} \times \truthset{M}{\lnot \chi})$, will be denoted rather as $\knonfu{M}{\chi}$.
\end{definicion}

Note how $\fullig{M}{\chi}$ describes the indistinguishability of an agent that has full \emph{uncertainty} about $\chi$ (worlds differing on $\chi$'s truth-value cannot be distinguished) while also having full \emph{certainty} about everything else (all other worlds can be distinguished). Analogously, $\knonfu{M}{\chi}$ describes the indistinguishability of an agent that has full \emph{certainty} about $\chi$ (worlds differing on $\chi$'s truth-value can be distinguished) while also having full \emph{uncertainty} about everything else (worlds agreeing on $\chi$'s truth-value are indistinguishable). In other words, while the relation $\fullig{M}{\chi}$ contains the pairs of worlds in $\dom{M} \times \dom{M}$ that would be indistinguishable if the available information allowed to tell apart any pair of formulas but $\chi$ and $\lnot\chi$, the relation $\knonfu{M}{\chi}$ contains the pairs of worlds in $\dom{M} \times \dom{M}$ that would be indistinguishable if the available information allowed to tell apart $\chi$ and $\lnot\chi$, and nothing else. Thus, while $\fullig{M}{\chi}$ can be seen as a relation of \emph{full ignorance on $\chi$}, $\knonfu{M}{\chi}$ can be seen as a relation of \emph{knowing only $\chi$ fully}.

\medskip

Here is, then, the definition of the operation for \SSE-communication.

\begin{definicion}[\SSE-communication]\label{def:SSEmo}
  Let $M = \tupla{W, \R, V}$ be in \marm; take $\sen \subseteq \ag$ and let $\chi$ be a formula. The relations in $\mopSSE{M}{\sen}{\chi} = \tuplan{W, \mopSSE{\R}{\sen}{\chi}, V}$ (a structure in \marm) are given, for each $\agi \in \ag$, as
  \[ \sbi{\mopSSE{\R}{\sen}{\chi}} := \sbi{\R} \setminus \bigcup_{\agj \in \sen} (\comp{\sbj{\R}} \cap \fullig{M}{\chi}). \]
\end{definicion}

In order to understand the intuition behind the definition, note how the set $\comp{\sbj{\R}} \cap \fullig{M}{\chi}$ can be seen as the \emph{epistemic contribution} of agent $\agj$ about formula $\chi$ in model $M$, as it contains those pairs in the $M$-uncertainty about $\chi$ (the set $\fullig{M}{\chi}$) that $\agj$ has already discarded (the set $\comp{\sbj{\R}}$). With this in mind, the definition states that agent $\agi$'s uncertainty after agents in $\sen$ share all their information \emph{on subject $\chi$} (i.e., her uncertainty in the new model, $\sbi{\mopSSE{\R}{\sen}{\chi}}$) is given by her previous uncertainty (i.e., $\sbi{\R}$) minus the sum ($\bigcup$) of the epistemic contribution (i.e., $\comp{\sbj{\R}} \cap \fullig{M}{\chi}$) of each agent $\agj$ in $\sen$.\footnote{The definition might be easier to grasp by taking a `knowledge' perspective. Intuitively, the knowledge of the agent after a conversation in which agents in $\sen$ share what they know about $\chi$ is her initial knowledge plus the knowledge any agent in $\sen$ has about $\chi$. The provided definition describes exactly the same idea, stating it in terms of the agent's uncertainty.}

\smallskip

Note also how, through the operation, agents in $\sen$ communicate all they know \emph{about} $\chi$: intuitively, they share all the information that has allowed them to discard \emph{any} uncertainty between $\chi$- and $\lnot \chi$-worlds, and thus only edges in $\fullig{M}{\chi}$ can be eliminated. This emphasises the fact that $\chi$ is taken to be the \emph{subject}/\emph{topic} of the conversation, with agents intuitively sharing what has allowed them to discard edges between worlds disagreeing on $\chi$'s truth-value.\footnote{There is a natural alternative in which $\chi$ is rather taken to be the \emph{content} of the conversation. In this asymmetric version, agents intuitively share only what has allowed them to discard $\lnot\chi$ as a possibility, and thus only edges \emph{pointing to $\lnot\chi$-worlds} can be eliminated.}

\smallskip

The relations in the model resulting from \SSE-communication can be described in a simpler way.

\begin{proposicion}\label{pro:SSEmo}
  Let $M = \tupla{W, \R, V}$ be in \marm; take $\sen \subseteq \ag$ and let $\chi$ be a formula. Then,
  \begin{ctabular}{c}
    $\sbi{\mopSSE{\R}{\sen}{\chi}} = \sbi{\R} \cap (\sbDs{\R}{\sen} \cup \knonfu{M}{\chi})$.
  \end{ctabular}
  \begin{proof}
    By definition, $\sbi{\mopSSE{\R}{\sen}{\chi}} := \sbi{\R} \setminus \bigcup_{\agj \in \sen} (\comp{\sbj{\R}} \cap \fullig{M}{\chi})$, Then, from the definitions of both set subtraction and $\knonfu{M}{\chi}$, the right-hand side becomes $\sbi{\R} \cap \bigcap_{\agj \in \sen} (\sbj{\R} \cup \knonfu{M}{\chi})$. On the latter, using $\bigcap_{\agj \in \sen} (\sbj{\R} \cup \knonfu{M}{\chi}) = \sbDs{\R}{\sen} \cup \knonfu{M}{\chi}$ produces the required $\sbi{\R} \cap (\sbDs{\R}{\sen} \cup \knonfu{M}{\chi})$.
  \end{proof}
\end{proposicion}

Before discussing the \SSE operation further, here are some examples.

\begin{ejemplo}\label{eje:SSE}
  Recall the model $M_1$ from \itemfautoref{eje:marl}{eje:marl:sym}, where each agent knows the truth-value of one atom ($\aga$ knows whether $p$, $\agb$ knows whether $q$, $\agc$ knows whether $r$) but does not know the truth-value of the others.

  \begin{compactenumerate}
    \renewcommand{\arraystretch}{2}
    \item\label{eje:SSE:atom} The diagrams below (reflexivity assumed) depict $M_1$ three times. They show, respectively, the partition generated by the (recall: equivalence) relations $\knonfu{M}{p}$, $\knonfu{M}{q}$ and $\knonfu{M}{r}$.
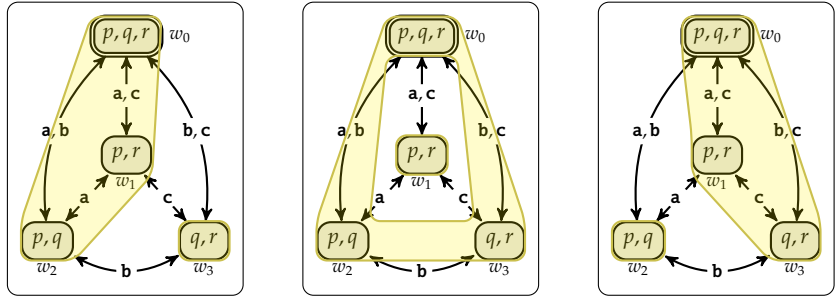
    \begin{ctabular}{@{}c@{\quad}c@{\quad}c@{}}
      \begin{tabular}{@{}c}
        \begin{tikzpicture}[node distance = 1.75em and 1em, frame rectangle]
          \node [mundo, label = {[etiqMundo, outer sep = 2pt]right:$w_0$}, double] (w0) {$p, q, r$};
          \node [mundo, label = {[etiqMundo]below:$w_1$}, below = 3em of w0] (w1) {$p,r$};
          \node [mundo, label = {[etiqMundo]below:$w_2$}, below left = of w1] (w2) {$p,q$};
          \node [mundo, label = {[etiqMundo]below:$w_3$}, below right = of w1] (w3) {$q,r$};

          \path (w0) edge [flecha, <->] node [etiqFlecha, within] {$\aga,\agc$} (w1)
                     edge [flecha, <->, bend right = 25] node [etiqFlecha, within] {$\aga, \agb$} (w2)
                     edge [flecha, <->, bend left = 25] node [etiqFlecha, within] {$\agb,\agc$} (w3)
                (w1) edge [flecha, <->] node [etiqFlecha, within] {$\aga\;$} (w2)
                     edge [flecha, <->] node [etiqFlecha, within] {$\;\agc$} (w3)
                (w2) edge [flecha, <->, bend right] node [etiqFlecha, within] {$\agb$} (w3);

          \draw[topicclass] (w0.north west) -- (w0.north east) -- (w1.south east) -- (w2.south east) -- (w2.south west) -- (w2.north west) -- cycle;
          \draw[topicclass] (w3.north east) -- (w3.south east) -- (w3.south west) -- (w3.north west) -- cycle;
        \end{tikzpicture}
      \end{tabular}
      &
      \begin{tabular}{c}
        \begin{tikzpicture}[node distance = 1.75em and 1em, frame rectangle]
          \node [mundo, label = {[etiqMundo, outer sep = 2pt]right:$w_0$}, double] (w0) {$p, q, r$};
          \node [mundo, label = {[etiqMundo]below:$w_1$}, below = 3em of w0] (w1) {$p,r$};
          \node [mundo, label = {[etiqMundo]below:$w_2$}, below left = of w1] (w2) {$p,q$};
          \node [mundo, label = {[etiqMundo]below:$w_3$}, below right = of w1] (w3) {$q,r$};

          \path (w0) edge [flecha, <->] node [etiqFlecha, within] {$\aga,\agc$} (w1)
                     edge [flecha, <->, bend right = 25] node [etiqFlecha, within] {$\aga, \agb$} (w2)
                     edge [flecha, <->, bend left = 25] node [etiqFlecha, within] {$\agb,\agc$} (w3)
                (w1) edge [flecha, <->] node [etiqFlecha, within] {$\aga\;$} (w2)
                     edge [flecha, <->] node [etiqFlecha, within] {$\;\agc$} (w3)
                (w2) edge [flecha, <->, bend right] node [etiqFlecha, within] {$\agb$} (w3);

          \filldraw[topicclass, even odd rule]
            (w0.north west) -- (w0.north east) -- (w3.north east) -- (w3.south east) -- (w3.south west) -- (w2.south east) -- (w2.south west) -- (w2.north west) -- cycle
            (w0.south west) -- (w0.south) -- (w0.south east) -- (w3.north west) -- (w2.north east) -- cycle;
          \draw[topicclass] (w1.north east) -- (w1.south east) -- (w1.south west) -- (w1.north west) -- cycle;
        \end{tikzpicture}
      \end{tabular}
      &
      \begin{tabular}{c@{}}
        \begin{tikzpicture}[node distance = 1.75em and 1em, frame rectangle]
          \node [mundo, label = {[etiqMundo, outer sep = 2pt]right:$w_0$}, double] (w0) {$p, q, r$};
          \node [mundo, label = {[etiqMundo]below:$w_1$}, below = 3em of w0] (w1) {$p,r$};
          \node [mundo, label = {[etiqMundo]below:$w_2$}, below left = of w1] (w2) {$p,q$};
          \node [mundo, label = {[etiqMundo]below:$w_3$}, below right = of w1] (w3) {$q,r$};

          \path (w0) edge [flecha, <->] node [etiqFlecha, within] {$\aga,\agc$} (w1)
                     edge [flecha, <->, bend right = 25] node [etiqFlecha, within] {$\aga, \agb$} (w2)
                     edge [flecha, <->, bend left = 25] node [etiqFlecha, within] {$\agb,\agc$} (w3)
                (w1) edge [flecha, <->] node [etiqFlecha, within] {$\aga\;$} (w2)
                     edge [flecha, <->] node [etiqFlecha, within] {$\;\agc$} (w3)
                (w2) edge [flecha, <->, bend right] node [etiqFlecha, within] {$\agb$} (w3);

          \draw[topicclass] (w0.north west) -- (w0.north east) -- (w3.north east) -- (w3.south east) -- (w3.south west) -- (w1.south west) -- (w1.north west) -- (w0.south west) -- cycle;
          \draw[topicclass] (w2.north east) -- (w2.south east) -- (w2.south west) -- (w2.north west) -- cycle;
        \end{tikzpicture}
      \end{tabular}
    \end{ctabular}
    The diagrams further below (reflexivity assumed) show the result of three communication acts, all with $\sen = \set{\aga, \agb, \agc}$: the first on topic $p$, the second on topic $q$ and the third on topic $r$. When building the relations of the new model, the operation looks \emph{at the original model}, focussing only on edges between worlds disagreeing on the topic's truth-value (edges across partition cells in the diagrams above) and leaving the rest as they are. For example, for topic $p$, the operation focuss on edges between worlds in $\set{w_0, w_1, w_2}$ and worlds in $\set{w_3}$, disregarding (i.e., not affecting) the rest. When all agent share, as in this case, the operation simply removes the edges \emph{under consideration} (i.e., across partition cells) that are not in $\sbDs{\R}{\ag}$.
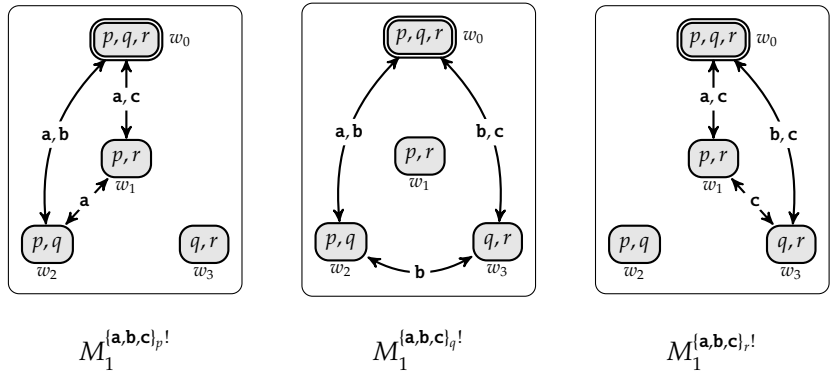
    \begin{ctabular}{@{}c@{\quad}c@{\quad}c@{}}
      \begin{tabular}{c}
        \begin{tikzpicture}[node distance = 1.75em and 1em, frame rectangle]
          \node [mundo, label = {[etiqMundo, outer sep = 2pt]right:$w_0$}, double] (w0) {$p, q, r$};
          \node [mundo, label = {[etiqMundo]below:$w_1$}, below = 3em of w0] (w1) {$p,r$};
          \node [mundo, label = {[etiqMundo]below:$w_2$}, below left = of w1] (w2) {$p,q$};
          \node [mundo, label = {[etiqMundo]below:$w_3$}, below right = of w1] (w3) {$q,r$};

          \path (w0) edge [flecha, <->] node [etiqFlecha, within] {$\aga,\agc$} (w1)
                     edge [flecha, <->, bend right = 25] node [etiqFlecha, within] {$\aga, \agb$} (w2)
                (w1) edge [flecha, <->] node [etiqFlecha, within] {$\aga\;$} (w2);
        \end{tikzpicture}
      \end{tabular}
      &
      \begin{tabular}{c}
        \begin{tikzpicture}[node distance = 1.75em and 1em, frame rectangle]
          \node [mundo, label = {[etiqMundo, outer sep = 2pt]right:$w_0$}, double] (w0) {$p, q, r$};
          \node [mundo, label = {[etiqMundo]below:$w_1$}, below = 3em of w0] (w1) {$p,r$};
          \node [mundo, label = {[etiqMundo]below:$w_2$}, below left = of w1] (w2) {$p,q$};
          \node [mundo, label = {[etiqMundo]below:$w_3$}, below right = of w1] (w3) {$q,r$};

          \path (w0) edge [flecha, <->, bend right = 25] node [etiqFlecha, within] {$\aga, \agb$} (w2)
                     edge [flecha, <->, bend left = 25] node [etiqFlecha, within] {$\agb,\agc$} (w3)
                (w2) edge [flecha, <->, bend right] node [etiqFlecha, within] {$\agb$} (w3);
        \end{tikzpicture}
      \end{tabular}
      &
      \begin{tabular}{c}
        \begin{tikzpicture}[node distance = 1.75em and 1em, frame rectangle]
          \node [mundo, label = {[etiqMundo, outer sep = 2pt]right:$w_0$}, double] (w0) {$p, q, r$};
          \node [mundo, label = {[etiqMundo]below:$w_1$}, below = 3em of w0] (w1) {$p,r$};
          \node [mundo, label = {[etiqMundo]below:$w_2$}, below left = of w1] (w2) {$p,q$};
          \node [mundo, label = {[etiqMundo]below:$w_3$}, below right = of w1] (w3) {$q,r$};

          \path (w0) edge [flecha, <->] node [etiqFlecha, within] {$\aga,\agc$} (w1)
                     edge [flecha, <->, bend left = 25] node [etiqFlecha, within] {$\agb,\agc$} (w3)
                (w1) edge [flecha, <->] node [etiqFlecha, within] {$\;\agc$} (w3);
        \end{tikzpicture}
      \end{tabular}
      \\
      $\mopSSE{M_1}{\set{\aga,\agb,\agc}}{p}$ & $\mopSSE{M_1}{\set{\aga,\agb,\agc}}{q}$ & $\mopSSE{M_1}{\set{\aga,\agb,\agc}}{r}$
    \end{ctabular}
    As the diagrams show, the operation behaves as expected. For example, a conversation among all agents about $p$ (leftmost model) benefits $\agb$ and $\agc$ (they get to know $p$'s truth-value) but does not benefit the only agent who knew $p$'s truth-value before, namely $\aga$.\footnote{To be more precise, the action does not give $\aga$ any \emph{factual} information. Yet, she gets information, as after the conversation she knows that both $\agb$ and $\agc$ know $p$'s truth-value.}

    \item\label{eje:SSE:disjunction} Again, below are three copies of $M_1$ (reflexivity assumed), this time showing (respectively) the partition generated by the relations $\knonfu{M}{p \land q}$, $\knonfu{M}{p \land r}$ and $\knonfu{M}{q \land r}$.
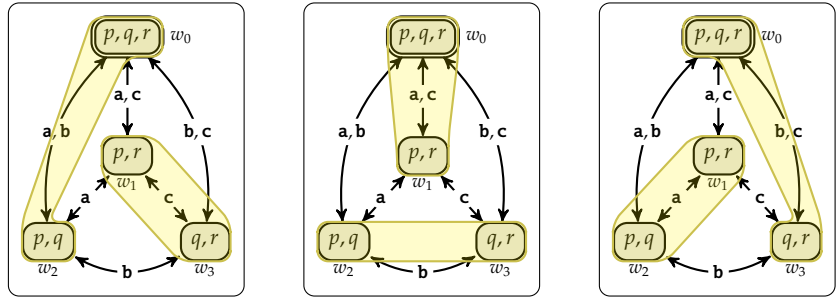
    \begin{ctabular}{@{}c@{\quad}c@{\quad}c@{}}
      \begin{tabular}{@{}c}
        \begin{tikzpicture}[node distance = 1.75em and 1em, frame rectangle]
          \node [mundo, label = {[etiqMundo, outer sep = 2pt]right:$w_0$}, double] (w0) {$p, q, r$};
          \node [mundo, label = {[etiqMundo]below:$w_1$}, below = 3em of w0] (w1) {$p,r$};
          \node [mundo, label = {[etiqMundo]below:$w_2$}, below left = of w1] (w2) {$p,q$};
          \node [mundo, label = {[etiqMundo]below:$w_3$}, below right = of w1] (w3) {$q,r$};

          \path (w0) edge [flecha, <->] node [etiqFlecha, within] {$\aga,\agc$} (w1)
                     edge [flecha, <->, bend right = 25] node [etiqFlecha, within] {$\aga, \agb$} (w2)
                     edge [flecha, <->, bend left = 25] node [etiqFlecha, within] {$\agb,\agc$} (w3)
                (w1) edge [flecha, <->] node [etiqFlecha, within] {$\aga\;$} (w2)
                     edge [flecha, <->] node [etiqFlecha, within] {$\;\agc$} (w3)
                (w2) edge [flecha, <->, bend right] node [etiqFlecha, within] {$\agb$} (w3);

          \draw[topicclass] (w0.north west) -- (w0.north) -- (w0.north east) -- (w0.east) -- (w0.south east) -- (w0.south) -- (w2.north) -- (w2.north east) -- (w2.east) -- (w2.south east) -- (w2.south west) -- (w2.north west) -- cycle;
          \draw[topicclass] (w3.north east) -- (w3.east) -- (w3.south east) -- (w3.south) -- (w3.south west) -- (w1.south west) -- (w1.west) -- (w1.north west) -- (w1.north) -- (w1.north east) -- cycle;
        \end{tikzpicture}
      \end{tabular}
      &
      \begin{tabular}{c}
        \begin{tikzpicture}[node distance = 1.75em and 1em, frame rectangle]
          \node [mundo, label = {[etiqMundo, outer sep = 2pt]right:$w_0$}, double] (w0) {$p, q, r$};
          \node [mundo, label = {[etiqMundo]below:$w_1$}, below = 3em of w0] (w1) {$p,r$};
          \node [mundo, label = {[etiqMundo]below:$w_2$}, below left = of w1] (w2) {$p,q$};
          \node [mundo, label = {[etiqMundo]below:$w_3$}, below right = of w1] (w3) {$q,r$};

          \path (w0) edge [flecha, <->] node [etiqFlecha, within] {$\aga,\agc$} (w1)
                     edge [flecha, <->, bend right = 25] node [etiqFlecha, within] {$\aga, \agb$} (w2)
                     edge [flecha, <->, bend left = 25] node [etiqFlecha, within] {$\agb,\agc$} (w3)
                (w1) edge [flecha, <->] node [etiqFlecha, within] {$\aga\;$} (w2)
                     edge [flecha, <->] node [etiqFlecha, within] {$\;\agc$} (w3)
                (w2) edge [flecha, <->, bend right] node [etiqFlecha, within] {$\agb$} (w3);

          \draw[topicclass] (w0.south west) -- (w0.west) -- (w0.north west) -- (w0.north) -- (w0.north east) -- (w0.east) -- (w0.south east) -- (w1.north east) -- (w1.east) -- (w1.south east) -- (w1.south) -- (w1.south west) -- (w1.west) -- cycle;
          \draw[topicclass] (w2.north) -- (w2.north east) -- (w3.north west) -- (w3.north) -- (w3.north east) -- (w3.east) -- (w3.south east) -- (w2.south east) -- (w2.south) -- (w2.south west) -- (w2.west) -- (w2.north west) -- cycle;
        \end{tikzpicture}
      \end{tabular}
      &
      \begin{tabular}{c@{}}
        \begin{tikzpicture}[node distance = 1.75em and 1em, frame rectangle]
          \node [mundo, label = {[etiqMundo, outer sep = 2pt]right:$w_0$}, double] (w0) {$p, q, r$};
          \node [mundo, label = {[etiqMundo]below:$w_1$}, below = 3em of w0] (w1) {$p,r$};
          \node [mundo, label = {[etiqMundo]below:$w_2$}, below left = of w1] (w2) {$p,q$};
          \node [mundo, label = {[etiqMundo]below:$w_3$}, below right = of w1] (w3) {$q,r$};

          \path (w0) edge [flecha, <->] node [etiqFlecha, within] {$\aga,\agc$} (w1)
                     edge [flecha, <->, bend right = 25] node [etiqFlecha, within] {$\aga, \agb$} (w2)
                     edge [flecha, <->, bend left = 25] node [etiqFlecha, within] {$\agb,\agc$} (w3)
                (w1) edge [flecha, <->] node [etiqFlecha, within] {$\aga\;$} (w2)
                     edge [flecha, <->] node [etiqFlecha, within] {$\;\agc$} (w3)
                (w2) edge [flecha, <->, bend right] node [etiqFlecha, within] {$\agb$} (w3);

          \draw[topicclass] (w0.north west) -- (w0.north) -- (w0.north east) -- (w3.north east) -- (w3.east) -- (w3.south east) -- (w3.south) -- (w3.south west) -- (w3.west) -- (w3.north west) -- (w3.north) -- (w0.south) -- (w0.south west) -- (w0.west) -- cycle;
          \draw[topicclass] (w1.north west) -- (w1.north) -- (w1.north east) -- (w1.east) -- (w1.south east) -- (w2.south east) -- (w2.south) -- (w2.south west) -- (w2.west) -- (w2.north west) -- cycle;
        \end{tikzpicture}
      \end{tabular}
    \end{ctabular}
    The diagrams further below (reflexivity assumed) show the result of three communication acts: $\symbSSE{\set{\aga,\agb}}{p \land q}$, $\symbSSE{\set{\aga,\agb}}{p \land r}$ and $\symbSSE{\set{\aga,\agb}}{q \land r}$.
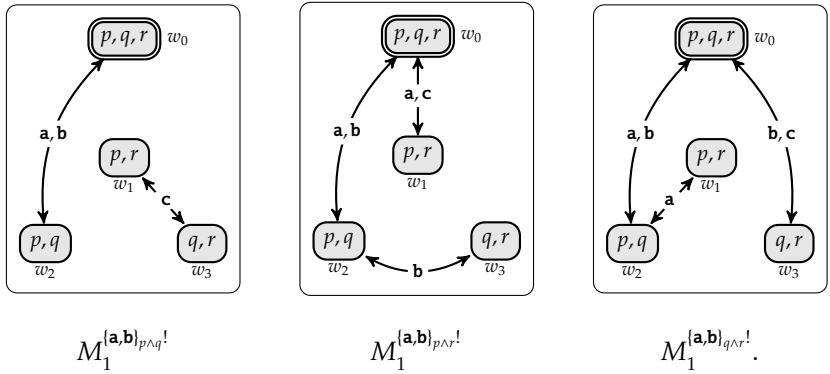
    \begin{ctabular}{@{}c@{\quad}c@{\quad}c@{}}
      \begin{tabular}{c}
        \begin{tikzpicture}[node distance = 1.75em and 1em, frame rectangle]
          \node [mundo, label = {[etiqMundo, outer sep = 2pt]right:$w_0$}, double] (w0) {$p, q, r$};
          \node [mundo, label = {[etiqMundo]below:$w_1$}, below = 3em of w0] (w1) {$p,r$};
          \node [mundo, label = {[etiqMundo]below:$w_2$}, below left = of w1] (w2) {$p,q$};
          \node [mundo, label = {[etiqMundo]below:$w_3$}, below right = of w1] (w3) {$q,r$};

          \path (w0) edge [flecha, <->, bend right = 25] node [etiqFlecha, within] {$\aga, \agb$} (w2)
                (w1) edge [flecha, <->] node [etiqFlecha, within] {$\;\agc$} (w3);
        \end{tikzpicture}
      \end{tabular}
      &
      \begin{tabular}{c}
        \begin{tikzpicture}[node distance = 1.75em and 1em, frame rectangle]
          \node [mundo, label = {[etiqMundo, outer sep = 2pt]right:$w_0$}, double] (w0) {$p, q, r$};
          \node [mundo, label = {[etiqMundo]below:$w_1$}, below = 3em of w0] (w1) {$p,r$};
          \node [mundo, label = {[etiqMundo]below:$w_2$}, below left = of w1] (w2) {$p,q$};
          \node [mundo, label = {[etiqMundo]below:$w_3$}, below right = of w1] (w3) {$q,r$};

          \path (w0) edge [flecha, <->] node [etiqFlecha, within] {$\aga,\agc$} (w1)
                     edge [flecha, <->, bend right = 25] node [etiqFlecha, within] {$\aga, \agb$} (w2)
                (w2) edge [flecha, <->, bend right] node [etiqFlecha, within] {$\agb$} (w3);
        \end{tikzpicture}
      \end{tabular}
      &
      \begin{tabular}{c}
        \begin{tikzpicture}[node distance = 1.75em and 1em, frame rectangle]
          \node [mundo, label = {[etiqMundo, outer sep = 2pt]right:$w_0$}, double] (w0) {$p, q, r$};
          \node [mundo, label = {[etiqMundo]below:$w_1$}, below = 3em of w0] (w1) {$p,r$};
          \node [mundo, label = {[etiqMundo]below:$w_2$}, below left = of w1] (w2) {$p,q$};
          \node [mundo, label = {[etiqMundo]below:$w_3$}, below right = of w1] (w3) {$q,r$};

          \path (w0) edge [flecha, <->, bend right = 25] node [etiqFlecha, within] {$\aga, \agb$} (w2)
                     edge [flecha, <->, bend left = 25] node [etiqFlecha, within] {$\agb,\agc$} (w3)
                (w1) edge [flecha, <->] node [etiqFlecha, within] {$\aga\;$} (w2);
        \end{tikzpicture}
      \end{tabular}
      \\
      $\mopSSE{M_1}{\set{\aga,\agb}}{p \land q}$ & $\mopSSE{M_1}{\set{\aga,\agb}}{p \land r}$ & $\mopSSE{M_1}{\set{\aga,\agb}}{q \land r}$.
    \end{ctabular}
    In all these cases, only agents in $\sen = \set{\aga,\agb}$ communicate. The operation again looks only at edges across partition cells, with each agent $\agi \in \sen$ keeping only those in $\sbDs{\R}{\sen}$ and each agent $\agi \notin \sen$ keeping only those in $\sbDs{\R}{\sen \cup \set{\agi}}$. For example, after $\set{\aga,\agb}$ exchange (recall: publicly) information about $p \land q$ (leftmost model), every agent will know the truth-value of both $p$ and $q$ (since $\agc$ additionally knew $r$'s, now she knows which the real situation is). But if $\aga$ and $\agb$'s conversation is rather about $q \land r$ (rightmost model), then less information is shared (neither of them knew whether $r$, and $\aga$ did not get the chance to talk about what she knew: $p$). In fact, afterwards no agent knows the truth-value of all atoms, and only two agents ($\aga$ and $\agc$) know the truth-value of two of them ($q$ and, respectively, $p$ and $r$).
  \end{compactenumerate}
\end{ejemplo}

After the examples, here are some observations.

\begin{itemize}
  \item As $\sbi{\mopSSE{\R}{\sen}{\chi}} = \sbi{\R} \cap (\sbDs{\R}{\sen} \cup \knonfu{M}{\chi}) \subseteq \sbi{\R}$, an \SSE act can only \emph{reduce} uncertainty.

  \item Since $\knonfu{M}{\chi} = \knonfu{M}{\lnot \chi}$, it follows that $\mopSSE{M}{\sen}{\chi} = \mopSSE{M}{\sen}{\lnot \chi}$.

  \item If $\agi \in \sen$, then $\sbi{\mopSSE{\R}{\sen}{\chi}} = \sbDs{\R}{\sen} \cup (\sbi{\R} \cap \knonfu{M}{\chi})$.

  \item An \SSE act is \emph{not} a generalisation of \SEE: there is no formula $\chi$ such that, for every model $M = \tupla{W, \R, V}$ and every set of agents $\sen$, an `$\sen$-conversation about $\chi$' (producing $\sbi{\mopSSE{\R}{\sen}{\chi}} = \sbi{\R} \cap (\sbDs{\R}{\sen} \cup \knonfu{M}{\chi})$) has the same effects as an `$\sen$-conversation' about everything (producing $\sbi{\mopSEE{\R}{\sen}} = \sbi{\R} \cap \sbDs{\R}{\sen}$). For that, one would need a $\chi$ such that $\sbi{\R} \cap (\sbDs{\R}{\sen} \cup \knonfu{M}{\chi}) = \sbi{\R} \cap \sbDs{\R}{\sen}$ (and thus, $\knonfu{M}{\chi} \subseteq \sbDs{\R}{\sen}$) \emph{for every $M$ and every $\sen$}.\footnote{There is no $\chi$ satisfying $\knonfu{M}{\chi} \subseteq \sbDs{\R}{\sen}$ for every $M$ and $\sen$: there are $M$ and $\sen$ with $\sbDs{\R}{\sen} = \emptyset$, and yet the (reflexive) relation $\knonfu{M}{\chi}$ is never empty (domains are non-empty). But even when working with equivalence relations, no $\chi$ satisfies the requirement \emph{for every $M$ and every $\sen$}.} Still, one can find specific situations (a model $M$, a set of agents $\sen$ and a formula $\chi$) an `$\sen$-conversation about $\chi$' has the same effects as an `$\sen$-conversation about everything'.

  \item If $\sbi{\R}$ and the relations in $\set{\sbj{\R} \mid \agj \in \sen}$ are all reflexive/symmetric, then so is the resulting $\sbi{\mopEEE{\R}}$.\footnote{For reflexivity, take any $w \in W$. By the assumption, $\sbi{\R}ww$ and $\sbDs{\R}{\sen}ww$; moreover, $w \knonfu{M}{\chi} w$. Thus, $\sbi{\mopEEE{\R}}ww$. For symmetry, if $\sbi{\mopSSE{\R}{\sen}{\chi}}wu$ then $\sbi{\R}wu$ and either $\sbDs{\R}{\sen}wu$ or $w \knonfu{M}{\chi} u$. But, by the assumptions and $\knonfu{M}{\chi}$'s symmetry, $\sbi{\R}uw$ and either $\sbDs{\R}{\sen}uw$ or $u \knonfu{M}{\chi} w$. Thus, $\sbi{\mopEEE{\R}}uw$.} However, the operation does not preserve transitivity and neither `Euclideanity' (see, e.g., \itemfautoref{eje:SSE}{eje:SSE:disjunction}, in particular the effects of the action $\symbSSE{\set{\aga,\agb}}{p \land r}$ on the relation for agent $\aga$).
\end{itemize}

Now, the modality.

\begin{definicion}[Modality $\modSSE{\sen}{\chi}$; language \LAdSSE]\label{def:SSEmm}
  ~\!\!\!The language \LAdSSE extends \LAd with modalities $\modSSE{\sen}{\chi}$ for each set of agents $\sen \subseteq \ag$ and each formula $\chi$. For their semantic interpretation,
  \begin{lbtabular}{l@{\qssidefq}l}
    $(M, w) \Vdash \mmSSE{\sen}{\chi}{\varphi}$ & $(\mopSSE{M}{\sen}{\chi}, w) \Vdash \varphi$.
  \end{lbtabular}
\end{definicion}

Using $\truthset{}{\cdot}$, note that $w \in \truthset{M}{\mmSSE{\sen}{\chi}{\varphi}}$ if and only if $w \in \truthset{\mopSSE{M}{\sen}{\chi}}{\varphi}$. Thus,
\[ \truthsetn{M}{\mmSSE{\sen}{\chi}{\varphi}} = \truthset{\mopSSE{M}{\sen}{\chi}}{\varphi}. \]

\bsparagraph{Properties} Here are some observations about the \SSE operation, starting again with a caveat.

\begin{proposicion}\label{pro:SSE:Moore}
  Let $(M, w)$ be a pointed \marm model; take $\sen \subseteq \ag$ and $\agi \in \ag$. Then, $(M, w) \Vdash \mmDs{\sen}{\varphi} \limp \mmSSE{\sen}{\varphi}{\mmKi{\varphi}}$ holds when $M$ and $\varphi$ are such that $\truthset{M}{\varphi} \subseteq \truthset{\mopSSE{M}{\sen}{\varphi}}{\varphi}$ \emph{and the relations in $M$ are reflexive}. However, $\not\Vdash \mmDs{\sen}{\varphi} \limp \mmSSE{\sen}{\varphi}{\mmKi{\varphi}}$.
  \begin{proof}
    Since $(M, w) \Vdash \mmDs{\sen}{\varphi}$, then $\sbDs{\R}{\sen}(w) \subseteq \truthset{M}{\varphi}$ and thus, by the assumption, $\sbDs{\R}{\sen}(w) \subseteq \truthset{\mopSSE{M}{\sen}{\varphi}}{\varphi}$. To obtain $(M, w) \Vdash \mmSSE{\sen}{\varphi}{\mmKi{\varphi}}$, one requires $(\mopSSE{M}{\sen}{\varphi}, w) \Vdash \mmKi{\varphi}$; since $\agi \in \sen$, this boils down to $(\sbDs{\R}{\sen} \cup (\sbi{\R} \cap \knonfu{M}{\varphi}))(w) \subseteq \truthset{\mopSSE{M}{\sen}{\varphi}}{\varphi}$, that is, to both $\sbDs{\R}{\sen}(w) \subseteq \truthset{\mopSSE{M}{\sen}{\varphi}}{\varphi}$ and $(\sbi{\R} \cap \knonfu{M}{\varphi})(w) \subseteq \truthset{\mopSSE{M}{\sen}{\varphi}}{\varphi}$. The first has been obtained. For the second, take any $u \in (\sbi{\R} \cap \knonfu{M}{\varphi})(w)$, so $w$ and $u$ coincide in $\varphi$'s truth-value in $M$. By reflexivity, $w \in \sbDs{\R}{\sen}(w)$, so $w \in \truthset{M}{\varphi}$ (the antecedent $\mmDs{\sen}{\varphi}$) and hence $u \in \truthset{M}{\varphi}$. Therefore, by the assumption, $u \in \truthset{\mopSSE{M}{\sen}{\varphi}}{\varphi}$, as required.

    \smallskip

    Both requirements are essential. In \autoref{hch:SSE:no-valid} it will be shown that reflexivity without $\truthset{M}{\varphi} \subseteq \truthset{\mopSSE{M}{\sen}{\varphi}}{\varphi}$ is not enough. For $\truthset{M}{\varphi} \subseteq \truthset{\mopSSE{M}{\sen}{\varphi}}{\varphi}$ without reflexivity, consider a model $M = \tupla{W, \R, V}$ with $W = \set{w_0, w_1, w_2}$, $\sba{\R} = \set{(w_0, w_2)}$, $\sbb{\R} = \sba{\R} \cup \set{(w_0, w_1)}$ and $V(p) = \set{w_2}$. Note how $(M, w_0) \Vdash \mmDs{\set{\aga, \agb}}{p}$; yet, $(M, w_0) \not\Vdash \mmSSE{\set{\aga, \agb}}{p}{\mmKb{p}}$ as $(\mopSSE{M}{\set{\aga, \agb}}{p}, w_0) \not\Vdash \mmKb{p}$ because $M$ and $\mopSSE{M}{\set{\aga, \agb}}{p}$ are identical.
  \end{proof}
\end{proposicion}

The following proposition provides the two useful observations.

\begin{proposicion}\label{pro:SSE:mo}
  Let $M = \tupla{W, \R, V}$ be in \marm; take $\sen, \sen_1, \sen_2 \subseteq \ag$ and let $\chi, \chi_1, \chi_2$ be formulas. Let $\mopSSE{M}{\sen}{\chi} = \tuplan{W, \mopSSE{\R}{\sen}{\chi}, V}$ and $\mopSSE{(\mopSSE{M}{\sen_1}{\chi_1})}{\sen_2}{\chi_2} = \tuplan{W, \mopSSE{(\mopSSE{\R}{\sen_1}{\chi_1})}{\sen_2}{\chi_2}, V}$ be as indicated in \autoref{def:SSEmo}. Then,
  \begin{enumerate}
    \item\label{pro:SSE:mo:D} $\sbDg{(\mopSSE{\R}{\sen}{\chi})} = \sbDs{\R}{\sen \cup \sag} \cup (\sbDg{\R} \cap \knonfu{M}{\chi})$ \, for every $\sag \subseteq \ag$;
    \item\label{pro:SSE:mo:SSE} $\sbi{\mopSSE{(\mopSSE{\R}{\sen_1}{\chi_1})}{\sen_2}{\chi_2}} =
    \sbi{\R}
    \cap
    \Bigg(
      \sbDs{\R}{\sen_1 \cup \sen_2}
      \cup
      \big(
        \sbDs{\R}{\sen_1} \cap \knonfu{M}{\mmSSE{\sen_1}{\chi_1}{\chi_2}}
      \big)
      \cup
      \big(
        \sbDs{\R}{\sen_2} \cap \knonfu{M}{\chi_1}
      \big)
      \cup
      \big(
        \knonfu{M}{\chi_1} \cap \knonfu{M}{\mmSSE{\sen_1}{\chi_1}{\chi_2}}
      \big)
    \Bigg)$
    \, for every $\agi \in \ag$.
  \end{enumerate}
  \begin{proof}
    ~
    \begin{enumerate}
      \item For any $\sag \subseteq \ag$, the relation $\sbDg{(\mopSSE{\R}{\sen}{\chi})}$ is $\bigcap_{\agk \in \sag} \sbk{\mopSSE{\R}{\sen}{\chi}}$ (def. of $\sbDg{(\mopSSE{\R}{\sen}{\chi})}$), equal to $\bigcap_{\agk \in \sag} \left( \sbk{\R} \cap (\sbDs{\R}{\sen} \cup \knonfu{M}{\chi}) \right)$ (def. of $\sbk{\mopSSE{\R}{\sen}{\chi}}$) and thus to $(\sbDs{\R}{\sen} \cup \knonfu{M}{\chi}) \cap \bigcap_{\agk \in \sag} \sbk{\R}$, which is $(\sbDs{\R}{\sen} \cup \knonfu{M}{\chi}) \cap \sbDg{\R}$. Then, the De Morgan's laws and $\sbDs{\R}{\sen} \cap \sbDg{\R} = \sbDs{\R}{\sen\cup\sag}$ yield the required $\sbDs{\R}{\sen \cup \sag} \cup (\sbDg{\R} \cap \knonfu{M}{\chi})$.

      \item By algebraic manipulations starting from
      \[
        \sbi{\mopSSE{(\mopSSE{\R}{\sen_1}{\chi_1})}{\sen_2}{\chi_2}}
        =
        \sbi{\mopSSE{\R}{\sen_1}{\chi_1}}
        \cap
        \left(
          \sbDs{(\mopSSE{\R}{\sen_1}{\chi_1})}{\sen_2}
          \cup
          \knonfu{\mopSSE{M}{\sen_1}{\chi_1}}{\chi_2}
        \right),
      \]
      using the fact that $\knonfu{\mopSSE{M}{\sen_1}{\chi_1}}{\chi_2} = \knonfu{M}{\mmSSE{\sen_1}{\chi_1}{\chi_2}}$.\footnote{Indeed, by definition, $\knonfu{\mopSSE{M}{\sen_1}{\chi_1}}{\chi_2} = \Big(\truthset{\mopSSE{M}{\sen_1}{\chi_1}}{\chi_2} \times \truthset{\mopSSE{M}{\sen_1}{\chi_1}}{\chi_2}\Big) \cup \Big(\truthset{\mopSSE{M}{\sen_1}{\chi_1}}{\lnot \chi_2} \times \truthset{\mopSSE{M}{\sen_1}{\chi_1}}{\lnot \chi_2} \Big)$. But recall: $\truthset{\mopSSE{M}{\sen}{\chi}}{\varphi} = \truthsetn{M}{\mmSSE{\sen}{\chi}{\varphi}}$; moreover, $\truthset{\mopSSE{M}{\sen}{\chi}}{\lnot\varphi} = W \setminus \truthset{\mopSSE{M}{\sen}{\chi}}{\varphi} = W \setminus \truthset{M}{\mmSSE{\sen}{\chi}{\varphi}} = \truthset{M}{\lnot \mmSSE{\sen}{\chi}{\varphi}}$. Then, $\knonfu{\mopSSE{M}{\sen_1}{\chi_1}}{\chi_2} = \left(\truthsetn{M}{\mmSSE{\sen_1}{\chi_1}{\chi_2}} \times \truthsetn{M}{\mmSSE{\sen_1}{\chi_1}{\chi_2}}\right) \cup \left(\truthsetn{M}{\lnot \mmSSE{\sen_1}{\chi_1}{\chi_2}} \times \truthsetn{M}{\lnot \mmSSE{\sen_1}{\chi_1}{\chi_2}}\right) = \knonfu{M}{\mmSSE{\sen_1}{\chi_1}{\chi_2}}$.}
    \end{enumerate}
  \end{proof}
\end{proposicion}

On the one hand, the second result of the previous proposition shows the indistinguishability relation of an agent $\agi$ after two \SSE operations in a row. By comparing it with the same relation after a single \SSE operation, $\sbi{\mopSSE{\R}{\sen}{\chi}} = \sbi{\R} \cap (\sbDs{\R}{\sen} \cup \knonfu{M}{\chi})$, one understands why, different from the two previous cases, two successive \SSE actions cannot be compressed into a single one.

\smallskip

On the other hand, the first result indicates that the relation for interpreting $\modDg$ in the new model, $\sbDg{(\mopSSE{\R}{\sen}{\chi})}$ can be described as $\sbDs{\R}{\sen \cup \sag} \cup (\sbDg{\R} \cap \knonfu{M}{\chi})$. This, together with the abbreviation
\[
  \mmDsr{\sag}{\chi}{\varphi}
  :=
  (\chi \limp\mmDs{\sag}{(\chi \limp \varphi)})
  \land
  (\lnot\chi \limp\mmDs{\sag}{(\lnot\chi \limp \varphi)})
\]
(so $(M, w) \Vdash \mmDsr{\sag}{\chi}{\varphi}$ if and only if $(\sbDg{\R} \cap \knonfu{M}{\chi})(w) \subseteq \truthset{M}{\varphi}$) provides the following validity, which will be useful for the axiomatisation.

\begin{proposicion}\label{pro:SSE}
  $\Vdash \mmSSE{\sen}{\chi}{\mmDg{\varphi}} \lldimp \big(\mmDs{\sen \cup \sag}{\mmSSE{\sen}{\chi}{\varphi}} \land \mmDsr{\sag}{\chi}{\mmSSE{\sen}{\chi}{\varphi}}\big)$.
  \begin{proof}
    Let $M = \tupla{W, \R, V}$ be in \marm. By definition, $w \in \truthset{M}{\mmSSE{\sen}{\chi}{\mmDg{\varphi}}}$ if and only if $w \in \truthset{\mopSSE{M}{\sen}{\chi}}{\mmDg{\varphi}}$, which holds if and only if $\sbDg{(\mopSSE{\R}{\sen}{\chi})}(w) \subseteq \truthset{\mopSSE{M}{\sen}{\chi}}{\varphi}$. But, from \itemfautoref{pro:SSE:mo}{pro:SSE:mo:D} and semantic interpretation, the latter is equivalent to $(\sbDs{\R}{\sen \cup \sag} \cup (\sbDg{\R} \cap \knonfu{M}{\chi}))(w) \subseteq \truthset{M}{\mmSSE{\sen}{\chi}{\varphi}}$, that is, to the conjunction $\sbDs{\R}{\sen \cup \sag}(w) \subseteq \truthset{M}{\mmSSE{\sen}{\chi}{\varphi}})$ and $(\sbDg{\R} \cap \knonfu{M}{\chi})(w) \subseteq \truthset{M}{\mmSSE{\sen}{\chi}{\varphi}})$, whose conjuncts are equivalent, respectively, to $w \in \truthset{M}{\mmDs{\sen \cup \sag}{\mmSSE{\sen}{\chi}{\varphi}}}$ and $w \in \truthsetn{M}{\mmDsr{\sag}{\chi}{\mmSSE{\sen}{\chi}{\varphi}}}$.
  \end{proof}
\end{proposicion}

This validity yields immediately the following one, characterising the knowledge of an agent after \SSE communication:
\[
  \Vdash
    \mmSSE{\sen}{\chi}{\mmKi{\varphi}}
    \lldimp
    \big(
      \mmDs{\sen \cup \set{\agi}}{\mmSSE{\sen}{\chi}{\varphi}}
      \land
      \mmDsr{\set{\agi}}{\chi}{\mmSSE{\sen}{\chi}{\varphi}}
    \big).
\]
Here is another useful validity, a consequence of an earlier observation:
\[
  \Vdash
    \mmSSE{\sen}{\chi}{\varphi}
    \lldimp
    \mmSSE{\sen}{\lnot \chi}{\varphi}.
\]


\medskip

In the actions of \autoref{sec:XEE}, each relation in the new model is defined in terms of relations in the initial one. However, the \SSE operation defines each `new' relation in terms of `old' ones \emph{and the relation of only knowing $\chi$ fully}, $\knonfu{}{\chi}$ (with $\chi$ the communication's topic). This and the fact that model operations can change the truth-set of a formula is what makes \SSE behaves less similar to the \EEE- and \SEE-actions, and more similar to public announcements.

\begin{hecho}\label{hch:SSE:no-valid}
  ~
  \begin{enumerate}
    \item $\not\Vdash \mmSSE{\sen}{\chi}{\mmSSE{\sen}{\chi}{\varphi}} \ldimp \mmSSE{\sen}{\chi}{\varphi}$: two successive \SSE-acts by the same group and on the same topic cannot be collapsed into a single one on the same topic. 

    \item $\not\Vdash \mmSSE{\sen_1}{\chi_1}{\mmSSE{\sen_2}{\chi_2}{\varphi}} \ldimp \mmSSE{\sen_2}{\chi_2}{\mmSSE{\sen_1}{\chi_1}{\varphi}}$: \SSE-acts do not commute. 

    \item $\not\Vdash \mmSSE{\sen_1}{\chi}{\mmSSE{\sen_2}{\chi}{\varphi}} \ldimp \mmSSE{(\sen_1 \cup \sen_2)}{\chi}{\varphi}$: successive \SSE-acts by different groups on the same topic cannot be `compressed' by using the union of the groups. 

    \item $\not\Vdash \mmSSE{\sen}{\chi_1}{\mmSSE{\sen}{\chi_2}{\varphi}} \ldimp \mmSSE{\sen}{\chi_1 \land \chi_2}{\varphi}$: successive \SSE-acts by the same group on different topic cannot be `compressed' by using the conjunction of the topics. 
  \end{enumerate}
  \begin{proof}
    In the model below (with equivalence relations), each world indicates the truth-value of atoms $m_{\aga}, m_{\agb}, m_{\agc}$ (in that order) by using ``$\bullet$'' (the atom holds) or ``$\circ$'' (the atom fails).\footnote{Thus, e.g., at 
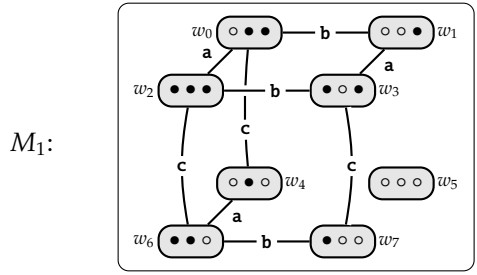
\begin{tikzpicture}[, baseline=-2pt, scale = 0.5]\node [mundo, outer sep = 0, inner sep = 2pt, font = \tiny, minimum height = 10pt] {$\bullet\circ\bullet$};\end{tikzpicture} atom $m_{\aga}$ is true, $m_{\agb}$ is false and $m_{\agc}$ is true.}
    \begin{ctabular}{cc}
      \begin{tabular}{r}
        $M_1$:
      \end{tabular}
      &
      \begin{tabular}{c}
        \begin{tikzpicture}[-, frame rectangle]
          \node (w0) at (-\ladocubo/2,  \ladocubo/2, -\ladocubo/2) [mundo, label = {[etiqMundo]left:$w_0$}] {$\circ\bullet\bullet$};
          \node (w1) at ( \ladocubo/2,  \ladocubo/2, -\ladocubo/2) [mundo, label = {[etiqMundo]right:$w_1$}] {$\circ\circ\bullet$};
          \node (w2) at (-\ladocubo/2,  \ladocubo/2,  \ladocubo/2) [mundo, label = {[etiqMundo]left:$w_2$}] {$\bullet\bullet\bullet$};
          \node (w3) at ( \ladocubo/2,  \ladocubo/2,  \ladocubo/2) [mundo, label = {[etiqMundo]right:$w_3$}] {$\bullet\circ\bullet$};
          \node (w4) at (-\ladocubo/2, -\ladocubo/2, -\ladocubo/2) [mundo, label = {[etiqMundo]right:$w_4$}] {$\circ\bullet\circ$};
          \node (w5) at ( \ladocubo/2, -\ladocubo/2, -\ladocubo/2) [mundo, label = {[etiqMundo]right:$w_5$}] {$\circ\circ\circ$};
          \node (w6) at (-\ladocubo/2, -\ladocubo/2,  \ladocubo/2) [mundo, label = {[etiqMundo]left:$w_6$}] {$\bullet\bullet\circ$};
          \node (w7) at ( \ladocubo/2, -\ladocubo/2,  \ladocubo/2) [mundo, label = {[etiqMundo]right:$w_7$}] {$\bullet\circ\circ$};

          \path (w0) edge [flecha] node [etiqFlecha, within] {$\agb$} (w1)
                     edge [flecha, bend right=5] node [etiqFlecha, within, pos = 0.7] {$\agc$} (w4)
                     edge [flecha] node [etiqFlecha, above, pos=0.75] {$\aga~~~$} (w2)
                (w1) edge [flecha] node [etiqFlecha, below, pos=0.1] {$~~~\aga$} (w3)
                (w4) 
                     edge [flecha] node [etiqFlecha, below, pos=0.1] {$~~~\aga$} (w6)
                (w2) edge [flecha, cross line] node [etiqFlecha, within, pos = 0.6] {$\agb$} (w3)
                     edge [flecha, bend right=10] node [etiqFlecha, within] {$\agc$} (w6)
                (w7) edge [flecha] node [etiqFlecha, within] {$\agb$} (w6)
                     edge [flecha, cross line, bend right=10] node [etiqFlecha, within] {$\agc$} (w3);
        \end{tikzpicture}
      \end{tabular}
    \end{ctabular}
    Define $\chi_{\agi} := \mmKi{m_{\agi}} \lor \mmKi{\lnot m_{\agi}}$ for each $\agi \in \ag = \set{\aga, \agb, \agc}$, stating that agent $\agi$ knows $m_{\agi}$'s truth-value. Define their disjunction $\chi^\lor := \chi_{\aga} \lor \chi_{\agb} \lor \chi_{\agc}$. Note how, at worlds in $\set{w_0, w_2, w_3, w_6}$, every agent $\agi$ has uncertainty about whether $m_{\agi}$ holds (i.e., $\chi_{\agi}$ fails for every $\agi$, and thus so does $\chi^\lor$). However at worlds in $\set{w_1, w_4, w_5, w_7}$, at least one agent $\agi$ knows whether $m_{\agi}$ (so $\chi^\lor$ is the case).\footnote{As a visual aid, in this model each atom $\chi_{\agi}$ holds in those worlds \emph{without} outgoing $\agi$-edges (other than the implicitly present reflexive ones).}
    \begin{compactenumerate}
      \item\label{hch:SSE:no-valid:i} Since $\truthset{M_1}{\chi^\lor} = \set{w_0, w_2, w_3, w_6}$, applying $\symbSSE{\ag}{\chi^\lor}$ to $M_1$ (note the generated partition) yields $M_2$ below. Then, since $\truthset{M_2}{\chi^\lor} = \set{w_2}$, a further application of $\symbSSE{\ag}{\chi^\lor}$ yields $M_3$.
      \begin{ctabular}{@{}c@{\,}c@{\,}c@{\,}c@{\,}c@{}}
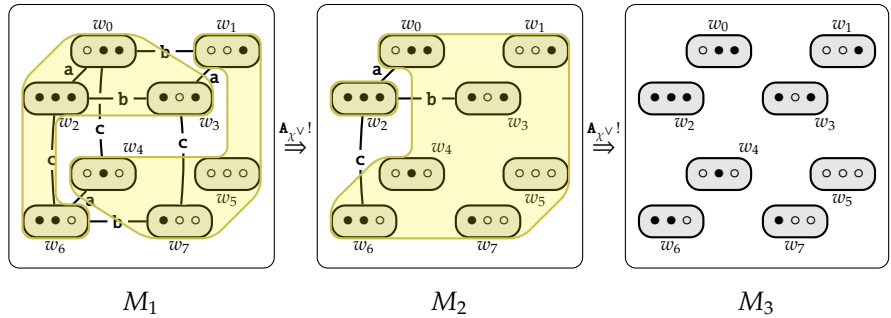
        \begin{tabular}{@{}c@{}}
          \begin{tikzpicture}[-, frame rectangle, scale=0.815]
            \node (w0) at (-\ladocubo/2,  \ladocubo/2, -\ladocubo/2) [mundo, label = {[etiqMundo]above:$w_0$}] {$\circ\bullet\bullet$};
            \node (w1) at ( \ladocubo/2,  \ladocubo/2, -\ladocubo/2) [mundo, label = {[etiqMundo]above:$w_1$}] {$\circ\circ\bullet$};
            \node (w2) at (-\ladocubo/2,  \ladocubo/2,  \ladocubo/2) [mundo, label = {[etiqMundo]below:$~~~~~~w_2$}] {$\bullet\bullet\bullet$};
            \node (w3) at ( \ladocubo/2,  \ladocubo/2,  \ladocubo/2) [mundo, label = {[etiqMundo]below right:$w_3$}] {$\bullet\circ\bullet$};
            \node (w4) at (-\ladocubo/2, -\ladocubo/2, -\ladocubo/2) [mundo, label = {[etiqMundo]above right:$w_4$}] {$\circ\bullet\circ$};
            \node (w5) at ( \ladocubo/2, -\ladocubo/2, -\ladocubo/2) [mundo, label = {[etiqMundo]below:$w_5$}] {$\circ\circ\circ$};
            \node (w6) at (-\ladocubo/2, -\ladocubo/2,  \ladocubo/2) [mundo, label = {[etiqMundo]below:$w_6$}] {$\bullet\bullet\circ$};
            \node (w7) at ( \ladocubo/2, -\ladocubo/2,  \ladocubo/2) [mundo, label = {[etiqMundo]below:$w_7$}] {$\bullet\circ\circ$};

            \path (w0) edge [flecha] node [etiqFlecha, within] {$\agb$} (w1)
                       edge [flecha, bend right=5] node [etiqFlecha, within, pos = 0.7] {$\agc$} (w4)
                       edge [flecha] node [etiqFlecha, above, pos=1.3] {$\aga$} (w2)
                  (w1) edge [flecha] node [etiqFlecha, below, pos=-0.3] {$\aga$} (w3)
                  (w4) edge [flecha] node [etiqFlecha, below, pos=-0.3] {$\aga$} (w6)
                  (w2) edge [flecha, cross line] node [etiqFlecha, within, pos = 0.6] {$\agb$} (w3)
                       edge [flecha, bend right=5] node [etiqFlecha, within] {$\agc$} (w6)
                  (w7) edge [flecha] node [etiqFlecha, within] {$\agb$} (w6)
                       edge [flecha, cross line, bend right=5] node [etiqFlecha, within, pos=0.7] {$\agc$} (w3);

            \draw[topicclass] (w0.north west) -- (w2.north west) -- (w2.south west) -- (w6.north west) -- (w6.south west) -- (w6.south east) -- (w6.north east) -- (w6.north) -- (w2.south) -- (w2.south east) -- (w3.south west) -- (w3.south east) -- (w3.north east) -- (w0.north east) -- cycle;

            \draw[topicclass] (w7.south east) -- (w5.south east) -- (w5.north east) -- (w1.south east) -- (w1.north east) -- (w1.north west) -- (w1.south west) -- (w1.south) -- (w5.north) -- (w4.north west) -- (w4.south west) -- (w7.south west) -- cycle;
          \end{tikzpicture}
        \end{tabular}
        &
        \begin{tabular}{@{}c@{}}
          {\small $\overset{\symbSSE{\ag}{\chi^\lor}}{\Rightarrow}$}
        \end{tabular}
        &
        \begin{tabular}{@{}c@{}}
          \begin{tikzpicture}[-, frame rectangle, scale=0.815]
            \node (w0) at (-\ladocubo/2,  \ladocubo/2, -\ladocubo/2) [mundo, label = {[etiqMundo]above:$w_0$}] {$\circ\bullet\bullet$};
            \node (w1) at ( \ladocubo/2,  \ladocubo/2, -\ladocubo/2) [mundo, label = {[etiqMundo]above:$w_1$}] {$\circ\circ\bullet$};
            \node (w2) at (-\ladocubo/2,  \ladocubo/2,  \ladocubo/2) [mundo, label = {[etiqMundo]below:$~~~~~~w_2$}] {$\bullet\bullet\bullet$};
            \node (w3) at ( \ladocubo/2,  \ladocubo/2,  \ladocubo/2) [mundo, label = {[etiqMundo]below right:$w_3$}] {$\bullet\circ\bullet$};
            \node (w4) at (-\ladocubo/2, -\ladocubo/2, -\ladocubo/2) [mundo, label = {[etiqMundo]above right:$w_4$}] {$\circ\bullet\circ$};
            \node (w5) at ( \ladocubo/2, -\ladocubo/2, -\ladocubo/2) [mundo, label = {[etiqMundo]below:$w_5$}] {$\circ\circ\circ$};
            \node (w6) at (-\ladocubo/2, -\ladocubo/2,  \ladocubo/2) [mundo, label = {[etiqMundo]below:$w_6$}] {$\bullet\bullet\circ$};
            \node (w7) at ( \ladocubo/2, -\ladocubo/2,  \ladocubo/2) [mundo, label = {[etiqMundo]below:$w_7$}] {$\bullet\circ\circ$};

            \path (w0) edge [flecha] node [etiqFlecha, above, pos=1.3] {$\aga$} (w2)
                  (w2) edge [flecha, cross line] node [etiqFlecha, within, pos = 0.6] {$\agb$} (w3)
                       edge [flecha, bend right=5] node [etiqFlecha, within] {$\agc$} (w6);

            \draw[topicclass] (w2.north west) -- (w2.north east) -- (w2.south east) -- (w2.south west) -- cycle;

            \draw[topicclass] (w0.south west) -- (w0.south) -- (w4.north) -- (w4.north west) -- (w6.north west) -- (w6.south west) -- (w7.south east) -- (w5.south east) -- (w1.north east) -- (w0.north west) -- cycle;
          \end{tikzpicture}
        \end{tabular}
        &
        \begin{tabular}{@{}c@{}}
          {\small $\overset{\symbSSE{\ag}{\chi^\lor}}{\Rightarrow}$}
        \end{tabular}
        &
        \begin{tabular}{@{}c@{}}
          \begin{tikzpicture}[-, frame rectangle, scale=0.815]
            \node (w0) at (-\ladocubo/2,  \ladocubo/2, -\ladocubo/2) [mundo, label = {[etiqMundo]above:$w_0$}] {$\circ\bullet\bullet$};
            \node (w1) at ( \ladocubo/2,  \ladocubo/2, -\ladocubo/2) [mundo, label = {[etiqMundo]above:$w_1$}] {$\circ\circ\bullet$};
            \node (w2) at (-\ladocubo/2,  \ladocubo/2,  \ladocubo/2) [mundo, label = {[etiqMundo]below:$~~~~~~w_2$}] {$\bullet\bullet\bullet$};
            \node (w3) at ( \ladocubo/2,  \ladocubo/2,  \ladocubo/2) [mundo, label = {[etiqMundo]below right:$w_3$}] {$\bullet\circ\bullet$};
            \node (w4) at (-\ladocubo/2, -\ladocubo/2, -\ladocubo/2) [mundo, label = {[etiqMundo]above right:$w_4$}] {$\circ\bullet\circ$};
            \node (w5) at ( \ladocubo/2, -\ladocubo/2, -\ladocubo/2) [mundo, label = {[etiqMundo]below:$w_5$}] {$\circ\circ\circ$};
            \node (w6) at (-\ladocubo/2, -\ladocubo/2,  \ladocubo/2) [mundo, label = {[etiqMundo]below:$w_6$}] {$\bullet\bullet\circ$};
            \node (w7) at ( \ladocubo/2, -\ladocubo/2,  \ladocubo/2) [mundo, label = {[etiqMundo]below:$w_7$}] {$\bullet\circ\circ$};
          \end{tikzpicture}
        \end{tabular}
        \\
        $M_1$ && $M_2$ && $M_3$ \\
      \end{ctabular}
      Thus, $(M_1, w_2) \Vdash \mmSSE{\ag}{\chi^\lor}{\mmSSE{\ag}{\chi^\lor}{\chi_{\aga}}}$ and yet $(M_1, w_2) \not\Vdash \mmSSE{\ag}{\chi^\lor}{\chi_{\aga}}$. Note also how $(M_1, w_2) \Vdash \mmDs{\ag}{\lnot \chi^\lor} \land \lnot \mmSSE{\ag}{\lnot \chi^\lor}{\mmKa{\lnot \chi^\lor}}$, thus showing that $\not\Vdash \mmDs{\sen}{\varphi} \limp \mmSSE{\sen}{\varphi}{\mmKi{\varphi}}$ for $\agi \in \sen$, even when the model is reflexive (cf. the discussion in the proof of \autoref{pro:SSE:Moore}).

      \item\label{hch:SSE:no-valid:ii} On the one hand, $\truthset{M_1}{\chi_{\aga}} = \set{w_5, w_7}$, so applying $\symbSSE{\set{\aga}}{\chi_{\aga}}$ to $M_1$ yields $M'_2$ below. Then, $\truthset{M'_2}{\chi_{\agc}} = \set{w_1, w_3, w_5, w_7}$ so a further $\symbSSE{\set{\agb,\agc}}{\chi_{\agc}}$ yields $M'_3$.
      \begin{ctabular}{@{}c@{\,}c@{\,}c@{\,}c@{\,}c@{}}
        \begin{tabular}{@{}c@{}}
          \begin{tikzpicture}[-, frame rectangle, scale=0.815]
            \node (w0) at (-\ladocubo/2,  \ladocubo/2, -\ladocubo/2) [mundo, label = {[etiqMundo]above:$w_0$}] {$\circ\bullet\bullet$};
            \node (w1) at ( \ladocubo/2,  \ladocubo/2, -\ladocubo/2) [mundo, label = {[etiqMundo]above:$w_1$}] {$\circ\circ\bullet$};
            \node (w2) at (-\ladocubo/2,  \ladocubo/2,  \ladocubo/2) [mundo, label = {[etiqMundo]below:$~~~~~~w_2$}] {$\bullet\bullet\bullet$};
            \node (w3) at ( \ladocubo/2,  \ladocubo/2,  \ladocubo/2) [mundo, label = {[etiqMundo]below right:$w_3$}] {$\bullet\circ\bullet$};
            \node (w4) at (-\ladocubo/2, -\ladocubo/2, -\ladocubo/2) [mundo, label = {[etiqMundo]above right:$w_4$}] {$\circ\bullet\circ$};
            \node (w5) at ( \ladocubo/2, -\ladocubo/2, -\ladocubo/2) [mundo, label = {[etiqMundo]below:$w_5$}] {$\circ\circ\circ$};
            \node (w6) at (-\ladocubo/2, -\ladocubo/2,  \ladocubo/2) [mundo, label = {[etiqMundo]below:$w_6$}] {$\bullet\bullet\circ$};
            \node (w7) at ( \ladocubo/2, -\ladocubo/2,  \ladocubo/2) [mundo, label = {[etiqMundo]below:$w_7$}] {$\bullet\circ\circ$};

            \path (w0) edge [flecha] node [etiqFlecha, within] {$\agb$} (w1)
                       edge [flecha, bend right=5] node [etiqFlecha, within, pos = 0.7] {$\agc$} (w4)
                       edge [flecha] node [etiqFlecha, above, pos=1.3] {$\aga$} (w2)
                  (w1) edge [flecha] node [etiqFlecha, below, pos=-0.3] {$\aga$} (w3)
                  (w4) edge [flecha] node [etiqFlecha, below, pos=-0.3] {$\aga$} (w6)
                  (w2) edge [flecha, cross line] node [etiqFlecha, within, pos = 0.6] {$\agb$} (w3)
                       edge [flecha, bend right=5] node [etiqFlecha, within] {$\agc$} (w6)
                  (w7) edge [flecha] node [etiqFlecha, within] {$\agb$} (w6)
                       edge [flecha, cross line, bend right=5] node [etiqFlecha, within, pos=0.7] {$\agc$} (w3);

            \draw[topicclass] (w5.north west) -- (w5.north east) -- (w5.south east) -- (w7.south east) -- (w7.south west) -- (w7.north west) -- cycle;

            \draw[topicclass] (w0.north west) -- (w1.north east) -- (w1.south east) -- (w3.south east) -- (w4.south east) -- (w6.south east) -- (w6.south west) -- (w2.north west) -- cycle;
          \end{tikzpicture}
        \end{tabular}
        &
        \begin{tabular}{@{}c@{}}
          {\small $\overset{\symbSSE{\ag}{\chi_{\aga}}}{\Rightarrow}$}
        \end{tabular}
        &
        \begin{tabular}{@{}c@{}}
          \begin{tikzpicture}[-, frame rectangle, scale=0.815]
            \node (w0) at (-\ladocubo/2,  \ladocubo/2, -\ladocubo/2) [mundo, label = {[etiqMundo]above:$w_0$}] {$\circ\bullet\bullet$};
            \node (w1) at ( \ladocubo/2,  \ladocubo/2, -\ladocubo/2) [mundo, label = {[etiqMundo]above:$w_1$}] {$\circ\circ\bullet$};
            \node (w2) at (-\ladocubo/2,  \ladocubo/2,  \ladocubo/2) [mundo, label = {[etiqMundo]below:$~~~~~~w_2$}] {$\bullet\bullet\bullet$};
            \node (w3) at ( \ladocubo/2,  \ladocubo/2,  \ladocubo/2) [mundo, label = {[etiqMundo]below right:$w_3$}] {$\bullet\circ\bullet$};
            \node (w4) at (-\ladocubo/2, -\ladocubo/2, -\ladocubo/2) [mundo, label = {[etiqMundo]above right:$w_4$}] {$\circ\bullet\circ$};
            \node (w5) at ( \ladocubo/2, -\ladocubo/2, -\ladocubo/2) [mundo, label = {[etiqMundo]below:$w_5$}] {$\circ\circ\circ$};
            \node (w6) at (-\ladocubo/2, -\ladocubo/2,  \ladocubo/2) [mundo, label = {[etiqMundo]below:$w_6$}] {$\bullet\bullet\circ$};
            \node (w7) at ( \ladocubo/2, -\ladocubo/2,  \ladocubo/2) [mundo, label = {[etiqMundo]below:$w_7$}] {$\bullet\circ\circ$};

            \path (w0) edge [flecha] node [etiqFlecha, within] {$\agb$} (w1)
                       edge [flecha, bend right=5] node [etiqFlecha, within, pos = 0.7] {$\agc$} (w4)
                       edge [flecha] node [etiqFlecha, above, pos=1.3] {$\aga$} (w2)
                  (w1) edge [flecha] node [etiqFlecha, below, pos=-0.3] {$\aga$} (w3)
                  (w4) edge [flecha] node [etiqFlecha, below, pos=-0.3] {$\aga$} (w6)
                  (w2) edge [flecha, cross line] node [etiqFlecha, within, pos = 0.6] {$\agb$} (w3)
                       edge [flecha, bend right=5] node [etiqFlecha, within] {$\agc$} (w6);

            \draw[topicclass] (w0.north west) -- (w0.north east) -- (w4.south east) -- (w6.south east) -- (w6.south west) -- (w2.north west) -- cycle;

            \draw[topicclass] (w1.north west) -- (w1.north east) -- (w5.south east) -- (w7.south east) -- (w7.south west) -- (w3.north west) -- cycle;
          \end{tikzpicture}
        \end{tabular}
        &
        \begin{tabular}{@{}c@{}}
          {\small $\overset{\symbSSE{\ag}{\chi_{\agc}}}{\Rightarrow}$}
        \end{tabular}
        &
        \begin{tabular}{@{}c@{}}
          \begin{tikzpicture}[-, frame rectangle, scale=0.815]
            \node (w0) at (-\ladocubo/2,  \ladocubo/2, -\ladocubo/2) [mundo, label = {[etiqMundo]above:$w_0$}] {$\circ\bullet\bullet$};
            \node (w1) at ( \ladocubo/2,  \ladocubo/2, -\ladocubo/2) [mundo, label = {[etiqMundo]above:$w_1$}] {$\circ\circ\bullet$};
            \node (w2) at (-\ladocubo/2,  \ladocubo/2,  \ladocubo/2) [mundo, label = {[etiqMundo]below:$~~~~~~w_2$}] {$\bullet\bullet\bullet$};
            \node (w3) at ( \ladocubo/2,  \ladocubo/2,  \ladocubo/2) [mundo, label = {[etiqMundo]below right:$w_3$}] {$\bullet\circ\bullet$};
            \node (w4) at (-\ladocubo/2, -\ladocubo/2, -\ladocubo/2) [mundo, label = {[etiqMundo]above right:$w_4$}] {$\circ\bullet\circ$};
            \node (w5) at ( \ladocubo/2, -\ladocubo/2, -\ladocubo/2) [mundo, label = {[etiqMundo]below:$w_5$}] {$\circ\circ\circ$};
            \node (w6) at (-\ladocubo/2, -\ladocubo/2,  \ladocubo/2) [mundo, label = {[etiqMundo]below:$w_6$}] {$\bullet\bullet\circ$};
            \node (w7) at ( \ladocubo/2, -\ladocubo/2,  \ladocubo/2) [mundo, label = {[etiqMundo]below:$w_7$}] {$\bullet\circ\circ$};

            \path (w0) edge [flecha, bend right=5] node [etiqFlecha, within, pos = 0.7] {$\agc$} (w4)
                       edge [flecha] node [etiqFlecha, above, pos=1.3] {$\aga$} (w2)
                  (w1) edge [flecha] node [etiqFlecha, below, pos=-0.3] {$\aga$} (w3)
                  (w4) edge [flecha] node [etiqFlecha, below, pos=-0.3] {$\aga$} (w6)
                  (w2) edge [flecha, bend right=5] node [etiqFlecha, within] {$\agc$} (w6);

          \end{tikzpicture}
        \end{tabular}
        \\
        $M_1$ && $M'_2$ && $M'_3$ \\
      \end{ctabular}
      On the other hand, $\truthset{M_1}{\chi_{\agc}} = \set{w_1, w_5}$, so applying $\symbSSE{\set{\agb,\agc}}{\chi_{\agc}}$ to $M_1$ yields $M''_2$ below. Then, $\truthset{M''_2}{\chi_{\aga}} = \set{w_1, w_3, w_5, w_7}$ so a further $\symbSSE{\set{\aga}}{\chi_{\aga}}$ yields $M''_3$.
      \begin{ctabular}{@{}c@{\,}c@{\,}c@{\,}c@{\,}c@{}}
        \begin{tabular}{@{}c@{}}
          \begin{tikzpicture}[-, frame rectangle, scale=0.815]
            \node (w0) at (-\ladocubo/2,  \ladocubo/2, -\ladocubo/2) [mundo, label = {[etiqMundo]above:$w_0$}] {$\circ\bullet\bullet$};
            \node (w1) at ( \ladocubo/2,  \ladocubo/2, -\ladocubo/2) [mundo, label = {[etiqMundo]above:$w_1$}] {$\circ\circ\bullet$};
            \node (w2) at (-\ladocubo/2,  \ladocubo/2,  \ladocubo/2) [mundo, label = {[etiqMundo]below:$~~~~~~w_2$}] {$\bullet\bullet\bullet$};
            \node (w3) at ( \ladocubo/2,  \ladocubo/2,  \ladocubo/2) [mundo, label = {[etiqMundo]below right:$w_3$}] {$\bullet\circ\bullet$};
            \node (w4) at (-\ladocubo/2, -\ladocubo/2, -\ladocubo/2) [mundo, label = {[etiqMundo]above right:$w_4$}] {$\circ\bullet\circ$};
            \node (w5) at ( \ladocubo/2, -\ladocubo/2, -\ladocubo/2) [mundo, label = {[etiqMundo]below:$w_5$}] {$\circ\circ\circ$};
            \node (w6) at (-\ladocubo/2, -\ladocubo/2,  \ladocubo/2) [mundo, label = {[etiqMundo]below:$w_6$}] {$\bullet\bullet\circ$};
            \node (w7) at ( \ladocubo/2, -\ladocubo/2,  \ladocubo/2) [mundo, label = {[etiqMundo]below:$w_7$}] {$\bullet\circ\circ$};

            \path (w0) edge [flecha] node [etiqFlecha, within] {$\agb$} (w1)
                       edge [flecha, bend right=5] node [etiqFlecha, within, pos = 0.7] {$\agc$} (w4)
                       edge [flecha] node [etiqFlecha, above, pos=1.3] {$\aga$} (w2)
                  (w1) edge [flecha] node [etiqFlecha, below, pos=-0.3] {$\aga$} (w3)
                  (w4) edge [flecha] node [etiqFlecha, below, pos=-0.3] {$\aga$} (w6)
                  (w2) edge [flecha, cross line] node [etiqFlecha, within, pos = 0.6] {$\agb$} (w3)
                       edge [flecha, bend right=5] node [etiqFlecha, within] {$\agc$} (w6)
                  (w7) edge [flecha] node [etiqFlecha, within] {$\agb$} (w6)
                       edge [flecha, cross line, bend left=5] node [etiqFlecha, within, pos=0.7] {$\agc$} (w3);

            \draw[topicclass] (w5.south east) -- (w5.north east) -- (w1.south east) -- (w1.north east) -- (w1.north west) -- (w1.south west) -- (w1.south) -- (w5.north) -- (w5.north west) -- (w5.south west) -- cycle;

            \draw[topicclass] (w0.north west) -- (w0.north east) -- (w3.north east) -- (w3.south east) -- (w3.south) -- (w7.north) -- (w7.north east) -- (w7.south east) -- (w6.south west) -- (w2.north west) -- cycle;
          \end{tikzpicture}
        \end{tabular}
        &
        \begin{tabular}{@{}c@{}}
          {\small $\overset{\symbSSE{\ag}{\chi_{\agc}}}{\Rightarrow}$}
        \end{tabular}
        &
        \begin{tabular}{@{}c@{}}
          \begin{tikzpicture}[-, frame rectangle, scale=0.815]
            \node (w0) at (-\ladocubo/2,  \ladocubo/2, -\ladocubo/2) [mundo, label = {[etiqMundo]above:$w_0$}] {$\circ\bullet\bullet$};
            \node (w1) at ( \ladocubo/2,  \ladocubo/2, -\ladocubo/2) [mundo, label = {[etiqMundo]above:$w_1$}] {$\circ\circ\bullet$};
            \node (w2) at (-\ladocubo/2,  \ladocubo/2,  \ladocubo/2) [mundo, label = {[etiqMundo]below:$~~~~~~w_2$}] {$\bullet\bullet\bullet$};
            \node (w3) at ( \ladocubo/2,  \ladocubo/2,  \ladocubo/2) [mundo, label = {[etiqMundo]below right:$w_3$}] {$\bullet\circ\bullet$};
            \node (w4) at (-\ladocubo/2, -\ladocubo/2, -\ladocubo/2) [mundo, label = {[etiqMundo]above right:$w_4$}] {$\circ\bullet\circ$};
            \node (w5) at ( \ladocubo/2, -\ladocubo/2, -\ladocubo/2) [mundo, label = {[etiqMundo]below:$w_5$}] {$\circ\circ\circ$};
            \node (w6) at (-\ladocubo/2, -\ladocubo/2,  \ladocubo/2) [mundo, label = {[etiqMundo]below:$w_6$}] {$\bullet\bullet\circ$};
            \node (w7) at ( \ladocubo/2, -\ladocubo/2,  \ladocubo/2) [mundo, label = {[etiqMundo]below:$w_7$}] {$\bullet\circ\circ$};

            \path (w0) edge [flecha, bend right=5] node [etiqFlecha, within, pos = 0.7] {$\agc$} (w4)
                       edge [flecha] node [etiqFlecha, above, pos=1.3] {$\aga$} (w2)
                  (w4) edge [flecha] node [etiqFlecha, below, pos=-0.3] {$\aga$} (w6)
                  (w2) edge [flecha, cross line] node [etiqFlecha, within, pos = 0.6] {$\agb$} (w3)
                       edge [flecha, bend right=5] node [etiqFlecha, within] {$\agc$} (w6)
                  (w7) edge [flecha] node [etiqFlecha, within] {$\agb$} (w6)
                       edge [flecha, cross line, bend right=5] node [etiqFlecha, within, pos=0.7] {$\agc$} (w3);

            \draw[topicclass] (w0.north west) -- (w0.north east) -- (w4.south east) -- (w6.south east) -- (w6.south west) -- (w2.north west) -- cycle;

            \draw[topicclass] (w1.north west) -- (w1.north east) -- (w5.south east) -- (w7.south east) -- (w7.south west) -- (w3.north west) -- cycle;
          \end{tikzpicture}
        \end{tabular}
        &
        \begin{tabular}{@{}c@{}}
          {\small $\overset{\symbSSE{\ag}{\chi_{\aga}}}{\Rightarrow}$}
        \end{tabular}
        &
        \begin{tabular}{@{}c@{}}
          \begin{tikzpicture}[-, frame rectangle, scale=0.815]
            \node (w0) at (-\ladocubo/2,  \ladocubo/2, -\ladocubo/2) [mundo, label = {[etiqMundo]above:$w_0$}] {$\circ\bullet\bullet$};
            \node (w1) at ( \ladocubo/2,  \ladocubo/2, -\ladocubo/2) [mundo, label = {[etiqMundo]above:$w_1$}] {$\circ\circ\bullet$};
            \node (w2) at (-\ladocubo/2,  \ladocubo/2,  \ladocubo/2) [mundo, label = {[etiqMundo]below:$~~~~~~w_2$}] {$\bullet\bullet\bullet$};
            \node (w3) at ( \ladocubo/2,  \ladocubo/2,  \ladocubo/2) [mundo, label = {[etiqMundo]below right:$w_3$}] {$\bullet\circ\bullet$};
            \node (w4) at (-\ladocubo/2, -\ladocubo/2, -\ladocubo/2) [mundo, label = {[etiqMundo]above right:$w_4$}] {$\circ\bullet\circ$};
            \node (w5) at ( \ladocubo/2, -\ladocubo/2, -\ladocubo/2) [mundo, label = {[etiqMundo]below:$w_5$}] {$\circ\circ\circ$};
            \node (w6) at (-\ladocubo/2, -\ladocubo/2,  \ladocubo/2) [mundo, label = {[etiqMundo]below:$w_6$}] {$\bullet\bullet\circ$};
            \node (w7) at ( \ladocubo/2, -\ladocubo/2,  \ladocubo/2) [mundo, label = {[etiqMundo]below:$w_7$}] {$\bullet\circ\circ$};

            \path (w0) edge [flecha, bend right=5] node [etiqFlecha, within, pos = 0.7] {$\agc$} (w4)
                       edge [flecha] node [etiqFlecha, above, pos=1.3] {$\aga$} (w2)
                  (w4) edge [flecha] node [etiqFlecha, below, pos=-0.3] {$\aga$} (w6)
                  (w2) edge [flecha, bend right=5] node [etiqFlecha, within] {$\agc$} (w6)
                  (w7) edge [flecha, cross line, bend right=5] node [etiqFlecha, within, pos=0.7] {$\agc$} (w3);

          \end{tikzpicture}
        \end{tabular}
        \\
        $M_1$ && $M''_2$ && $M''_3$ \\
      \end{ctabular}
      Thus, $(M_1, w_3) \Vdash \mmSSE{\set{\aga}}{\chi_{\aga}}{\mmSSE{\set{\agb, \agc}}{\chi_{\agc}}{\chi_{\agc}}}$ and yet $(M_1, w_3) \not\Vdash \mmSSE{\set{\agb, \agc}}{\chi_{\agc}}{\mmSSE{\set{\aga}}{\chi_{\aga}}{\chi_{\agc}}}$.

      \item\label{hch:SSE:no-valid:iii} As it can be seen from \autoref{hch:SSE:no-valid:i}, $(M_1, w_2) \Vdash \mmSSE{\set{\aga, \agb}}{\chi^\lor}{\mmSSE{\set{\agc}}{\chi^\lor}{\chi_{\aga}}}$ (the model that results from $\symbSSE{\set{\aga,\agb}}{\chi^\lor}$ is also $M_2$, and a further application of $\symbSSE{\set{\agc}}{\chi^\lor}$ yields a model $M'''_3$ differing from $M_3$ in having an additional $\agc$-edge between $w_2$ and $w_6$), and yet $(M_1, w_2) \not\Vdash \mmSSE{\set{\aga, \agb, \agc}}{\chi^\lor}{\chi_{\aga}}$.

      \item\label{hch:SSE:no-valid:iv} As it can be seen from \autoref{hch:SSE:no-valid:ii}, $(M_1, w_3) \Vdash \mmSSE{\ag}{\chi_{\aga}}{\mmSSE{\ag}{\chi_{\agc}}{\chi_{\agb}}}$ (the model that results from $\symbSSE{\ag}{\chi_{\aga}}$ is also $M'_2$, and a further $\symbSSE{\ag}{\chi_{\agc}}$ yields $M'_3$ too). However, $\truthset{M_1}{\chi_{\aga} \land \chi_{\agc}} = \set{w_5}$, so applying $\symbSSE{\ag}{(\chi_{\aga} \land \chi_{\agc})}$ on $M_1$ has no effect, and thus $(M_1, w_3) \not\Vdash \mmSSE{\ag}{(\chi_{\aga} \land \chi_{\agc})}{\chi_{\agb}}$.
    \end{compactenumerate}

    The provided counterexamples show that the given formulas are not valid, even under equivalence relations.
  \end{proof}
\end{hecho}

\bsparagraph{Axiom system} Once again, the axiom system for $\LAdSSE$ relies on reduction axioms \autorefp{tbl:LOdSSE}. Axioms \dsaSSE{p}, \dsaSSE{\lnot}, \dsaSSE{\land} and rule \dsreSSE are similar to their previous matching cases. Axiom \dsaSSE{\opgDist} is the important one, indicating that a group $\sag$ knows $\varphi$ distributively after the action ($\mmSSE{\sen}{\chi}{\mmDg{\varphi}}$) if and only if the group $\sen \cup \sag$ knew, distributively, that $\varphi$ would hold after the action ($\mmDs{\sen \cup \sag}{\mmSSE{\sen}{\chi}{\varphi}}$) and the group $\sag$ knew, distributively \emph{and relative to similarity on $\chi$}, that $\varphi$ would hold after the action ($\mmDsr{\sag}{\chi}{\mmSSE{\sen}{\chi}{\varphi}}$).

\begin{table}[ht]
  \begin{smallctabular}{c}
    \toprule
    \begin{tabular}[t]{@{}r@{\;\;}l@{\qquad\quad}r@{\;\;}l@{}}
      \dsaSSE{p}:        & $\vdashLOdSSE \mmSSE{\sen}{\chi}{p} \lldimp p$ &
      \dsaSSE{\opgDist}: & $\vdashLOdSSE \mmSSE{\sen}{\chi}{\mmDg{\varphi}} \lldimp (\mmDs{\sen \cup \sag}{\mmSSE{\sen}{\chi}{\varphi}} \land \mmDsr{\sag}{\chi}{\mmSSE{\sen}{\chi}{\varphi}})$ \\
      \dsaSSE{\lnot}:    & $\vdashLOdSSE \mmSSE{\sen}{\chi}{\lnot \varphi} \lldimp \lnot \mmSSE{\sen}{\chi}{\varphi}$ &
      \dsreSSE:          & If $\vdashLOdSSE \varphi_1 \ldimp \varphi_2$ then $\vdashLOdSSE \mmSSE{\sen}{\chi}{\varphi_1} \ldimp \mmSSE{\sen}{\chi}{\varphi_2}$ \\
      \dsaSSE{\land}:    & \multicolumn{3}{@{}l}{$\vdashLOdSSE \mmSSE{\sen}{\chi}{(\varphi \land \psi)} \lldimp (\mmSSE{\sen}{\chi}{\varphi} \land \mmSSE{\sen}{\chi}{\psi})$} \\
    \end{tabular} \\
    \bottomrule
  \end{smallctabular}
  \caption{Additional axioms and rules for \LOdSSE, which characterises the formulas in \LAdSSE that are valid on models in \marm.}
  \label{tbl:LOdSSE}
\end{table}

\begin{teorema}\label{teo:LOdSSE}
  The axiom system \LOdSSE (\LOd[\autoref{tbl:LOd}]+\autoref{tbl:LOdSSE}) is sound and strongly complete for formulas in \LAdSSE valid over models in \marm.
\end{teorema}

\section{Coda: sharing with \emph{somebody}}\label{sec:XXS}

\autoref{sec:XEE} introduced two forms of inter-agent communication: \EEE, through which \emph{all} agents share \emph{all} their information with \emph{everybody}, and \SEE, through which \emph{some} agents share \emph{all} their information with \emph{everybody}. Then, \autoref{sec:XXS} introduced \SEE, a form of communication through which \emph{some} agents share \emph{some} of their information with \emph{everybody}. The next natural step is defining a form of communication through which \emph{some} agents share \emph{some} of their information with \emph{somebody}. This operation, which could be denoted by \SSS, will not be studied in this manuscript. Instead, the text will only discuss some of the modelling options that arise.

\medskip

First, when defining the \SSS operation, one might decide that the shared information is received only by agents in a given group $\rec \subseteq \ag$. This would represent a scenario similar to a round table with a potentially distracted audience: the people on `the table' get to talk about a given topic, with some members of the audience listening and the distracted ones missing the information. However, one could also think of a different scenario, one in which every sharing agent $\agi$ has a specific set of listeners $\rec_{\agi}$. This is closer to what happens in some online social networks, where only `friends' or `followers' can receive what a given agent sends (cf. with \citealp{BaltagSmets2020}, discussed in \autoref{sbs:XEE:others}). In fact, the scenario with a single set $\rec$ can be seen as the special case of the scenario with $\set{\rec_{\agi}}_{\agi \in \ag}$ in which all sets $\rec_{\agi}$ are the same.

A further possibility is to take the social networks idea seriously, and work with models that represent explicitly the social connections each agent has. These structures have been used by the logic community (\citealp{SeligmanEtAl2011} is one of the earliest proposals), either for studying information flow in social networks (peer pressure: \citealp{ZhenSeligman2011}; reflexive influence: \citealp{ChristoffEtAl2016}; diffusion and prediction: \citealp{BaltagEtAl2019}) or for studying social network formation (as a side effect of peer pressure: \citealp{PedersenSlavkovik2017}; by similarity: \citealp{SmetsVelazquez2018,SmetsVelazquez2017th,SmetsVelazquez2018dali-si}).

\smallskip

A second decision to make when defining an \SSS operation is the following: what information, if any, obtain the agents that do not `hear' the communication? On one extreme, these agents might be oblivious, not only to the content of the communication, but also to the fact that a communication took place. On the technical side, representing this form of `private' communication requires an operation that changes not only the model's indistinguishability relations, but also its domain. Indeed, while the `real' part of the model would change as the communication takes place (the receiving agents might learn something), for non-receiving agents one should keep `a copy' of the original model, to indicate that they see no change at all.\footnote{This is how the \emph{action models} of \citet{BaltagMossSolecki1998,BaltagMossSolecki1999} deal with private announcements.} On the other extreme, non-receiving agents might not hear the content of the communication, but they might notice that the communication takes place (the \emph{semi-public} form of communication from \citealp{AgotnesWang2017} and \citealp{BaltagSmets2020} discussed in \autoref{sbs:XEE:others}). Even more, they might know the \emph{topic} of the conversation. Combining this with some previous knowledge about what the communicating agents know, the non-receivers might get to know part of what is being shared, and thus they might get to know (part of) the new epistemic state of the communicating agents.

\section{Summary and further research lines}\label{sec:end}

This paper has discussed three communication actions. Different from most epistemic acts present in the literature, they truly describe processes of inter-agent communication: those in which the information that is being shared is information some of the agents already have. Through the first two actions, \EEE and \SEE, all/some agents share all their information with everybody. Through the third, \SSE, the core of this contribution, some agents share \emph{some} of their information with everybody. The text has presented examples of these actions at work, discussing some of their basic properties and providing, in all cases, a sound and complete axiom system for modalities describing their effects.

\medskip

There are several research lines that arise from the current proposal. Here are some of them.
\begin{compactitemize}
  \item Distributed knowledge has played a crucial role in this manuscript. Yet, the equally important notion of \emph{common knowledge} (referring to the infinite iteration including everybody knows, everybody knows that everybody knows, and so on; \citealp{Lewis1969,sep-common-knowledge}) has been absent. Common knowledge is an important piece when discussing communication between agents, as one would like to know not only what each individual agent gets from the action (their individual knowledge), but also what a group as a whole learns from it (the group's common knowledge). Adding the common knowledge operator to the studied languages involves the use of additional tools, in particular for the axiomatisation.

  \item In a \SSE action, what each sharing agent communicates is given by a formula $\chi$, understood as the \emph{subject}/\emph{topic} of the conversation. This is the reason why the operation only considers for elimination edges between worlds differing in $\chi$'s truth-value. However, as mentioned, there are other alternatives. One can understand $\chi$ as the \emph{content} of the conversation instead; in such case, the operation should consider for elimination only edges \emph{pointing to $\lnot\chi$-worlds}. One could also take a \emph{set} of formulas. If this represents the set of topics then, under some reasonable restrictions on the class of models, such variation could turn out to be a generalisation of one of the other actions: an \SSE-communication \emph{on all topics} could be equivalent to an \SEE communication. In any of these alternatives, one can also look for a less uniform treatment, assigning to each sharing agent (and even to sets of of them) a potentially different (set of) formula(s) (cf. \citealp{vanBenthemMinica2012,BaltagBoddySmets2018}). All these options deserve a proper exploration.

  \item In the communication actions studied in this text, each agent $\agi$ communicates by sharing (some of) the possibilities she has already discarded (those in $\ov{\sbi{\R}}$). Thus, it has been implicitly assumed that, while agents might not share `everything that they know', what they share is `something that they do know' (in other words, they `communicate' a set $S \subseteq \ov{\sbi{\R}}$). But one can also find agents that share `more than what they know' (they communicate a set $S \supset \ov{\sbi{\R}}$), and even only `things they do not know' (they communicate a set $S$ with $S \cap \ov{\sbi{\R}} = \emptyset$). These alternatives can be used for studying acts of \emph{lying}, which so far have been modelled only by adapting tools from public announcement logic \citep{vanDitmarschLying2013}.

  \item The success of public announcement logic comes from the fact that it provides formal tools for studying different epistemic change phenomena. Among them, one can find 
  studies on \emph{arbitrary announcements}. Indeed, some proposals (e.g., \citealp{BalbianiEtAl2008,GalimullinAgotnes2021}) have worked with modalities of the form $\mmDia{\star!}{\varphi}$, read as ``there is a formula that can be truthfully announced, and after doing so $\varphi$ will be the case''. The setting presented here offers a further alternative for quantifying over information change: one can define an operation through which a given agent (and eventually a set of them) shares an \emph{arbitrary} set $S \subseteq \ov{\sbi{\R}}$. Thus, one can study what a set of agents can get to know (i.e., a form of \emph{knowability}) by sharing their information.
\end{compactitemize}

\appendix
\section{Appendix}\label{sec:proofs}

In the proofs, IH abbreviates ``inductive hypothesis''.

\subsection{Proof of \autoref{teo:LOdEEE} (System \texorpdfstring{\LOdEEE}{LOdEEE})}\label{proof:teo:LOdEEE}

\nsparagraph{Soundness} The soundness of axioms and rules in \LOd is well-known. For those in \autoref{tbl:LOdEEE}, the case of \dsaEEE{p} is straightforward (the operation does not affect atomic valuations), the cases for \dsaEEE{\lnot} and \dsaEEE{\land} follow from the inductive hypotheses, and \dsaEEE{\opgDist} has been proved valid already \autorefp{pro:EEE}. For \dsreEEE, suppose $\Vdash \varphi_1 \ldimp \varphi_2$, that is, suppose $\truthset{M}{\varphi_1} = \truthset{M}{\varphi_2}$ for every $M$ in \marm. Take any pointed model $(M, w)$ with $M = \tupla{W, \R, V}$. Then, $(M,w) \Vdash \mmEEE{\varphi_1}$ if and only if $w \in \truthset{\mopEEE{M}}{\varphi_1}$. But $\mopEEE{M}$ is a model in \marm, so the latter holds if and only if $w \in \truthset{\mopEEE{M}}{\varphi_2}$, which holds if and only if $(M,w) \Vdash \mmEEE{\varphi_2}$. Thus, $(M,w) \Vdash \mmEEE{\varphi_1} \ldimp \mmEEE{\varphi_2}$.

\bsparagraph{Completeness} The argument relies on a translation from \LAdEEE to \LAd, which is defined as follows.

\begin{definicion}[Translation]\label{def:trans:EEE}
  The translation $\tr$ is given by
  \begingroup
    \small
    \[
      \renewcommand{\arraystretch}{1.4}
      \begin{array}{c@{\qquad\qquad}c}
        \begin{array}{@{}r@{\;:=\;}l@{}}
          \tr(p)                         & p, \\
          \tr(\lnot\varphi)              & \lnot \tr(\varphi), \\
          \tr(\varphi_1 \land \varphi_2) & \tr(\varphi_1) \land \tr(\varphi_2), \\
          \tr(\mmDg{\varphi})            & \mmDg{\tr(\varphi)}, \\
          \multicolumn{2}{@{}r}{}
        \end{array}
        &
        \begin{array}{@{}r@{\;:=\;}l@{}}
          \tr(\mmEEE{p})                           & \tr(p), \\
          \tr(\mmEEE{\lnot \varphi})               & \tr(\lnot \mmEEE{\varphi}), \\
          \tr(\mmEEE{(\varphi_1 \land \varphi_2)}) & \tr(\mmEEE{\varphi_1} \land \mmEEE{\varphi_2}), \\
          \tr(\mmEEE{\mmDg{\varphi}})              & \tr(\mmDs{\ag}{\mmEEE{\varphi}}), \\
          \tr(\mmEEE{\mmEEE{\varphi}})             & \tr(\mmEEE{\tr(\mmEEE{\varphi})}). \\
        \end{array}
      \end{array}
    \]
  \endgroup
\end{definicion}

This translation works with formulas of the form $\mmEEE{\mmEEE{\varphi}}$ in an ``inside-out' fashion, dealing first with the deepest occurrence of $\modEEE$ (i.e., translating $\mmEEE{\varphi}$) before dealing with the rest (i.e., before translating $\mmEEE{\tr(\mmEEE{\varphi})}$). Because of this, the strategy proving strong completeness is as follows (cf. \citealp{Plaza1989} and \citealp[Section 4.4]{Gerbrandy1999}): \begin{inlineenum} \item show that $\tr$ is a proper recursive translation that returns formulas in \LAd \autorefp{pro:tr-LAdEEE-LAd:proper}, \item show that a formula and its translation are both provably and semantically equivalent \autorefp{pro:tr-LAdEEE-LAd:works}, and \item use $\tr$ and the completeness of \LOd to show that, if $\Psi \cup \set{\varphi} \subseteq \LAdEEE$, then from $\Psi \Vdash \varphi$ it follows that $\varphi$ is derivable from $\Psi$ in \LOdEEE\end{inlineenum}.

\medskip

It is clear that the domain of $\tr$ is \LAdEEE: cases on the leftmost column take care of formulas that do not start with $\modEEE$, and cases on the rightmost column take care of formulas that start with $\modEEE$. Still, one needs to show not only that, if $\varphi$ is in $\LAdEEE$, then $\tr(\varphi)$ is in \LAd, but also that the calculation of $\tr(\varphi)$ actually ends. In doing so, the following notion of complexity will play a crucial role.

\begin{definicion}[Complexity for \LAdEEE]\label{def:comp:EEE}
  The functions $\nscom:\LAdEEE \to \Nat\setminus\set{0}$ (nested `static' complexity, focussing on operators in \LAd) and $\ndcom:\LAdEEE \to \Nat$ (nested `dynamic' complexity, focussing on $\modEEE$) are defined as follows.
  \begin{smallcbtabular}{@{}c@{\qquad}c@{}}
    \begin{tabular}{@{}r@{\;:=\;}l@{}}
      $\nscom(p)$                         & $1$, \\
      $\nscom(\lnot \varphi)$             & $1 + \nscom(\varphi)$, \\
      $\nscom(\varphi_1 \land \varphi_2)$ & $1 + \max\set{\nscom(\varphi_1), \nscom(\varphi_2)}$, \\
      $\nscom(\mmDg{\varphi})$            & $1 + \nscom(\varphi)$, \\
      $\nscom(\mmEEE{\varphi})$           & $2\nscom(\varphi)$. \\
    \end{tabular}
    &
    \begin{tabular}{@{}r@{\;:=\;}l@{}}
      $\ndcom(p)$                         & $0$, \\
      $\ndcom(\lnot \varphi)$             & $\ndcom(\varphi)$, \\
      $\ndcom(\varphi_1 \land \varphi_2)$ & $\max\set{\ndcom(\varphi_1), \ndcom(\varphi_2)}$, \\
      $\ndcom(\mmDg{\varphi})$            & $\ndcom(\varphi)$, \\
      $\ndcom(\mmEEE{\varphi})$           & $1 + \ndcom(\varphi)$. \\
    \end{tabular}
  \end{smallcbtabular}
  Then, for any $\varphi_1, \varphi_2 \in \LAdEEE$, write $\com(\varphi_1) > \com(\varphi_2)$ if and only if
  \begin{ctabular}{c}
    $\ndcom(\varphi_1) > \ndcom(\varphi_2)$
    \qtnq{or}
    $\ndcom(\varphi_1) = \ndcom(\varphi_2)$ and $\nscom(\varphi_1) > \nscom(\varphi_2)$.
  \end{ctabular}
\end{definicion}

Thus, while $\nscom$ counts a formula's nested Boolean and modal operators, $\ndcom$ counts a formula's nested dynamic operators. The complexity ordering $\com$ relies on both $\nscom$ and $\ndcom$, with the latter taking precedence: $\varphi_1$ is more complex than $\varphi_2$ (i.e., $\com(\varphi_1) > \com(\varphi_2)$) if and only if either $\varphi_1$'s `dynamic' complexity is greater than that of $\varphi_2$ (i.e., $\ndcom(\varphi_1) > \ndcom(\varphi_2)$), or else both have the same `dynamic' complexity but $\varphi_1$'s `static' complexity is greater than that of $\varphi_2$ (i.e., $\ndcom(\varphi_1) = \ndcom(\varphi_2)$ and $\nscom(\varphi_1) > \nscom(\varphi_2)$).

\smallskip

Relying on the complexity ordering $\com$, the following proposition states that $\tr$ is a proper recursive translation from \LAdEEE \emph{to \LAd}.

\begin{proposicion}\label{pro:tr-LAdEEE-LAd:proper}
  For every $\varphi \in \LAdEEE$,
  \begin{enumeratetr}
    \item\label{pro:tr-LAdEEE-LAd:proper:i} if $\tr(\varphi)$ is defined in terms of $\tr(\psi)$, then $\com(\varphi) > \com(\psi)$. Thus, the calculation of $\tr(\varphi)$ will eventually end.
    \item\label{pro:tr-LAdEEE-LAd:proper:ii} $\tr(\varphi) \in \LAd$.
  \end{enumeratetr}
  \begin{proof}
    The following lemma will be useful.

    \begin{lema}\label{lem:tr-LAdEEE-LAd:proper:ssub}
      Define the \emph{strict subformula} function $\ssub: \LAdEEE \to \power{\LAdEEE}$ as
      \begin{smallctabular}{@{}c@{\qquad}c@{}}
        \begin{tabular}{@{}r@{\,:=\,}l@{}}
          $\ssub(p)$                         & $\emptyset$, \\
          $\ssub(\lnot \varphi)$             & $\set{\varphi} \cup \ssub(\varphi)$, \\
          $\ssub(\varphi_1 \land \varphi_2)$ & $\set{\varphi_1, \varphi_2} \cup \ssub(\varphi_1) \cup \ssub(\varphi_2)$, \\
        \end{tabular}
        &
        \begin{tabular}{@{}r@{\,:=\,}l@{}}
          $\ssub(\mmDg{\varphi})$  & $\set{\varphi} \cup \ssub(\varphi)$, \\
          $\ssub(\mmEEE{\varphi})$ & $\set{\varphi} \cup \ssub(\varphi)$. \\
        \end{tabular}
      \end{smallctabular}
      Let $\varphi$ be an \LAdEEE-formula. Then, $\com(\varphi) > \com(\psi)$ for every $\psi \in \ssub(\varphi)$.
      \begin{proof}
        The proof is by structural induction on $\varphi$. Here are the cases.
        \begin{compactitemize}
          \item \tb{Base case ($\bs{p}$)}. Immediate, as $\ssub(p) = \emptyset$.

          \item \tb{Inductive case ($\bs{\lnot \varphi}$)}. Note that, by definition,
          \begin{multicols}{2}
            \begin{enumerate}
              \item\label{itm:lem:tr-LAdEEE-LAd:not:ndcom} $\ndcom(\lnot \varphi) = \ndcom(\varphi)$,
              \item\label{itm:lem:tr-LAdEEE-LAd:not:nscom} $\nscom(\lnot \varphi) > \nscom(\varphi)$.
            \end{enumerate}
          \end{multicols}
          Take any $\psi \in \ssub(\lnot \varphi) = \set{\varphi} \cup \ssub(\varphi)$, and consider the cases.
          \begin{compactitemize}
            \item \ti{Case $\psi = \varphi$}. By \autoref{itm:lem:tr-LAdEEE-LAd:not:ndcom} and \autoref{itm:lem:tr-LAdEEE-LAd:not:nscom}, it follows that $\com(\lnot \varphi) > \com(\psi)$.

            \item \ti{Case $\psi \in \ssub(\varphi)$}. By IH, $\com(\varphi) > \com(\psi)$ for any such $\psi$. By definition of $\com$, this implies either
            \begin{compactitemize}
              \item $\ndcom(\varphi) > \ndcom(\psi)$, so $\ndcom(\lnot \varphi) > \ndcom(\psi)$ (\autoref{itm:lem:tr-LAdEEE-LAd:not:ndcom}) hence $\com(\lnot \varphi) > \com(\psi)$, or

              \item $\ndcom(\varphi) = \ndcom(\psi)$ and $\nscom(\varphi) > \nscom(\psi)$, so $\ndcom(\lnot \varphi) = \ndcom(\psi)$ (by \autoref{itm:lem:tr-LAdEEE-LAd:not:ndcom}) and $\nscom(\lnot \varphi) > \nscom(\psi)$ (by \autoref{itm:lem:tr-LAdEEE-LAd:not:nscom}); hence, $\com(\lnot \varphi) > \com(\psi)$.
            \end{compactitemize}
          \end{compactitemize}

          \item \tb{Inductive case ($\bs{\varphi_1 \land \varphi_2}$)}. Note that, by definition, for $i \in \set{1,2}$,
          \begin{multicols}{2}
            \begin{enumerate}
              \item\label{itm:lem:tr-LAdEEE-LAd:and:ndcom} $\ndcom(\varphi_1 \land \varphi_2) \geqslant \ndcom(\varphi_i)$,
              \item\label{itm:lem:tr-LAdEEE-LAd:and:nscom} $\nscom(\varphi_1 \land \varphi_2) > \nscom(\varphi_i)$.
            \end{enumerate}
          \end{multicols}
          Take any $\psi \in \ssub(\varphi_1 \land \varphi_2) = \set{\varphi_1, \varphi_2} \cup \ssub(\varphi_1) \cup \ssub(\varphi_2)$.
          \begin{compactitemize}
            \item \ti{Case $\psi = \varphi_i$}. By \autoref{itm:lem:tr-LAdEEE-LAd:and:ndcom}, $\ndcom(\varphi_1 \land \varphi_2) \geqslant \ndcom(\psi)$. If $\ndcom(\varphi_1 \land \varphi_2) > \ndcom(\psi)$, then $\com(\varphi_1 \land \varphi_2) > \com(\psi)$ follows immediately; otherwise, $\ndcom(\varphi_1 \land \varphi_2) = \ndcom(\psi)$ and $\nscom(\varphi_1 \land \varphi_2) > \nscom(\psi)$ (\autoref{itm:lem:tr-LAdEEE-LAd:and:nscom}), so $\com(\varphi_1 \land \varphi_2) > \com(\psi)$ again.

            \item \ti{Case $\psi \in \ssub(\varphi_i)$}. By IH, $\com(\varphi_i) > \com(\psi)$ for any such $\psi$. Thus, either
            \begin{compactitemize}
              \item $\ndcom(\varphi_i) > \ndcom(\psi)$, so $\ndcom(\varphi_1 \land \varphi_2) > \ndcom(\psi)$ (by \autoref{itm:lem:tr-LAdEEE-LAd:and:ndcom}) and hence $\com(\varphi_1 \land \varphi_2) > \com(\psi)$, or

              \item $\ndcom(\varphi_i) = \ndcom(\psi)$ and $\nscom(\varphi_i) > \nscom(\psi)$. By \autoref{itm:lem:tr-LAdEEE-LAd:and:ndcom}, either $\ndcom(\varphi_1 \land \varphi_2) > \ndcom(\varphi_i)$ or $\ndcom(\varphi_1 \land \varphi_2) = \ndcom(\varphi_i)$. In the first case, $\ndcom(\varphi_1 \land \varphi_2) > \ndcom(\psi)$ and thus $\com(\varphi_1 \land \varphi_2) > \com(\psi)$; in the second case, $\ndcom(\varphi_1 \land \varphi_2) = \ndcom(\psi)$ and $\nscom(\varphi_1 \land \varphi_2) > \nscom(\psi)$ (using \autoref{itm:lem:tr-LAdEEE-LAd:and:nscom}), so $\com(\varphi_1 \land \varphi_2) > \com(\psi)$.
            \end{compactitemize}
          \end{compactitemize}

          \item \tb{Inductive case ($\bs{\mmDg{\varphi}}$)}. Note that, by definition,
          \begin{multicols}{2}
            \begin{enumerate}
              \item\label{itm:lem:tr-LAdEEE-LAd:box:ndcom} $\ndcom(\mmDg{\varphi}) = \ndcom(\varphi)$,
              \item\label{itm:lem:tr-LAdEEE-LAd:box:nscom} $\nscom(\mmDg{\varphi}) > \nscom(\varphi)$.
            \end{enumerate}
          \end{multicols}
          Take any $\psi \in \ssub(\mmDg{\varphi}) = \set{\varphi} \cup \ssub(\varphi)$. Thus, there are two cases.
          \begin{compactitemize}
            \item \ti{Case $\psi = \varphi$}. By \autoref{itm:lem:tr-LAdEEE-LAd:box:ndcom} and \autoref{itm:lem:tr-LAdEEE-LAd:box:nscom}, it follows that $\com(\mmDg{\varphi}) > \com(\psi)$.

            \item \ti{Case $\psi \in \ssub(\varphi)$}. By IH, $\com(\varphi) > \com(\psi)$ for any such $\psi$. This implies either
            \begin{compactitemize}
              \item $\ndcom(\varphi) {>} \ndcom(\psi)$, so $\ndcom(\mmDg{\varphi}) > \ndcom(\psi)$ (\autoref{itm:lem:tr-LAdEEE-LAd:box:ndcom}) then $\com(\mmDg{\varphi}) > \com(\psi)$, or

              \item $\ndcom(\varphi) = \ndcom(\psi)$ and $\nscom(\varphi) > \nscom(\psi)$, so $\ndcom(\mmDg{\varphi}) = \ndcom(\psi)$ (by \autoref{itm:lem:tr-LAdEEE-LAd:box:ndcom}) and $\nscom(\mmDg{\varphi}) > \nscom(\psi)$ (by \autoref{itm:lem:tr-LAdEEE-LAd:box:nscom}); hence, $\com(\mmDg{\varphi}) > \com(\psi)$.
            \end{compactitemize}
          \end{compactitemize}

          \item \tb{Inductive case ($\bs{\mmEEE{\varphi}}$)}. Note that, by definition,
          \begin{enumerate}
            \item\label{itm:lem:tr-LAdEEE-LAd:EEE:ndcom} $\ndcom(\mmEEE{\varphi}) > \ndcom(\varphi)$.
          \end{enumerate}
          Take any $\psi \in \ssub(\mmEEE{\varphi}) = \set{\varphi} \cup \ssub(\varphi)$. Thus, there are two cases.
          \begin{compactitemize}
            \item \ti{Case $\psi = \varphi$}. By \autoref{itm:lem:tr-LAdEEE-LAd:EEE:ndcom}, it follows that $\com(\mmEEE{\varphi}) > \com(\psi)$.

            \item \ti{Case $\psi \in \ssub(\varphi)$}. By IH, $\com(\varphi) > \com(\psi)$ for any such $\psi$. Then, either
            \begin{compactitemize}
              \item $\ndcom(\varphi) > \ndcom(\psi)$, so $\ndcom(\mmEEE{\varphi}) > \ndcom(\psi)$ (\autoref{itm:lem:tr-LAdEEE-LAd:EEE:ndcom}) then $\com(\mmEEE{\varphi}) > \com(\psi)$, or

              \item $\ndcom(\varphi) = \ndcom(\psi)$ and $\nscom(\varphi) > \nscom(\psi)$, so $\ndcom(\mmEEE{\varphi}) > \ndcom(\psi)$ (by \autoref{itm:lem:tr-LAdEEE-LAd:EEE:ndcom}) and hence $\com(\mmEEE{\varphi}) > \com(\psi)$.
            \end{compactitemize}
          \end{compactitemize}

        \end{compactitemize}
      \end{proof}
    \end{lema}

    Now, for the actual proposition, the proof proceeds by induction on $\com(\varphi)$, the complexity of $\varphi$, with both \autoref{pro:tr-LAdEEE-LAd:proper:i} and \autoref{pro:tr-LAdEEE-LAd:proper:ii} proved simultaneously.
    \begin{compactitemize}
      \item \tb{Base case ($\bs{\varphi}$ such that $\bs{\com(\varphi)}$ is minimum)}. A $\varphi$ with the minimum $\com$ should be minimum at both $\ndcom$ (i.e., $\ndcom(\varphi) = 0$, so any $\varphi$ without observation operators) and $\nscom$ (i.e., $\nscom(\varphi) = 1$). The only such formula is $p$. Proving \autoref{pro:tr-LAdEEE-LAd:proper:i} is immediate, as the definition of $\tr(p)$ does not use $\tr$; proving \autoref{pro:tr-LAdEEE-LAd:proper:ii} is also immediate, as $\tr(p) = p$ is a formula in \LAd.

      \item \tb{Inductive case ($\bs{\varphi}$ such that $\bs{\com(\varphi)}$ is not minimum), case $\bs{\lnot \varphi}$}. For \autoref{pro:tr-LAdEEE-LAd:proper:i}, the definition of $\tr(\lnot\varphi)$ uses $\tr(\varphi)$. But $\varphi \in \ssub(\lnot\varphi)$ so, by \autoref{lem:tr-LAdEEE-LAd:proper:ssub}, $\com(\lnot\varphi) > \com(\varphi)$. For \autoref{pro:tr-LAdEEE-LAd:proper:ii}, the same $\com(\lnot\varphi) > \com(\varphi)$ implies that, by IH, $\tr(\varphi) \in \LAd$; hence, so is $\lnot \tr(\varphi) = \tr(\lnot\varphi)$.

      \item \tb{Inductive case ($\bs{\varphi}$ such that $\bs{\com(\varphi)}$ is not minimum), case $\bs{\varphi_1 \land \varphi_2}$}. For \autoref{pro:tr-LAdEEE-LAd:proper:i}, the definition of $\tr(\varphi_1 \land \varphi_2)$ uses both $\tr(\varphi_1)$ and $\tr(\varphi_2)$. But $\varphi_i \in \ssub(\varphi_1 \land \varphi_2)$ for $i \in \set{1,2}$ so, by \autoref{lem:tr-LAdEEE-LAd:proper:ssub}, $\com(\varphi_1 \land \varphi_2) > \com(\varphi_i)$. For \autoref{pro:tr-LAdEEE-LAd:proper:ii}, the same $\com(\varphi_1 \land \varphi_2) > \com(\varphi_i)$ implies that, by IH, $\tr(\varphi_i) \in \LAd$; hence, so is $\tr(\varphi_1) \land \tr(\varphi_2) = \tr(\varphi_1 \land \varphi_2)$.

      \item \tb{Inductive case ($\bs{\varphi}$ such that $\bs{\com(\varphi)}$ is not minimum), case $\bs{\mmDg{\varphi}}$}. For \autoref{pro:tr-LAdEEE-LAd:proper:i}, the definition of $\tr(\mmDg{\varphi})$ uses $\tr(\varphi)$. But $\varphi \in \ssub(\mmDg{\varphi})$ so, by \autoref{lem:tr-LAdEEE-LAd:proper:ssub}, $\com(\mmDg{\varphi}) > \com(\varphi)$. For \autoref{pro:tr-LAdEEE-LAd:proper:ii}, the same $\com(\mmDg{\varphi}) > \com(\varphi)$ implies that, by IH, $\tr(\varphi) \in \LAd$; hence, so is $\mmDg{\tr(\varphi)} = \tr(\mmDg{\varphi})$.

      \item \tb{Inductive case ($\bs{\varphi}$ such that $\bs{\com(\varphi)}$ is not minimum), case $\bs{\mmEEE{p}}$}. For \autoref{pro:tr-LAdEEE-LAd:proper:i}, $\tr(\mmEEE{p})$ uses $\tr(p)$. But $p \in \ssub(\mmEEE{p})$, so $\com(\mmEEE{p}) > \com(p)$ (\autoref{lem:tr-LAdEEE-LAd:proper:ssub}). For \autoref{pro:tr-LAdEEE-LAd:proper:ii}, from the same $\com(\mmEEE{p}) > \com(p)$ and IH, $\tr(p) = \tr(\mmEEE{p}) \in \LAd$.

      \item \tb{Inductive case ($\bs{\varphi}$ such that $\bs{\com(\varphi)}$ is not minimum), case $\bs{\mmEEE{\lnot \varphi}}$}. For \autoref{pro:tr-LAdEEE-LAd:proper:i}, the definition of $\tr(\mmEEE{\lnot \varphi})$ uses $\tr(\lnot \mmEEE{\varphi})$. Now, on the one hand,
      \begin{multicols}{2}
        \begin{itemizecom}
          \item $\ndcom(\mmEEE{\lnot \varphi}) 
          = \bs{1 + \ndcom(\varphi)}$,
          \item $\ndcom(\lnot \mmEEE{\varphi}) 
          = \bs{1 + \ndcom(\varphi)}$,
        \end{itemizecom}
      \end{multicols}
      but, on the other hand,
      \begin{multicols}{2}
        \begin{itemizecom}
          \item $\nscom(\mmEEE{\lnot \varphi}) 
          = \bs{2 + 2\nscom(\varphi)}$,
          \item $\nscom(\lnot \mmEEE{\varphi}) 
          = \bs{1 + 2\nscom(\varphi)}$.
        \end{itemizecom}
      \end{multicols}
      Thus, $\com(\mmEEE{\lnot\varphi}) > \com(\lnot \mmEEE{\varphi})$. For \autoref{pro:tr-LAdEEE-LAd:proper:ii}, take the just obtained $\com(\mmEEE{\lnot \varphi}) > \com(\lnot \mmEEE{\varphi})$; then, by IH, $\tr(\lnot \mmEEE{\varphi}) = \tr(\mmEEE{\lnot\varphi}) \in \LAd$.

      \item \tb{Inductive case ($\bs{\varphi}$ such that $\bs{\com(\varphi)}$ is not minimum), case $\bs{\mmEEE{(\varphi_1 \land \varphi_2)}}$}. For \autoref{pro:tr-LAdEEE-LAd:proper:i}, the definition of $\tr(\mmEEE{(\varphi_1 \land \varphi_2)})$ uses $\tr(\mmEEE{\varphi_1} \land \mmEEE{\varphi_2})$. Now note that, on the one hand, by taking $\max\set{\ndcom(\varphi_1), \ndcom(\varphi_2)} = \ndcom(\varphi_i)$,
      \begin{multicols}{2}
        \begin{itemizecom}
          \item $\ndcom(\mmEEE{(\varphi_1 \land \varphi_2)}) 
          = \bs{1 + \ndcom(\varphi_i)}$,
          \item $\ndcom(\mmEEE{\varphi_1} \land \mmEEE{\varphi_2}) 
          = \bs{1 + \ndcom(\varphi_i)}$,
        \end{itemizecom}
      \end{multicols}
      but, on the other hand, by taking $\max\set{\nscom(\varphi_1), \nscom(\varphi_2)} = \nscom(\varphi_i)$,
      \begin{multicols}{2}
        \begin{itemizecom}
          \item $\nscom(\mmEEE{(\varphi_1 \land \varphi_2)}) 
          = \bs{2 + 2\nscom(\varphi_i)}$,
          \item $\nscom(\mmEEE{\varphi_1} \land \mmEEE{\varphi_2}) 
          = \bs{1 + 2\nscom(\varphi_i)}$.
        \end{itemizecom}
      \end{multicols}
      Thus, $\com(\mmEEE{(\varphi_1 \land \varphi_2)}) > \com(\mmEEE{\varphi_1} \land \mmEEE{\varphi_2})$. For \autoref{pro:tr-LAdEEE-LAd:proper:ii}, take the same $\com(\mmEEE{(\varphi_1 \land \varphi_2)}) > \com(\mmEEE{\varphi_1} \land \mmEEE{\varphi_2})$; then, by IH, $\tr(\mmEEE{\varphi_1} \land \mmEEE{\varphi_2}) = \tr(\mmEEE{(\varphi_1 \land \varphi_2)}) \in \LAd$.

      \item \tb{Inductive case ($\bs{\varphi}$ such that $\bs{\com(\varphi)}$ is not minimum), case $\bs{\mmEEE{\mmDg{\varphi}}}$}. For \autoref{pro:tr-LAdEEE-LAd:proper:i}, $\tr(\mmEEE{\mmDg{\varphi}})$ uses $\tr(\mmDs{\ag}{\mmEEE{\varphi}})$. Now, on the one hand,
      \begin{multicols}{2}
        \begin{itemizecom}
          \item $\ndcom(\mmEEE{\mmDg{\varphi}}) 
          = \bs{1 + \ndcom(\varphi)}$,
          \item $\ndcom(\mmDs{\ag}{\mmEEE{\varphi}}) 
          = \bs{1 + \ndcom(\varphi)}$,
        \end{itemizecom}
      \end{multicols}
      but, on the other hand,
      \begin{multicols}{2}
        \begin{itemizecom}
          \item $\nscom(\mmEEE{\mmDg{\varphi}}) 
          = \bs{2 + 2\nscom(\varphi)}$,
          \item $\nscom(\mmDs{\ag}{\mmEEE{\varphi}}) 
          = \bs{1 + 2\nscom(\varphi)}$.
        \end{itemizecom}
      \end{multicols}
      Thus, $\com(\mmEEE{\mmDg{\varphi}}) > \com(\mmDs{\ag}{\mmEEE{\varphi}})$. For \autoref{pro:tr-LAdEEE-LAd:proper:ii}, the just obtained $\com(\mmEEE{\mmDg{\varphi}}) > \com(\mmDs{\ag}{\mmEEE{\varphi}})$ implies that, by IH, $\tr(\mmDs{\ag}{\mmEEE{\varphi}}) = \tr(\mmEEE{\mmDg{\varphi}}) \in \LAd$.

      \item \tb{Inductive case ($\bs{\varphi}$ such that $\bs{\com(\varphi)}$ is not minimum), case $\bs{\mmEEE{\mmEEE{\varphi}}}$}. For \autoref{pro:tr-LAdEEE-LAd:proper:i}, the definition of $\tr(\mmEEE{\mmEEE{\varphi}})$ uses two instances of $\tr$, namely $\tr(\mmEEE{\varphi})$ and $\tr(\mmEEE{\tr(\mmEEE{\varphi}}))$. For the first, $\mmEEE{\varphi} \in \ssub(\mmEEE{\mmEEE{\varphi}})$ so, by \autoref{lem:tr-LAdEEE-LAd:proper:ssub}, $\com(\mmEEE{\mmEEE{\varphi}}) > \com(\mmEEE{\varphi})$. For the second, note that
      \begin{itemizecom}
        \item $\ndcom(\mmEEE{\mmEEE{\varphi}}) 
        = \bs{2 + \ndcom(\varphi)}$,
        \item $\ndcom(\mmEEE{\tr(\mmEEE{\varphi})}) = 1 + \ndcom(\tr(\mmEEE{\varphi}))$. But, as it has been shown, $\com(\mmEEE{\mmEEE{\varphi}}) > \com(\mmEEE{\varphi})$; thus, by IH, $\tr(\mmEEE{\varphi}) \in \LAd$ and therefore $\ndcom(\mmEEE{\varphi}) = 0$. Hence, $\ndcom(\mmEEE{\tr(\mmEEE{\varphi})}) = \bs{1}$.
      \end{itemizecom}
      Thus, $\com(\mmEEE{\mmEEE{\varphi}}) > \com(\mmEEE{\tr(\mmEEE{\varphi})})$. For \autoref{pro:tr-LAdEEE-LAd:proper:ii}, the just obtained $\com(\mmEEE{\mmEEE{\varphi}}) > \com(\mmEEE{\tr(\mmEEE{\varphi})})$ and IH imply $\tr(\mmEEE{\tr(\mmEEE{\varphi})}) = \tr(\mmEEE{\mmEEE{\varphi}}) \in \LAd$.
    \end{compactitemize}
  \end{proof}
\end{proposicion}

Using the ordering $\com$, the proposition below shows that a formula $\varphi \in \LAdEEE$ and its translation $\tr(\varphi) \in \LAd$ are both provably and semantically equivalent.

\begin{proposicion}\label{pro:tr-LAdEEE-LAd:works}
  For every $\varphi \in \LAdEEE$,
  \begin{multicols}{2}
    \begin{enumeratetr}
      \item\label{pro:tr-LAdEEE-LAd:works:i} $\vdash \varphi \ldimp \tr(\varphi)$ under \LOdEEE,
      \item\label{pro:tr-LAdEEE-LAd:works:ii} $\Vdash \varphi \ldimp \tr(\varphi)$
    \end{enumeratetr}
  \end{multicols}
  \begin{proof}
    By induction on $\com(\varphi)$. The following rule will be useful.
    \begin{lema}\label{lem:reDg}
      Let $\varphi_1, \varphi_2$ be formulas in \LAdEEE. Then, when using the system \LOd,
      \begin{ctabular}{c}
        if $\vdash \varphi_1 \ldimp \varphi_2$ then $\vdash \mmDg{\varphi_1} \ldimp \mmDg{\varphi_2}$.
      \end{ctabular}
      \begin{proof}
        Suppose $\vdash \varphi_1 \ldimp \varphi_2$. By propositional reasoning and modus ponens, $\vdash \varphi_1 \limp \varphi_2$ and $\vdash \varphi_2 \limp \varphi_1$, so $\vdash \mmDg{(\varphi_1 \limp \varphi_2)}$ and $\vdash \mmDg{(\varphi_2 \limp \varphi_1)}$ (by rule \ax{G}{}{\opgDist}). Then, from \ax{K}{}{\opgDist} and modus ponens, $\vdash \mmDg{\varphi_1} \limp \mmDg{\varphi_2}$ and $\vdash \mmDg{\varphi_2} \limp \mmDg{\varphi_1}$. Hence, by propositional reasoning and modus ponens, $\vdash \mmDg{\varphi_1} \ldimp \mmDg{\varphi_2}$.
      \end{proof}
    \end{lema}

    Here is the proof of the proposition.
    \begin{enumeratetr}
      \item Here are the cases.
      \begin{compactitemize}
        \item \tb{Base case ($\bs{p}$)}. By propositional reasoning, $\vdash p \ldimp p$. But $\tr(p) = p$ , so the required $\vdash p \ldimp \tr(p)$ follows.

        \item \tb{Inductive case ($\bs{\lnot \varphi}$)}. Since $\com(\lnot\varphi) > \com(\varphi)$ (same case in \autoref{pro:tr-LAdEEE-LAd:proper}), from IH it follows that $\vdash \varphi \ldimp \tr(\varphi)$. Then $\vdash \lnot \varphi \ldimp \lnot \tr(\varphi)$ (propositional reasoning) and thus, by $\tr$'s definition, $\vdash \lnot \varphi \ldimp \tr(\lnot \varphi)$.

        \item \tb{Inductive case ($\bs{\varphi_1 \land \varphi_2}$)}. Since $\com(\varphi_1 \land \varphi_2) > \com(\varphi_i)$ for $i \in \set{1,2}$ (same case in \autoref{pro:tr-LAdEEE-LAd:proper}), from IH it follows that $\vdash \varphi_i \ldimp \tr(\varphi_i)$. Then $\vdash (\varphi_1 \land \varphi_2) \ldimp (\tr(\varphi_1) \land \tr(\varphi_2))$ (propositional reasoning) and thus, by $\tr$'s definition, $\vdash (\varphi_1 \land \varphi_2) \ldimp \tr(\varphi_1 \land \varphi_2)$.

        \item \tb{Inductive case ($\bs{\mmDg{\varphi}}$)}. Since $\com(\mmDg{\varphi}) > \com(\varphi)$ (same case in \autoref{pro:tr-LAdEEE-LAd:proper}), from IH it follows that $\vdash \varphi \ldimp \tr(\varphi)$. Then $\vdash \mmDg{\varphi} \ldimp \mmDg{\tr(\varphi)}$ (\autoref{lem:reDg}, since \LOd is a subsystem of \LOdEEE) and thus, by $\tr$'s definition, $\vdash \mmDg{\varphi} \ldimp \tr(\mmDg{\varphi})$.

        \item \tb{Inductive case ($\bs{\mmEEE{p}}$)}. Since $\com(\mmEEE{p}) > \com(p)$ (see same case in \autoref{pro:tr-LAdEEE-LAd:proper}), from IH it follows that $\vdash p \ldimp \tr(p)$. But $\vdash \mmEEE{p} \ldimp p$ (axiom \dsaEEE{p}) so, by propositional reasoning, $\vdash \mmEEE{p} \ldimp \tr(p)$. Hence, by $\tr$'s definition, $\vdash \mmEEE{p} \ldimp \tr(\mmEEE{p})$.

        \item \tb{Inductive case ($\bs{\mmEEE{\lnot \varphi}}$)}. Since $\com(\mmEEE{\lnot \varphi}) > \com(\lnot \mmEEE{\varphi})$ (same case in \autoref{pro:tr-LAdEEE-LAd:proper}), from IH it follows that $\vdash \lnot \mmEEE{\varphi} \ldimp \tr(\lnot \mmEEE{\varphi})$. But $\vdash \mmEEE{\lnot\varphi} \ldimp \lnot \mmEEE{\varphi}$ (axiom \dsaEEE{\lnot}) so, by propositional reasoning, $\vdash \mmEEE{\lnot\varphi} \ldimp \tr(\lnot \mmEEE{\varphi})$. Hence, by $\tr$'s definition, $\vdash \mmEEE{\lnot\varphi} \ldimp \tr(\mmEEE{\lnot\varphi})$.

        \item \tb{Inductive case ($\bs{\mmEEE{(\varphi_1 \land \varphi_2)}}$)}. Since $\com(\mmEEE{(\varphi_1 \land \varphi_2)}) > \com(\mmEEE{\varphi_1} \land \mmEEE{\varphi_2})$ (same case in \autoref{pro:tr-LAdEEE-LAd:proper}), from IH it follows that $\vdash (\mmEEE{\varphi_1} \land \mmEEE{\varphi_2}) \ldimp \tr(\mmEEE{\varphi_1} \land \mmEEE{\varphi_2})$. But $\vdash \mmEEE{(\varphi_1 \land \varphi_2)} \ldimp (\mmEEE{\varphi_1} \land \mmEEE{\varphi_2})$ (axiom \dsaEEE{\land}) so, by propositional reasoning, $\vdash \mmEEE{(\varphi_1 \land \varphi_2)} \ldimp \tr(\mmEEE{\varphi_1} \land \mmEEE{\varphi_2})$. Hence, by $\tr$'s definition, $\vdash \mmEEE{(\varphi_1 \land \varphi_2)} \ldimp \tr(\mmEEE{(\varphi_1 \land \varphi_2)})$.

        \item \tb{Inductive case ($\bs{\mmEEE{\mmDg{\varphi}}}$)}. Since $\com(\mmEEE{\mmDg{\varphi}}) > \com(\mmDs{\ag}{\mmEEE{\varphi}})$ (same case in \autoref{pro:tr-LAdEEE-LAd:proper}), from IH it follows that $\vdash \mmDs{\ag}{\mmEEE{\varphi}} \ldimp \tr(\mmDs{\ag}{\mmEEE{\varphi}})$. But $\vdash \mmEEE{\mmDg{\varphi}} \ldimp \mmDs{\ag}{\mmEEE{\varphi}}$ (axiom \dsaEEE{\opgDist}) so, by propositional reasoning, $\vdash \mmEEE{\mmDg{\varphi}} \ldimp \tr(\mmDs{\ag}{\mmEEE{\varphi}})$. Hence, by $\tr$'s definition, $\vdash \mmEEE{\mmDg{\varphi}} \ldimp \tr(\mmEEE{\mmDg{\varphi}})$.

        \item \tb{Inductive case ($\bs{\mmEEE{\mmEEE{\varphi}}}$)}. Since $\com(\mmEEE{\mmEEE{\varphi}}) > \com(\mmEEE{\varphi})$ and $\com(\mmEEE{\mmEEE{\varphi}}) > \com(\mmEEE{\tr(\mmEEE{\varphi})})$ (same case in \autoref{pro:tr-LAdEEE-LAd:proper}), from IH it follows that $\vdash \mmEEE{\varphi} \ldimp \tr(\mmEEE{\varphi})$ and $\vdash \mmEEE{\tr(\mmEEE{\varphi})} \ldimp \tr(\mmEEE{\tr(\mmEEE{\varphi})})$. From the first and rule \dsreEEE, it follows that $\vdash \mmEEE{\mmEEE{\varphi}} \ldimp \mmEEE{\tr(\mmEEE{\varphi})}$. Hence, from the last two and propositional reasoning, $\vdash \mmEEE{\mmEEE{\varphi}} \ldimp \tr(\mmEEE{\tr(\mmEEE{\varphi})})$ and thus, by $\tr$'s definition, $\vdash \mmEEE{\mmEEE{\varphi}} \ldimp \tr(\mmEEE{\mmEEE{\varphi}})$.
      \end{compactitemize}

      \item By the previous item, $\vdash \varphi \ldimp \tr(\varphi)$. But, as it has been shown, \LOdEEE is sound for pointed \marm-models; therefore, $\Vdash \varphi \ldimp \tr(\varphi)$.
    \end{enumeratetr}
  \end{proof}
\end{proposicion}

Finally, the argument for strong completeness, which has three steps.\label{teo:compl:LOdEEE:three-steps}

\begin{compactenumerate}
  \item Take $\Psi \cup \set{\varphi} \subseteq \LAdEEE$ and suppose $\Psi \Vdash \varphi$, i.e., suppose that, for every pointed \marm-model $(M, w)$, if $(M, w) \Vdash \Psi$ then $(M, w) \Vdash \varphi$ or, in other words, for every such $(M, w)$,
  \[
    (M, w) \Vdash \psi \;\text{for all}\; \psi \in \Psi
    \qquad\text{implies}\qquad
    (M, w) \Vdash \varphi.
  \]
  Since $\Vdash \varphi' \ldimp \tr(\varphi')$ for every $\varphi' \in \LAdEEE$ \itemfautorefp{pro:tr-LAdEEE-LAd:works}{pro:tr-LAdEEE-LAd:works:ii}, it follows that, for every $(M, w)$,
  \begin{compactitemize}
    \item $(M, w) \Vdash \tr(\psi)$ for all $\psi \in \Psi$ if and only if $(M, w) \Vdash \psi$ for all $\psi \in \Psi$, and
    \item $(M, w) \Vdash \varphi$ if and only if $(M, w) \Vdash \tr(\varphi)$.
  \end{compactitemize}
  Thus, for every $(M, w)$,
  \[
    (M, w) \Vdash \tr(\psi) \;\text{for all}\; \psi \in \Psi
    \qquad\text{implies}\qquad
    (M, w) \Vdash \tr(\varphi).
  \]
  By defining $\tr(\Psi) := \set{\tr(\psi) \mid \psi \in \Psi}$, it follows that $(M, w) \Vdash \tr(\Psi)$ implies $(M, w) \Vdash \tr(\varphi)$ for every $(M, w)$; in other words, $\tr(\Psi) \Vdash \tr(\varphi)$.

  \item Since $\tr(\varphi') \in \LAd$ for every $\varphi' \in \LAdEEE$ \itemfautorefp{pro:tr-LAdEEE-LAd:proper}{pro:tr-LAdEEE-LAd:proper:ii}, it follows that $\tr(\Psi) \cup \set{\tr(\varphi)} \subseteq \LAd$; therefore, the just obtained $\tr(\Psi) \Vdash \tr(\varphi)$ and \autoref{teo:LOd} imply $\tr(\Psi) \vdash \tr(\varphi)$ under \LOd. Since \LOd is a subsystem of \LOdEEE, it follows that $\tr(\Psi) \vdash \tr(\varphi)$ under \LOdEEE.

  \item Since $\tr(\Psi) \vdash \tr(\varphi)$ under \LOdEEE, there are $\psi'_1, \ldots, \psi'_n \in \tr(\Psi)$ such that
  \[ \vdash \left(\psi'_1 \land \cdots \land \psi'_n\right) \limp \tr(\varphi). \]
  under \LOdEEE. Then, from the definition of $\tr(\Psi)$, there are $\psi_1, \ldots, \psi_n \in \Psi$ such that
  \[ \vdash \left(\tr(\psi_1) \land \cdots \land \tr(\psi_n)\right) \limp \tr(\varphi). \]
  under \LOdEEE. But $\vdash \varphi' \ldimp \tr(\varphi')$ for every $\varphi' \in \LAdEEE$ \itemfautorefp{pro:tr-LAdEEE-LAd:works}{pro:tr-LAdEEE-LAd:works:i}; hence, from $\Psi \cup \set{\varphi} \subseteq \LAdEEE$ (and using propositional reasoning for the first),
  \[
    \vdash \left(\psi_1 \land \cdots \land \psi_n\right) \limp \left(\tr(\psi_1) \land \cdots \land \tr(\psi_n)\right)
    \quad\text{and}\quad
    \vdash \tr(\varphi) \rightarrow \varphi.
  \]
  under \LOdEEE. Therefore,
  \[ \vdash \left(\psi_1 \land \cdots \land \psi_n\right) \limp \varphi \]
  and hence $\Psi \vdash \varphi$ under \LOdEEE, as required.
\end{compactenumerate}

\subsection{Proof of \autoref{teo:LOdSEE} (System \texorpdfstring{\LOdSEE}{LOdSEE})}\label{proof:teo:LOdSEE}

\nsparagraph{Soundness} Again, the soundness of axioms and rules in \LOd is well-known. For those in \autoref{tbl:LOdSEE}, the soundness of \dsaSEE{p}, \dsaSEE{\lnot}, \dsaSEE{\land} and \dsreSEE is as in the \EEE case (for the latter, recall that $\mopSEE{M}{\sen}$ is a model in \marm), and \dsaSEE{\opgDist} has been proved valid already \autorefp{pro:SEE}.

\bsparagraph{Completeness} The argument relies on the following translation.

\begin{definicion}[Translation]\label{def:trans:SEE}
  The translation $\tr$ is given by
  \begingroup
    \small
    \[
      \renewcommand{\arraystretch}{1.4}
      \begin{array}{c@{\qquad\qquad}c}
        \begin{array}{@{}r@{\;:=\;}l@{}}
          \tr(p)                         & p, \\
          \tr(\lnot\varphi)              & \lnot \tr(\varphi), \\
          \tr(\varphi_1 \land \varphi_2) & \tr(\varphi_1) \land \tr(\varphi_2), \\
          \tr(\mmDg{\varphi})            & \mmDg{\tr(\varphi)}, \\
          \multicolumn{2}{@{}r}{}
        \end{array}
        &
        \begin{array}{@{}r@{\;:=\;}l@{}}
          \tr(\mmSEE{\sen}{p})                           & \tr(p), \\
          \tr(\mmSEE{\sen}{\lnot \varphi})               & \tr(\lnot \mmSEE{\sen}{\varphi}), \\
          \tr(\mmSEE{\sen}{(\varphi_1 \land \varphi_2)}) & \tr(\mmSEE{\sen}{\varphi_1} \land \mmSEE{\sen}{\varphi_2}), \\
          \tr(\mmSEE{\sen}{\mmDg{\varphi}})              & \tr(\mmDs{\sen\cup\sag}{\mmSEE{\sen}{\varphi}}), \\
          \tr(\mmSEE{\sen_1}{\mmSEE{\sen_2}{\varphi}})   & \tr(\mmSEE{\sen_1}{\tr(\mmSEE{\sen_2}{\varphi})}). \\
        \end{array}
      \end{array}
    \]
  \endgroup
\end{definicion}

This translation works again in an ``inside-out' fashion, dealing first with the deepest occurrence of $\modSEE{\sen}$ before dealing with the rest. The strategy for proving strong completeness is as in the \EEE case: \begin{inlineenum} \item show that $\tr$ is a proper recursive translation that returns formulas in \LAd \autorefp{pro:tr-LAdSEE-LAd:proper}, \item show that a formula and its translation are both provably and semantically equivalent \autorefp{pro:tr-LAdSEE-LAd:works}, and \item use $\tr$ and the completeness of \LOd to show that, if $\Psi \cup \set{\varphi} \subseteq \LAdSEE$, then from $\Psi \Vdash \varphi$ it follows that $\varphi$ is derivable from $\Psi$ in \LOdSEE\end{inlineenum}.

\medskip

Here is the notion of complexity on which the proofs of \autoref{pro:tr-LAdSEE-LAd:proper} and \autoref{pro:tr-LAdSEE-LAd:works} rely.

\begin{definicion}[Complexity for \LAdSEE]\label{def:comp:SEE}
  The functions $\nscom:\LAdSEE \to \Nat\setminus\set{0}$ and $\ndcom:\LAdSEE \to \Nat$ are defined, for atoms, Boolean operators and the modality $\modDg$, as in the \EEE case \autorefp{def:comp:EEE}. For the dynamic operator $\modSEE{\sen}$, the cases are as for $\modEEE$:
  \begin{smallcbtabular}{@{}c@{\qquad\qquad}c@{}}
    \begin{tabular}{@{}r@{\;:=\;}l@{}}
      $\nscom(\mmSEE{\sen}{\varphi})$ & $2\nscom(\varphi)$, \\
    \end{tabular}
    &
    \begin{tabular}{@{}r@{\;:=\;}l@{}}
      $\ndcom(\mmSEE{\sen}{\varphi})$ & $1 + \ndcom(\varphi)$. \\
    \end{tabular}
  \end{smallcbtabular}
  Then define $\com$ as before: given $\varphi_1, \varphi_2 \in \LAdSEE$, write $\com(\varphi_1) > \com(\varphi_2)$ if and only if
  \begin{ctabular}{c}
    $\ndcom(\varphi_1) > \ndcom(\varphi_2)$
    \qtnq{or}
    $\ndcom(\varphi_1) = \ndcom(\varphi_2)$ and $\nscom(\varphi_1) > \nscom(\varphi_2)$.
  \end{ctabular}
\end{definicion}

\smallskip

First, $\tr$ is a proper recursive translation from \LAdSEE \emph{to \LAd}.

\begin{proposicion}\label{pro:tr-LAdSEE-LAd:proper}
  For every $\varphi \in \LAdSEE$,
  \begin{enumeratetr}
    \item\label{pro:tr-LAdSEE-LAd:proper:i} if $\tr(\varphi)$ is defined in terms of $\tr(\psi)$, then $\com(\varphi) > \com(\psi)$.
    \item\label{pro:tr-LAdSEE-LAd:proper:ii} $\tr(\varphi) \in \LAd$.
  \end{enumeratetr}
  \begin{proof}
    The following lemma will be useful.
    \begin{lema}\label{lem:tr-LAdSEE-LAd:proper:ssub}
      Let $\ssub:\LAdSEE \to \power{\LAdSEE}$ be the \emph{strict subformula} function for the language \LAdSEE (defined in the expected way); let $\varphi$ be an \LAdSEE-formula. Then, $\com(\varphi) > \com(\psi)$ for every $\psi \in \ssub(\varphi)$.
      \begin{proof}
        The proof is by structural induction on $\varphi$. Given that $\nscom$, $\ndcom$ and $\com$ are defined as for the \EEE instance, all cases here are exactly as in their \EEE counterpart \autorefp{lem:tr-LAdEEE-LAd:proper:ssub}.
      \end{proof}
    \end{lema}

    The proof of the proposition is by induction on $\com(\varphi)$ relying on \autoref{lem:tr-LAdSEE-LAd:proper:ssub}, just as in the \EEE case. Again, \autoref{pro:tr-LAdSEE-LAd:proper:i} and \autoref{pro:tr-LAdSEE-LAd:proper:ii} are proved simultaneously. Given that $\nscom$, $\ndcom$, $\com$ \emph{and $\tr$} are defined as for the \EEE instance, all cases here are exactly as in their \EEE counterpart \autorefp{pro:tr-LAdEEE-LAd:proper}.
  \end{proof}
\end{proposicion}

Then, a formula $\varphi \in \LAdSEE$ and its translation $\tr(\varphi) \in \LAd$ are both provably and semantically equivalent.

\begin{proposicion}\label{pro:tr-LAdSEE-LAd:works}
  For every $\varphi \in \LAdSEE$,
  \begin{multicols}{2}
    \begin{enumeratetr}
      \item\label{pro:tr-LAdSEE-LAd:works:i} $\vdash \varphi \ldimp \tr(\varphi)$,
      \item\label{pro:tr-LAdSEE-LAd:works:ii} $\Vdash \varphi \ldimp \tr(\varphi)$
    \end{enumeratetr}
  \end{multicols}
  \begin{proof}
    Here are the arguments.
    \begin{enumeratetr}
      \item The proof proceeds by induction on $\com(\varphi)$. Given \autoref{pro:tr-LAdSEE-LAd:proper} and the fact that $\tr$ is defined as in the \EEE instance, the base case ($\bs{p}$) and inductive cases $\bs{\lnot \varphi}$, $\bs{\varphi_1 \land \varphi_2}$ and $\bs{\mmDg{\varphi}}$ are as in \itemfautoref{pro:tr-LAdEEE-LAd:works}{pro:tr-LAdEEE-LAd:works:i}, the latter using \autoref{lem:reDg} and the fact that \LOd is a subsystem of \LOdSEE. The inductive cases $\bs{\mmSEE{\sen}{p}}$, $\bs{\mmSEE{\sen}{\lnot \varphi}}$, $\bs{\mmSEE{\sen}{(\varphi_1 \land \varphi_2)}}$, $\bs{\mmSEE{\sen}{\mmDg{\varphi}}}$ and $\bs{\mmSEE{\sen_1}{\mmSEE{\sen_2}{\varphi}}}$ are also as in \itemfautoref{pro:tr-LAdEEE-LAd:works}{pro:tr-LAdEEE-LAd:works:i} (relying on \autoref{pro:tr-LAdSEE-LAd:proper} and $\tr$'s definition), this time using axioms \dsaSEE{p}, \dsaSEE{\lnot}, \dsaSEE{\land}, \dsaSEE{\opgDist} and rule \dsreSEE, respectively.

      \item Exactly as in \itemfautoref{pro:tr-LAdEEE-LAd:works}{pro:tr-LAdEEE-LAd:works:ii}.
    \end{enumeratetr}
  \end{proof}
\end{proposicion}

Finally, the argument for strong completeness is as the \EEE case (\autopageref{teo:compl:LOdEEE:three-steps}), relying on \autoref{pro:tr-LAdSEE-LAd:proper} and \autoref{pro:tr-LAdSEE-LAd:works} instead.

\subsection{Proof of \autoref{teo:LOdSSE} (System \texorpdfstring{\LOdSSE}{LOdSSE})}\label{proof:teo:LOdSSE}

\nsparagraph{Soundness} The soundness of axioms and rules in \LOd is well-known. For those in \autoref{tbl:LOdSSE}, soundness of \dsaSSE{p}, \dsaSSE{\lnot}, \dsaSSE{\land} and \dsreSSE is as in the previous cases (for the latter, recall that $\mopSSE{M}{\sen}{\chi}$ is a model in \marm), and \dsaSSE{\opgDist} has been proved valid already \autorefp{pro:SSE}.

\bsparagraph{Completeness} The argument uses the following ``inside-out'' translation.

\begin{definicion}[Translation]\label{def:trans:SSE}
  The translation $\tr$ is given by
  \begingroup
    \small
    \[
      \renewcommand{\arraystretch}{1.4}
      \begin{array}{c@{\qquad}c}
        \begin{array}{@{}r@{\;:=\;}l@{}}
          \tr(p)                         & p, \\
          \tr(\lnot\varphi)              & \lnot \tr(\varphi), \\
          \tr(\varphi_1 \land \varphi_2) & \tr(\varphi_1) \land \tr(\varphi_2), \\
          \tr(\mmDg{\varphi})            & \mmDg{\tr(\varphi)}, \\
          \multicolumn{2}{@{}r}{}
        \end{array}
        &
        \begin{array}{@{}r@{\;:=\;}l@{}}
          \tr(\mmSSE{\sen}{\chi}{p})                           & \tr(p), \\
          \tr(\mmSSE{\sen}{\chi}{\lnot \varphi})               & \tr(\lnot \mmSSE{\sen}{\chi}{\varphi}), \\
          \tr(\mmSSE{\sen}{\chi}{(\varphi_1 \land \varphi_2)}) & \tr(\mmSSE{\sen}{\chi}{\varphi_1} \land \mmSSE{\sen}{\chi}{\varphi_2}), \\
          \tr(\mmSSE{\sen}{\chi}{\mmDg{\varphi}})              & \tr(\mmDs{\sen \cup \sag}{\mmSSE{\sen}{\chi}{\varphi}} \land \mmDsr{\sag}{\chi}{\mmSSE{\sen}{\chi}{\varphi}}), \\
          \tr(\mmSSE{\sen_1}{\chi_1}{\mmSSE{\sen_2}{\chi_2}{\varphi}}) & \tr(\mmSSE{\sen_1}{\chi_1}{\tr(\mmSSE{\sen_2}{\chi_2}{\varphi})}). \\
        \end{array}
      \end{array}
    \]
  \endgroup
\end{definicion}

Here is the notion of complexity on which the proofs of \autoref{pro:tr-LAdSSE-LAd:proper} and \autoref{pro:tr-LAdSSE-LAd:works} will rely.

\begin{definicion}[Complexity for \LAdSEE]\label{def:comp:SSE}
  The functions $\nscom:\LAdSSE \to \Nat\setminus\set{0}$ and $\ndcom:\LAdSSE \to \Nat$ are defined, for atoms, Boolean operators and the modality $\modDg$, as in the \EEE case \autorefp{def:comp:EEE}. The case of dynamic operator $\modSSE{\sen}{\chi}$, though, is different:
  \begin{smallcbtabular}{@{}c@{\qquad\qquad}c@{}}
    \begin{tabular}{@{}r@{\;:=\;}l@{}}
      $\nscom(\mmSSE{\sen}{\chi}{\varphi})$ & $\big(8 + \nscom(\chi)\big)\nscom(\varphi)$, \\
    \end{tabular}
    &
    \begin{tabular}{@{}r@{\;:=\;}l@{}}
      $\ndcom(\mmSSE{\sen}{\chi}{\varphi})$ & $1 + \ndcom(\chi) + \ndcom(\varphi)$. \\
    \end{tabular}
  \end{smallcbtabular}
  Define $\com$ as before: given $\varphi_1, \varphi_2 \in \LAdSSE$, write $\com(\varphi_1) > \com(\varphi_2)$ if and only if
  \begin{ctabular}{c}
    $\ndcom(\varphi_1) > \ndcom(\varphi_2)$
    \qtnq{or}
    $\ndcom(\varphi_1) = \ndcom(\varphi_2)$ and $\nscom(\varphi_1) > \nscom(\varphi_2)$.
  \end{ctabular}
\end{definicion}

\smallskip

First, $\tr$ is a proper recursive translation from \LAdSSE \emph{to \LAd}.

\begin{proposicion}\label{pro:tr-LAdSSE-LAd:proper}
  For every $\varphi \in \LAdSSE$,
  \begin{enumeratetr}
    \item\label{pro:tr-LAdSSE-LAd:proper:i} if $\tr(\varphi)$ is defined in terms of $\tr(\psi)$, then $\com(\varphi) > \com(\psi)$.
    \item\label{pro:tr-LAdSSE-LAd:proper:ii} $\tr(\varphi) \in \LAd$.
  \end{enumeratetr}
  \begin{proof}
    The following lemma will be useful.
    \begin{lema}\label{lem:tr-LAdSSE-LAd:proper:ssub}
      Let $\ssub:\LAdSSE \to \power{\LAdSSE}$ be the \emph{strict subformula} function for the language \LAdSSE (defined for atoms, Boolean operators and the $\modDg$ modality as before, and for $\modSSE{\sen}{\chi}$ as $\ssub(\mmSSE{\sen}{\chi}{\varphi}) := \set{\chi, \varphi} \cup \ssub(\chi) \cup \ssub(\varphi)$); let $\varphi$ be an \LAdSSE-formula. Then, $\com(\varphi) > \com(\psi)$ for every $\psi \in \ssub(\varphi)$.
      \begin{proof}
        The proof is by structural induction on $\varphi$. For atoms, Boolean operators and the $\modDg$ modality, the functions $\nscom$, $\ndcom$ and $\com$ are all defined as for the \EEE instance; thus, the cases for $p$, $\lnot \varphi$, $\varphi_1 \land \varphi_2$ and $\mmDg{\varphi}$ are exactly as in \autoref{lem:tr-LAdEEE-LAd:proper:ssub}. For the inductive case $\mmSSE{\sen}{\chi}{\varphi}$ note that, by definition,
        \begin{enumerate}
          \item\label{itm:lem:tr-LAdSSE-LAd:SSE:ndcom} $\ndcom(\mmSSE{\sen}{\chi}{\varphi}) > \ndcom(\chi)$ \;and\; $\ndcom(\mmSSE{\sen}{\chi}{\varphi}) > \ndcom(\varphi)$.
        \end{enumerate}
        Thus, take any $\psi \in \ssub(\mmSSE{\sen}{\chi}{\varphi}) = \set{\chi, \varphi} \cup \ssub(\chi) \cup \ssub(\varphi)$, and consider the cases.
        \begin{compactitemize}
          \item \ti{Cases $\psi = \varphi$ and $\psi = \chi$}. By \autoref{itm:lem:tr-LAdSSE-LAd:SSE:ndcom}, it follows that $\com(\mmSSE{\sen}{\chi}{\varphi}) > \com(\psi)$.

          \item \ti{Case $\psi \in \ssub(\chi)$}. By IH, $\com(\chi) > \com(\psi)$ for any such $\psi$. Then, by definition of $\com$, either
          \begin{compactitemize}
            \item $\ndcom(\chi) > \ndcom(\psi)$, so $\ndcom(\mmSSE{\sen}{\chi}{\varphi}) > \ndcom(\psi)$ (by \autoref{itm:lem:tr-LAdSSE-LAd:SSE:ndcom}) and therefore $\com(\mmSSE{\sen}{\chi}{\varphi}) > \com(\psi)$, or

            \item $\ndcom(\chi) = \ndcom(\psi)$ and $\nscom(\chi) > \nscom(\psi)$, so $\ndcom(\mmSSE{\sen}{\chi}{\chi}) > \ndcom(\psi)$ (by \autoref{itm:lem:tr-LAdSSE-LAd:SSE:ndcom}) and hence $\com(\mmSSE{\sen}{\chi}{\varphi}) > \com(\psi)$.
          \end{compactitemize}

          \item \ti{Case $\psi \in \ssub(\varphi)$}. Exactly as the previous one.
        \end{compactitemize}

      \end{proof}
    \end{lema}

    The proof of the proposition is by induction on $\com(\varphi)$ relying on \autoref{lem:tr-LAdSSE-LAd:proper:ssub}, analogous to the \EEE case. Again, \autoref{pro:tr-LAdSSE-LAd:proper:i} and \autoref{pro:tr-LAdSSE-LAd:proper:ii} are proved simultaneously. For atoms, Boolean operators and formulas of the form $\mmDg{\varphi}$, the functions $\nscom$, $\ndcom$, $\com$ \emph{and the translation $\tr$} are defined as for the \EEE instance; thus, cases $p$, $\lnot \varphi$, $\varphi_1 \land \varphi_2$ and $\mmDg{\varphi}$ are exactly as in \autoref{pro:tr-LAdEEE-LAd:proper}. Here are the remaining cases.

    \begin{compactitemize}
      \item \tb{Inductive case ($\bs{\varphi}$ such that $\bs{\com(\varphi)}$ is not minimum), case $\bs{\mmSSE{\sen}{\chi}{p}}$}. For \autoref{pro:tr-LAdSSE-LAd:proper:i}, the definition of $\tr(\mmSSE{\sen}{\chi}{p})$ uses $\tr(p)$. But $p \in \ssub(\mmSSE{\sen}{\chi}{p})$ so, by \autoref{lem:tr-LAdSSE-LAd:proper:ssub}, $\com(\mmSSE{\sen}{\chi}{p}) > \com(p)$. For \autoref{pro:tr-LAdSSE-LAd:proper:ii}, take the just obtained $\com(\mmSSE{\sen}{\chi}{p}) > \com(p)$; then, by IH, $\tr(p) = \tr(\mmSSE{\sen}{\chi}{p}) \in \LAd$.

      \item \tb{Inductive case ($\bs{\varphi}$ such that $\bs{\com(\varphi)}$ is not minimum), case $\bs{\mmSSE{\sen}{\chi}{\lnot \varphi}}$}. For \autoref{pro:tr-LAdSSE-LAd:proper:i}, the definition of $\tr(\mmSSE{\sen}{\chi}{\lnot \varphi})$ uses $\tr(\lnot \mmSSE{\sen}{\chi}{\varphi})$. Now note that, on the one hand,
      \begin{itemizecom}
        \item $\ndcom(\mmSSE{\sen}{\chi}{\lnot \varphi}) 
        = \bs{1 + \ndcom(\chi) + \ndcom(\varphi)}$,
        \item $\ndcom(\lnot \mmSSE{\sen}{\chi}{\varphi}) 
        = \bs{1 + \ndcom(\chi) + \ndcom(\varphi)}$,
      \end{itemizecom}
      but, on the other hand,
      \begin{itemizecom}
        \item $\nscom(\mmSSE{\sen}{\chi}{\lnot \varphi}) 
        = \bs{8 + 8\nscom(\varphi) + \nscom(\chi) + \nscom(\chi)\nscom(\varphi)}$,
        \item $\nscom(\lnot \mmSSE{\sen}{\chi}{\varphi}) 
        = \bs{1 + 8\nscom(\varphi) + \nscom(\chi)\nscom(\varphi)}$.
      \end{itemizecom}
      Thus, $\com(\mmSSE{\sen}{\chi}{\lnot\varphi}) > \com(\lnot \mmSSE{\sen}{\chi}{\varphi})$. For \autoref{pro:tr-LAdSSE-LAd:proper:ii}, the just obtained $\com(\mmSSE{\sen}{\chi}{\lnot \varphi}) > \com(\lnot \mmSSE{\sen}{\chi}{\varphi})$ implies, by IH, $\tr(\lnot \mmSSE{\sen}{\chi}{\varphi}) = \tr(\mmSSE{\sen}{\chi}{\lnot\varphi}) \in \LAd$.

      \item \tb{Inductive case ($\bs{\varphi}$ such that $\bs{\com(\varphi)}$ is not minimum), case $\bs{\mmSSE{\sen}{\chi}{(\varphi_1 \land \varphi_2)}}$}. For \autoref{pro:tr-LAdSSE-LAd:proper:i}, the definition of $\tr(\mmSSE{\sen}{\chi}{(\varphi_1 \land \varphi_2)})$ uses $\tr(\mmSSE{\sen}{\chi}{\varphi_1} \land \mmSSE{\sen}{\chi}{\varphi_2})$. Now, on the one hand, by taking $\max\set{\ndcom(\varphi_1), \ndcom(\varphi_2)} = \ndcom(\varphi_i)$,
      \begin{itemizecom}
        \item $\ndcom(\mmSSE{\sen}{\chi}{(\varphi_1 \land \varphi_2)}) 
        = \bs{1 + \ndcom(\chi) + \ndcom(\varphi_i)}$,
        \item $\ndcom(\mmSSE{\sen}{\chi}{\varphi_1} \land \mmSSE{\sen}{\chi}{\varphi_2}) 
        = \bs{1 + \ndcom(\chi) + \ndcom(\varphi_i)}$,
      \end{itemizecom}
      but, on the other hand, by taking $\max\set{\nscom(\varphi_1), \nscom(\varphi_2)} = \nscom(\varphi_i)$
      \begin{itemizecom}
        \item $\nscom(\mmSSE{\sen}{\chi}{(\varphi_1 \land \varphi_2)}) 
        = \bs{8 + 8\nscom(\varphi_i) + \nscom(\chi) + \nscom(\chi)\nscom(\varphi_i)}$,

        \item $\nscom(\mmSSE{\sen}{\chi}{\varphi_1} \land \mmSSE{\sen}{\chi}{\varphi_2}) 
        = \bs{1 + 8\nscom(\varphi_i) + \nscom(\chi)\nscom(\varphi_i)}$.
      \end{itemizecom}
      Thus, $\com(\mmSSE{\sen}{\chi}{(\varphi_1 \land \varphi_2)}) > \com(\mmSSE{\sen}{\chi}{\varphi_1} \land \mmSSE{\sen}{\chi}{\varphi_2})$. For \autoref{pro:tr-LAdSSE-LAd:proper:ii}, take the just obtained $\com(\mmSSE{\sen}{\chi}{(\varphi_1 \land \varphi_2)}) > \com(\mmSSE{\sen}{\chi}{\varphi_1} \land \mmSSE{\sen}{\chi}{\varphi_2})$; then, by IH, $\tr(\mmSSE{\sen}{\chi}{\varphi_1} \land \mmSSE{\sen}{\chi}{\varphi_2}) = \tr(\mmSSE{\sen}{\chi}{(\varphi_1 \land \varphi_2)}) \in \LAd$.

      \item \tb{Inductive case ($\bs{\varphi}$ such that $\bs{\com(\varphi)}$ is not minimum), case $\bs{\mmSSE{\sen}{\chi}{\mmDg{\varphi}}}$}. For \autoref{pro:tr-LAdSSE-LAd:proper:i}, the definition of $\tr(\mmSSE{\sen}{\chi}{\mmDg{\varphi}})$ uses $\tr(\mmDs{\sen \cup \sag}{\mmSSE{\sen}{\chi}{\varphi}} \land \mmDsr{\sag}{\chi}{\mmSSE{\sen}{\chi}{\varphi}})$. Now note that, on the one hand,
      \begin{itemizecom}
        \item $\ndcom(\mmSSE{\sen}{\chi}{\mmDg{\varphi}}) 
         = \bs{1 + \ndcom(\chi) + \ndcom(\varphi)}$,

         \item $\ndcom(\mmDs{\sen \cup \sag}{\mmSSE{\sen}{\chi}{\varphi}} \land \mmDsr{\sag}{\chi}{\mmSSE{\sen}{\chi}{\varphi}})
          = \bs{1 + \ndcom(\chi) + \ndcom(\varphi)}$,
      \end{itemizecom}
      but, on the other hand,
      \begin{itemizecom}
        \item $\nscom(\mmSSE{\sen}{\chi}{\mmDg{\varphi}}) 
        = \bs{8 + 8\nscom(\varphi) + \nscom(\chi) + \nscom(\chi)\nscom(\varphi)}$,

        \item $\nscom(\mmDs{\sen \cup \sag}{\mmSSE{\sen}{\chi}{\varphi}} \land \mmDsr{\sag}{\chi}{\mmSSE{\sen}{\chi}{\varphi}})
          = \bs{8 + 8\nscom(\varphi) + \nscom(\chi)\nscom(\varphi)}$.
      \end{itemizecom}
      Thus, $\com(\mmSSE{\sen}{\chi}{\mmDg{\varphi}}) > \com(\mmDs{\ag}{\mmSSE{\sen}{\chi}{\varphi}})$. This also yields \autoref{pro:tr-LAdEEE-LAd:proper:ii}, as from it and IH it follows that $\tr(\mmDs{\ag}{\mmSSE{\sen}{\chi}{\varphi}}) = \tr(\mmSSE{\sen}{\chi}{\mmDg{\varphi}}) \in \LAd$.

      \item \tb{Inductive case ($\bs{\varphi}$ such that $\bs{\com(\varphi)}$ is not minimum), case $\bs{\mmSSE{\sen_1}{\chi_1}{\mmSSE{\sen_2}{\chi_2}{\varphi}}}$}. For \autoref{pro:tr-LAdSSE-LAd:proper:i}, the definition of $\tr(\mmSSE{\sen_1}{\chi_1}{\mmSSE{\sen_2}{\chi_2}{\varphi}})$ uses two instances of $\tr$, namely $\tr(\mmSSE{\sen_2}{\chi_2}{\varphi})$ and $\tr(\mmSSE{\sen_1}{\chi_1}{\tr(\mmSSE{\sen_2}{\chi_2}{\varphi})})$. For the first, $\mmSSE{\sen_2}{\chi_2}{\varphi} \in \ssub(\mmSSE{\sen_1}{\chi_1}{\mmSSE{\sen_2}{\chi_2}{\varphi}})$ so, by \autoref{lem:tr-LAdSSE-LAd:proper:ssub}, $\com(\mmSSE{\sen_1}{\chi_1}{\mmSSE{\sen_2}{\chi_2}{\varphi}}) > \com(\mmSSE{\sen_2}{\chi_2}{\varphi})$. For the second, note that
      \begin{itemizecom}
        \item $\ndcom(\mmSSE{\sen_1}{\chi_1}{\mmSSE{\sen_2}{\chi_2}{\varphi}}) 
        = \bs{2 + \ndcom(\chi_1) + \ndcom(\chi_2) + \ndcom(\varphi)}$,

        \item $\ndcom(\mmSSE{\sen_1}{\chi_1}{\tr(\mmSSE{\sen_2}{\chi_2}{\varphi})}) = 1 + \ndcom(\chi_1) + \ndcom(\tr(\mmSSE{\sen_2}{\chi_2}{\varphi}))$. But, as it has been shown, $\com(\mmSSE{\sen_1}{\chi_1}{\mmSSE{\sen_2}{\chi_2}{\varphi}}) > \com(\mmSSE{\sen_2}{\chi_2}{\varphi})$; thus, by IH, $\tr(\mmSSE{\sen_2}{\chi_2}{\varphi}) \in \LAd$ and therefore $\ndcom(\mmSSE{\sen_2}{\chi_2}{\varphi}) = 0$. Hence, $\ndcom(\mmSSE{\sen_1}{\chi_1}{\tr(\mmSSE{\sen_2}{\chi_2}{\varphi})}) = \bs{1 + \ndcom(\chi_1)}$.
      \end{itemizecom}
      Thus, $\com(\mmSSE{\sen_1}{\chi_1}{\mmSSE{\sen_2}{\chi_2}{\varphi}}) > \com(\mmSSE{\sen_1}{\chi_1}{\tr(\mmSSE{\sen_2}{\chi_2}{\varphi})})$. This also yields \autoref{pro:tr-LAdSSE-LAd:proper:ii} as, from it and IH, $\tr(\mmSSE{\sen_1}{\chi_1}{\tr(\mmSSE{\sen_2}{\chi_2}{\varphi})}) = \tr(\mmSSE{\sen_1}{\chi_1}{\mmSSE{\sen_2}{\chi_2}{\varphi}}) \in \LAd$.
    \end{compactitemize}
  \end{proof}
\end{proposicion}

Then, a formula $\varphi \in \LAdSSE$ and its translation $\tr(\varphi) \in \LAd$ are both provably and semantically equivalent.

\begin{proposicion}\label{pro:tr-LAdSSE-LAd:works}
  For every $\varphi \in \LAdSSE$,
  \begin{multicols}{2}
    \begin{enumeratetr}
      \item\label{pro:tr-LAdSSE-LAd:works:i} $\vdash \varphi \ldimp \tr(\varphi)$,
      \item\label{pro:tr-LAdSSE-LAd:works:ii} $\Vdash \varphi \ldimp \tr(\varphi)$
    \end{enumeratetr}
  \end{multicols}
  \begin{proof}
    Here are the arguments.
    \begin{enumeratetr}
      \item The proof proceeds by induction on $\com(\varphi)$. Given \autoref{pro:tr-LAdSSE-LAd:proper} and the fact that $\tr$ is defined as in the \EEE instance, the base case ($\bs{p}$) and inductive cases $\bs{\lnot \varphi}$, $\bs{\varphi_1 \land \varphi_2}$ and $\bs{\mmDg{\varphi}}$ are as in \itemfautoref{pro:tr-LAdEEE-LAd:works}{pro:tr-LAdEEE-LAd:works:i}, the latter using \autoref{lem:reDg} and the fact that \LOd is a subsystem of \LOdSSE. The inductive cases $\bs{\mmSSE{\sen}{\chi}{p}}$, $\bs{\mmSSE{\sen}{\chi}{\lnot \varphi}}$, $\bs{\mmSSE{\sen}{\chi}{(\varphi_1 \land \varphi_2)}}$, $\bs{\mmSSE{\sen}{\chi}{\mmDg{\varphi}}}$ and $\bs{\mmSSE{\sen_1}{\chi_1}{\mmSSE{\sen_2}{\chi_2}{\varphi}}}$ are also as in \itemfautoref{pro:tr-LAdEEE-LAd:works}{pro:tr-LAdEEE-LAd:works:i} (relying on \autoref{pro:tr-LAdSSE-LAd:proper} and $\tr$'s definition), this time using axioms \dsaSSE{p}, \dsaSSE{\lnot}, \dsaSSE{\land}, \dsaSSE{\opgDist} and rule \dsreSSE, respectively.

      \item Exactly as in \itemfautoref{pro:tr-LAdEEE-LAd:works}{pro:tr-LAdEEE-LAd:works:ii}.
    \end{enumeratetr}
  \end{proof}
\end{proposicion}

Finally, the argument for strong completeness is as the \EEE case (\autopageref{teo:compl:LOdEEE:three-steps}), relying on \autoref{pro:tr-LAdSSE-LAd:proper} and \autoref{pro:tr-LAdSSE-LAd:works} instead.

\settocbibname{References}
\bibliographystyle{../../myabbrvnatnew}
\bibliography{../../biblio(en)}

\end{document}